\documentclass[prd,nofootinbib,floatfix,preprintnumbers, reprint, superscriptaddress, showpacs]{revtex4-1}%
\pdfoutput=1 

\usepackage{amssymb,amsmath}
\usepackage{color}
\usepackage{graphicx}
\usepackage{comment}
\usepackage{epsfig}
\usepackage{latexsym} 
\usepackage{multirow}
\usepackage{slashed}
\usepackage{rotating} 
\usepackage{dcolumn}

\definecolor{refkey}{rgb}{0,0,.75}
\definecolor{labelkey}{rgb}{0,0.75,0}

\newcommand{\pref}{\vec{p}_{\text{ref}}}
\newcommand{\LG}{\mathrm{LG}}

\begin{document}

\preprint{JLAB-THY-12-1504}
\preprint{TCDMATH 12-03}

\title{$S$ and $D$-wave phase shifts in isospin-2 $\pi\pi$ scattering from lattice QCD}

\author{Jozef~J.~Dudek}
\email{dudek@jlab.org}
\affiliation{Jefferson Laboratory, 12000 Jefferson Avenue,  Newport News, VA 23606, USA}
\affiliation{Department of Physics, Old Dominion University, Norfolk, VA 23529, USA}

\author{Robert~G.~Edwards}
\email{edwards@jlab.org}
\affiliation{Jefferson Laboratory, 12000 Jefferson Avenue,  Newport News, VA 23606, USA}

\author{Christopher~E.~Thomas}
\email{thomasc@maths.tcd.ie}
\affiliation{School of Mathematics, Trinity College, Dublin 2, Ireland}

\collaboration{for the Hadron Spectrum Collaboration}

\begin{abstract}

The isospin-2 $\pi\pi$ system provides a useful testing ground for determining elastic hadron scattering parameters from finite-volume spectra obtained using lattice QCD computations. A reliable determination of the excited state spectrum of two pions in a cubic box follows from variational analysis of correlator matrices constructed using a large basis of operators. A general operator construction is presented which respects the symmetries of a multi-hadron system in flight. This is applied to the case of $\pi\pi$ and allows for the determination of the scattering phase-shifts at a large number of kinematic points, in both $S$-wave and $D$-wave, within the elastic region. The technique is demonstrated with a calculation at a pion mass of 396 MeV, where the elastic scattering is found to be well described by a scattering length parameterisation.
\end{abstract}


\maketitle

\section{Introduction}\label{sec:intro}

QCD is the accepted underlying theory of the strong interactions, and the properties of the spectrum and interactions of hadrons should be calculable from it using a suitable regularization scheme for the quark and gluon fields. A particularly convenient approach is to consider the theory on a finite lattice of space-time points so as to admit a numerical method of solution.
While significant progress has been made recently in determining the single particle spectrum of hadrons, describing the resonances seen in scattering experiments in terms of eigenstates of QCD has remained a challenge to lattice calculations. Direct access to the matrix elements related to decays is missing in the Euclidean formulations of lattice QCD. In principle, the relevant hadronic matrix elements can be inferred indirectly through a detailed study of the spectrum in a finite-volume lattice box \cite{DeWitt:1956be,Luscher:1991cf}. Within this approach, one can map the discrete spectrum of eigenstates of the finite volume theory to the infinite volume scattering parameters, and if present, observe resonant behavior. 

Crucial to this approach is the high-precision determination of multiple excited eigenstate energies with a given quantum number. Determination of the discrete spectrum of finite-volume eigenstates follows from analysis of the time-dependence of two-point correlation functions featuring operators of the desired quantum numbers constructed from quark and gluon fields. For creation and annihilation at time $0$ and $t$ respectively we have
\begin{equation}
C_{ij}(t) = \big\langle 0 \big| {\cal O}_i(t) {\cal O}_j^\dag(0) \big| 0
\big\rangle . \nonumber
\end{equation}
Inserting a complete set of eigenstates of the Hamiltonian, this correlator has a spectral decomposition
\begin{equation}
C_{ij}(t) = \sum_\mathfrak{n}  \big\langle 0  \big|  {\cal O}_i \big| \mathfrak{n} \big\rangle
 \big\langle \mathfrak{n} \big| {\cal O}_j^\dag \big| 0 \big\rangle\, e^{- E_\mathfrak{n} t} , \label{spectral}
\end{equation}
where the sum is over all states that have the same quantum numbers as the interpolating operators $ {\cal O}_i, \, \mathcal{O}_j$.  Note that in a finite volume, this yields a discrete set of energies, $E_\mathfrak{n}$. It is these finite-volume energies that are related to infinite volume scattering amplitudes through the L\"uscher method \cite{Luscher:1991cf}.

A relatively straightforward sector in which to study hadron scattering in finite-volume is $\pi\pi$ in isospin-2. At low energies, this channel is experimentally observed to be non-resonant in low partial-waves \cite{Hoogland:1977kt,Cohen:1973yx,Zieminski:1974ex,Durusoy:1973aj} and this lack of resonances ensures a slow variation of phase-shifts with energy. This makes the problem of determining the phase-shift as a function of energy somewhat easier. A difficulty of this choice of channel is that the interaction between pions in isospin 2 is weak so that the discrete energies in finite-volume are shifted relatively little from the values relevant for non-interacting pions. This will require us to make precision measurements of the energy spectrum in order to resolve the differences. Within the field-theory, the correlators for this channel do not contain any annihilation contributions; the only Wick contractions featuring are those in which the four quark fields in the creation operator (at $t=0$) propagate in time to the annihilation operator (at $t$). The absence of quark propagation from $t$ to $t$ reduces the computational overhead for the calculation.

In a previous publication \cite{Dudek:2010ew} we presented the first lattice QCD study of the energy-dependence of $S$ and $D$-wave $\pi\pi$ scattering in isospin-2. We limited ourselves to the $\pi\pi$ system overall at rest and found only a handful of points below the $4\pi$ inelastic threshold on the lattice volumes considered. In this paper we will also consider the $\pi\pi$ system ``in-flight", that is with an overall momentum (satisfying the periodic boundary conditions of the finite cubic lattice). This allows us to determine the phase-shifts at a larger number of discrete energies below the $4\pi$ inelastic threshold and to map out the energy dependence of the scattering in more detail. The price to be paid is that the relevant symmetry group in the lattice calculation is significantly reduced. At rest the lattice has a cubic symmetry whose irreducible representations (``irreps") contain multiple angular momenta, e.g. the ``scalar" representation, $A_1^+$ contains as well as $\ell=0$, also $\ell=4$ and higher. In-flight, with two pions having total momentum, $\vec{P}$, the symmetry is restricted to rotations and reflections which leave the cubic lattice and the axis defined by $\vec{P}$ invariant. The irreps of this symmetry group are even less sparse in $\pi\pi$ scattering angular momentum; the ``scalar" representations typically contain $\ell=0,2,4\ldots$. In this work we will consider the effect these higher partial waves have on the determination of scattering phase-shifts for the lowest $\ell$ values.  

In \cite{Dudek:2010ew}, we used only the simplest $\bar{\psi}\gamma_5\psi$ interpolators in construction of $\pi\pi$ correlators. Single-pion correlators constructed with these operators are saturated by the ground state only at rather large times, and similarly the $\pi\pi$ correlators receive significant contributions from excited $\pi^\star$ states. The need to consider correlators at large times increases the degree to which we feel the systematic effect of the finite temporal extent of the lattice ($T$). Limited account was taken of these effects in \cite{Dudek:2010ew}. In this paper we take steps to address finite-$T$ effects, firstly by using ``optimised" pion operators which are saturated by the ground state pion at earlier times, and secondly by explicitly attempting to remove the leading effects of finite-$T$ from the measured correlators. While these effects are small in absolute terms, determination of the rather weak $I=2$ interaction relies upon precise measurement of small energy shifts, and as such it is important to account for even small systematic effects.

Our approach to determining the finite-volume spectrum is to use a large basis of operators in each symmetry channel with which we form a matrix of correlation functions having all relevant operators at the source \emph{and} sink. This matrix can be analysed variationally\cite{Michael:1985ne,Luscher:1990ck,Blossier:2009kd}, extracting a spectrum of energy eigenstates which are orthogonal in the space of operators used. This orthogonality is particularly useful in cases where levels are close to degenerate and to extract relatively high-lying states whose contribution to any single correlation function may be small relative to the ground state. The excited single-hadron spectrum of isovector and isoscalar mesons \cite{Dudek:2009qf,Dudek:2010wm,Dudek:2011tt} and baryons \cite{Edwards:2011jj,Dudek:2012ag} has been extracted with some success using this procedure. In the present case we require a basis of operators capable of interpolating a pair of pions from the vacuum, constructed to transform irreducibly in the relevant symmetry group. The fact that $I=2$ is expected to have only relatively weak inter-pion interaction strength suggests a natural basis might be one resembling pairs of non-interacting pions, i.e. pions of definite momentum.

In general our $\pi\pi$ creation operators have the form
\begin{equation}
	\big( \pi\pi \big)_{\vec{P}, \Lambda,\mu}^{[\vec{k}_1, \vec{k}_2]\dag} = \sum
	_{\substack{\vec{k}_1, \vec{k}_2 \\ \vec{k}_1 + \vec{k}_2 = \vec{P} }}  \mathcal{C}(\vec{P},\Lambda,\mu; \vec{k}_1; \vec{k}_2 )\; \pi^\dag(\vec{k}_1)\, \pi^\dag(\vec{k}_2) \nonumber
\end{equation}
here $\mathcal{C}$ are the Clebsch-Gordon coefficients for combining the two pion operators of definite momentum $\vec{k}_1$, $\vec{k}_2$  so that the operator overall transforms in the irrep $\Lambda$ of the relevant symmetry group for total momentum $\vec{P} = \vec{k}_1 + \vec{k}_2$. This involves summing over multiple values of momenta $\vec{k}_1$, $\vec{k}_2$ with the same magnitudes, $|\vec{k}_1|$, $|\vec{k}_2|$ and related by allowed lattice rotations.  The basis is built up out of different magnitudes of pion momenta that can sum to give the same $\vec{P}$. Much greater detail will be presented later in this paper.

Using this basis we compute correlators within various irreps $\Lambda$ for various $\vec{P}$,
\begin{equation}
C^{\vec{P}, \Lambda, \mu}_{[\vec{k}'_1,\vec{k}'_2],[\vec{k}_1,\vec{k}_2]}\big( t \big)  = \left\langle \big(\pi\pi\big)_{\vec{P},\Lambda,\mu}^{[\vec{k}'_1, \vec{k}'_2]}(t) \cdot \big(\pi\pi\big)_{\vec{P},\Lambda,\mu}^{[\vec{k}_1, \vec{k}_2]\dag}(0) \right\rangle \nonumber
\end{equation}
and for a fixed $\vec{P},\Lambda,\mu$ we perform variational analysis in a basis of operators labeled by $[\vec{k}_1,\vec{k}_2]$ leading to a finite-volume spectrum, $E_\mathfrak{n}(\vec{P},\Lambda; L)$. This spectrum, determined in the rest frame of the lattice, corresponds to a discrete set of scattering momenta, $p_\mathsf{cm}$, in the center-of-momentum frame.


The finite-volume spectrum so obtained is related through the L\"uscher formalism \cite{Luscher:1990ck,Luscher:1991cf} (as extended in \cite{Rummukainen:1995vs,Kim:2005gf,Christ:2005gi} to the case of moving frames) to the phase-shifts, $\delta_\ell(p_\mathsf{cm})$, for elastic $\pi\pi$ scattering in partial waves of angular momentum, $\ell$. As discussed earlier, a given irrep $\Lambda$ of momentum $\vec{P}$, contains multiple angular momenta, $\ell$, and the formalism relates the finite-volume spectrum to the scattering amplitudes for all relevant $\ell$ though the following formula:
\begin{equation}
\det\left[ e^{2i \boldsymbol{\delta}(p_\mathsf{cm})}- \mathbf{U}^{(\vec{P},\Lambda)}\big( p_\mathsf{cm}\tfrac{L}{2\pi} \big) \right] = 0 \label{luescher_intro}
\end{equation}
Here $\mathbf{U}^{(\vec{P},\Lambda)}\big( p_\mathsf{cm}\tfrac{L}{2\pi} \big)$ is a matrix of known functions and $e^{2i \boldsymbol{\delta}(p_\mathsf{cm})}$ is a diagonal matrix featuring the scattering phase-shifts $\{ \delta_\ell \}$. In both cases the rows and columns of the matrices are labelled by the angular momenta, $\ell$, relevant for the irrep $(\vec{P},\Lambda)$. These matrices are formally infinite, but we may take advantage  of the hierarchy $\delta_0 \gg \delta_2 \gg \delta_4 \ldots$ relevant at low energies\footnote{near threshold, angular momentum conservation requires $\delta_\ell \sim p_\mathsf{cm}^{2\ell + 1}$}, that tends to reduce the effect of higher $\ell$ in Equation \ref{luescher_intro}.

We will explore two methods to extract the phase shifts. The first method, similar to the one used in \cite{Dudek:2010ew}, exploits the above hierarchy to determine the phase-shift in the lowest contributing partial wave and estimates a systematic uncertainty from plausible variation of the higher partial waves.\footnote{we note that Ref.~\cite{Leskovec:2012gb} has recently discussed a similar approach} The second method parameterizes the momentum-dependence of the phase-shifts in $\ell=0,2\ldots$ using effective range expansions, then by performing a global fit which attempts to describe many finite-volume momentum points in many irreps, finds the values of the effective range expansion parameters.

Computations were performed on anisotropic lattices with three dynamical flavors of Clover fermions~\cite{Edwards:2008ja,Lin:2008pr} with spatial lattice spacing $a_s\sim 0.12\,$fm, and a temporal lattice spacing approximately $3.5$ times smaller, corresponding to a temporal scale $a_t^{-1}\sim 5.6$ GeV. This fine temporal lattice spacing has proven useful in determining the spectrum of mesons and baryons, as well as the previous $\pi\pi$ $I=2$ results~\cite{Dudek:2010ew}. In this work, results are presented for the light quark $I=2$ spectrum at quark mass parameter $a_t m_l=-0.0840$ and $a_t m_s=-0.0743$ corresponding to a pion mass of $396$ MeV, and at lattice sizes of $16^3\times 128$, $20^3\times 128$ and $24^3\times 128$ with corresponding spatial extents $L \sim 2\,$fm, $\sim 2.5\,$fm and $\sim 3\,$fm.  Some details of the lattices and propagators used for correlation constructions are provided in Table~\ref{tab:lattices}. 

Recently, the NPLQCD collaboration \cite{Beane:2011sc} has determined the $\ell=0$ scattering phase-shift on the same ensembles as used in this study, plus an additional larger lattice volume $\sim 4 \,\mathrm{fm}$. Their calculation is limited in scope by the fact that their approach does not project pion operators of definite relative momentum at the source. We will compare the results of the different approaches later in this paper. Other studies of $\pi\pi$ $I=2$ scattering in lattice QCD (\cite{Beane:2007xs, Sasaki:2008sv, Feng:2009ij}) have largely limited themselves to the threshold behavior of the scattering amplitude in $S$-wave, as expressed by the scattering length.

Readers who are not concerned with the details of the calculation can skip to Section \ref{sec:results} where the results for elastic scattering are presented. The remainder of the paper is structured as follows:

Section \ref{sec:multi} outlines the construction of a basis of irreducible $\pi\pi$ operators at rest and in-flight from products of pion operators of definite momentum. Section \ref{sec:dist} describes the construction of correlators using the distillation framework. Section \ref{sec:projection} presents ``optimised" single pion operators constructed as linear combinations of composite QCD operators with pion quantum numbers. Section \ref{sec:dispersion} discusses the determination of the pion mass and anisotropy from measurements of the pion dispersion relation. Section \ref{sec:finiteT} considers the effects of the finite temporal extent of the lattice on $\pi\pi$ correlators and presents mechanisms for reducing the role of these effects in the determination of the discrete energy spectrum. Section \ref{sec:spectrum} presents the finite-volume spectrum obtained on three volumes. Section \ref{sec:luescher} discusses the extraction of elastic scattering phase-shifts from finite-volume spectra using the L\"uscher formalism including a study of the possible effect of sub-leading partial waves within a toy model. Section \ref{sec:results} presents our results for $\delta_{0,2}$ in the region of elastic scattering for 396 MeV pions. Section \ref{sec:summary} summarises our approach and results and suggests future applications of the methodology.

\begin{table}[t]
  \begin{tabular}{c|ccc}
    $(L/a_s)^3 \times (T/a_t)$   &$N_{\mathrm{cfgs}}$ & $N_{\mathrm{t_{srcs}}}$ & $N_{\mathrm{vecs}}$ \\
    \hline 
         $16^3\times 128$ & 479 & 12 & 64 \\
     $20^3\times 128$ & 601 & $\substack{5\; (\vec{P}=\vec{0}) \\ 3\; (\vec{P}\neq\vec{0})  }$ & 128\\
     $24^3\times 128$ & 553 & 3 & 162 \\
  \end{tabular}
  \caption{The lattice ensembles and propagators used in this paper. The light and strange quark mass are $a_t m_l=-0.0840$ and $a_t m_s=-0.0743$ described in Ref.~\cite{Lin:2008pr}, corresponding to a pion mass of $396$ MeV. The lattice size and number of configurations are listed, as well as the number of time-sources and the number of distillation vectors $N_{\mathrm{vecs}}$ (to be described in Sec \ref{sec:dist}) featuring in the correlator construction.}
\label{tab:lattices}
\end{table}


\section{Operator construction}\label{sec:multi}

In order to calculate scattering amplitudes we must extract multi-hadron energy levels with high precision and so need interpolating operators that efficiently interpolate these multi-hadron states.  To achieve this we consider operators constructed from the product of two hadron operators projected onto definite total momentum, $\vec{P}$, and transforming as a definite irreducible representation of the appropriate symmetry group, \emph{lattice irrep}\footnote{we use `lattice irrep' to refer to the octahedral group irrep for a particle at rest and the irrep of the appropriate little group, discussed later, for a particle at non-zero momentum} $\Lambda$, with irrep row, $\mu$,
\begin{eqnarray}
\left[\mathbb{O}_{\Lambda\mu}(\vec{P})\right]^{\dagger} &=& 
\sum_{\substack{\mu_1,\mu_2 \\ \vec{k}_1 , \vec{k}_2 \\ \vec{k}_1 + \vec{k}_2 = \vec{P}}}
~\mathcal{C}(\vec{P}\Lambda\mu; \vec{k}_1\Lambda_1\mu_1; \vec{k}_2\Lambda_2\mu_2 ) \nonumber \\
 &&\quad \times \left[\mathbb{O}_{\Lambda_1\mu_1}(\vec{k}_1)\right]^{\dagger}~
\left[\mathbb{O}_{\Lambda_2\mu_2}(\vec{k}_2)\right]^{\dagger} ~. \label{lat_two_part}
\end{eqnarray}
Here $\mathbb{O}_{\Lambda_1,\mu_1}(\vec{k}_1)$ and $\mathbb{O}_{\Lambda_2,\mu_2}(\vec{k}_2)$ are hadron operators (for example, fermion bilinear operators), each projected onto definite momentum, irrep and irrep row.  The Clebsch-Gordan coefficients, $\mathcal{C}$, and the momenta appearing in the sum over $\vec{k}_1$ and $\vec{k}_2$ will be discussed later.  

A conventional infinite volume continuum analogue of this construction (for total momentum zero with $\vec{p} = \vec{k}_1 = -\vec{k}_2$) would be
\begin{eqnarray}
\left[\mathcal{O}^{[S,\ell]}_{J,M}\right]^\dagger &\sim& \sum_{\lambda_1 \lambda_2} \vspace{-4mm} \int d\hat{p} ~\; C(J \ell S M; \vec{p}\,S_1 \lambda_1; -\vec{p}\,S_2 \lambda_2) \nonumber \\
 && \quad \times \left[\mathcal{O}^{S_1\lambda_1}(\vec{p})\right]^\dagger \left[\mathcal{O}^{S_2\lambda_2}(-\vec{p})\right]^\dagger  ~ ,
\label{equ:cont_two_part}
\end{eqnarray}
with
\begin{equation*}
C = \big\langle S_1\lambda_1; S_2 -\!\!\lambda_2 \big| S\lambda\big\rangle \big\langle \ell0;S\lambda\big|J\lambda\big\rangle\, D_{M \lambda}^{(J)*}(\hat{p}) ~,
\end{equation*}
where $D(\hat{p})$ is a Wigner-$D$ matrix and $S_{1,2}$ and $\lambda_{1,2}$ are respectively the spins and helicities of hadron 1,2.  The spins are coupled to $S = S_1 \otimes S_2$, $\ell$ is the partial wave, $J = \ell \otimes S$ is the total angular momentum and $M$ is its $z$ component.  However, in all but the simplest cases, multi-hadron operators constructed by subducing Eq.~(\ref{equ:cont_two_part}) into irreducible representations of the lattice symmetry can mix single-hadron operators transforming in different lattice irreps.  Therefore, we prefer Eq.~(\ref{lat_two_part}) where such mixings do not occur.  The single-hadron operators transforming in definite lattice irreps can be optimised variationally, as shown for the pion in Section \ref{sec:projection}. 

Here we concentrate on the operators to be used to study two-pion states; the generalisation to other multi-hadron states is given in Appendix \ref{app:operators}.  The flavor structure of the operators, for example the projection of $\pi\pi$ onto definite overall isospin, $I$, determines which combinations of Wick contractions appear in the calculation of the correlators (Section \ref{sec:dist}).  Because this flavor structure generally factorises from the spin and spatial structure we will not discuss it in detail here.  However, because we are considering two identical pions, Bose symmetry requires the overall wavefunction to be symmetric under the interchange of the two pions.  Therefore, in the $I=2$ case we are considering here or $I=0$, the symmetric flavor part requires a symmetric spatial part (even partial waves with positive parity).  In contrast, $I=1$ requires an antisymmetric spatial piece (odd partial waves with negative parity).  In addition, these operators have definite charge-conjugation parity, $C = +1$, for neutral combinations, generalising to $G$-parity for charged combinations; for brevity, in the following we omit the $C$-parity labels.

\subsection{Single-hadron operators}
\label{sec:singleops}

Respecting the reduced symmetry of a finite cubic lattice, the $J^{P} = 0^-$ pion at rest \emph{subduces} onto the one-dimensional $\Lambda^{P} = A_1^-$ irrep of the double-cover octahedral group with parity, $\text{O}^{\text{D}}_h$.  In Refs.~\cite{Dudek:2009qf,Dudek:2010wm} we discussed how operators with a definite continuum $J^P$ and $J_z$-component $M$, $\mathcal{O}^{J^P,M}(\vec{k}=\vec{0})$, can be constructed out of fermion bilinears featuring gauge-covariant derivatives and Dirac gamma matrices; the extension to baryons was described in Ref.~\cite{Edwards:2011jj}.  The appropriate lattice operators were formed by \emph{subducing} these continuum operators into octahedral group irreps.  Table \ref{table:latticeirreps} summarises how different integer continuum $J$ subduce into octahedral group irreps -- here we focus on the irreps relevant for mesons but the discussion applies equally to the irreps appropriate for half-integer spin.  In the case of a $J^P=0^-$ operator subducing to $\Lambda^P=A_1^-$ this subduction is trivial, $$\mathcal{O}^{[0^-]}_{A_1^-}(\vec{0}) = \mathcal{O}^{0^-}(\vec{0}) ~.$$

At non-zero momentum, $\vec{k}$, the symmetry is reduced further: the relevant symmetry group is the \emph{little group}, the subgroup of allowed transformations which leave $\vec{k}$ invariant~\cite{Moore:2005dw}.  In an infinite volume continuum the little group is the same for each $\vec{k}$; with only the constraints arising from rotational symmetry, states are now labelled by the magnitude of helicity, $|\lambda|$, rather than $J$.  On a finite cubic lattice with periodic boundary conditions the allowed momenta are quantised, $\vec{k} = \tfrac{2\pi}{L}(n,m,p)$ where $n,m,p$ are integers, and in general there are different little groups for different types of momentum.  We denote the little group for $\vec{k}$ by $\LG(\vec{k})$ and for convenience define $\LG(\vec{0}) = \text{O}^{\text{D}}_h$.  The pion subduces onto the one-dimensional $\Lambda = A_2$ irrep of the appropriate little group (at least for all $|\vec{k}|^2 < 14 \left(\tfrac{2\pi}{L}\right)^2$).  Table \ref{table:latticeirreps} shows the pattern of subductions of the helicities into the little group irreps.  In Ref.~\cite{Thomas:2011rh} we presented a method to construct subduced helicity operators, $\mathbb{O}^{[J^P,|\lambda|]}_{\Lambda,\mu}(\vec{k})$, and showed that these are useful for studying mesons with non-zero momentum on the lattice.  For a $J^P=0^-$ operator subduced into the $A_2$ irrep the construction is again trivial,
$$\mathbb{O}^{[0^-,0]}_{A_2}(\vec{k}) = \mathcal{O}^{0^-}(\vec{k}) ~.$$

\begin{table}
\begin{ruledtabular}
\begin{tabular}{c c | c l}
$\vec{P}$ & $\LG(\vec{P})$ & $\Lambda^{P}$ & \multicolumn{1}{c}{$J^{P}$}  \\
\hline \hline
\multirow{5}{*}{$[0,0,0]$} & \multirow{5}{*}{$\text{O}^{\text{D}}_h$} 
   & $A_1^{\pm}$ & $0^{\pm},~ 4^{\pm},~ \ldots$ \\
 & & $T_1^{\pm}$ & $1^{\pm},~ 3^{\pm},~ 4^{\pm},~ \ldots$ \\
 & & $T_2^{\pm}$ & $2^{\pm},~ 3^{\pm},~ 4^{\pm},~ \ldots$ \\
 & & $E^{\pm}$   & $2^{\pm},~ 4^{\pm},~ \ldots$ \\
 & & $A_2^{\pm}$ & $3^{\pm},~ \ldots$ \\
\end{tabular}
\end{ruledtabular}

\vspace{.5cm}

\begin{ruledtabular}
\begin{tabular}{c c | c l}
$\vec{P}$ & $\LG(\vec{P})$ & $\Lambda$ & \multicolumn{1}{c}{$|\lambda|^{(\tilde{\eta})}$} \\
\hline \hline
\multirow{5}{*}{$[0,0,n]$} & \multirow{5}{*}{$\text{Dic}_4$} 
   & $A_1$ & $0^+,~ 4,~ \ldots$ \\
 & & $A_2$ & $0^-,~ 4,~ \ldots$ \\
 & & $E_2$ & $1,~ 3,~ \ldots$ \\
 & & $B_1$ & $2,~ \ldots$ \\
 & & $B_2$ & $2,~ \ldots$ \\
\hline
\multirow{4}{*}{$[0,n,n]$} & \multirow{4}{*}{$\text{Dic}_2$} 
   & $A_1$ & $0^+,~ 2,~ 4,~ \ldots$ \\
 & & $A_2$ & $0^-,~ 2,~ 4,~ \ldots$ \\
 & & $B_1$ & $1,~ 3,~ \ldots$ \\
 & & $B_2$ & $1,~ 3,~ \ldots$ \\
\hline
\multirow{3}{*}{$[n,n,n]$} & \multirow{3}{*}{$\text{Dic}_3$} 
   & $A_1$ & $0^+,~ 3,~ \ldots$ \\
 & & $A_2$ & $0^-,~ 3,~ \ldots$ \\
 & & $E_2$ & $1,~ 2,~ 4,~ \ldots$ \\
\hline
$[n,m,0]$  & \multirow{2}{*}{$\text{C}_4$} & $A_1$    & $0^+,~ 1,~ 2,~ 3,~ 4,~ \ldots$ \\
 $[n,n,m]$ & & $A_2$ & $0^-,~ 1,~ 2,~ 3,~ 4,~ \ldots$ \\
\end{tabular}
\end{ruledtabular}
\caption{The pattern of subductions of the continuum spin, $J \le 4$, (for $\vec{P} = \vec{0}$) and helicity, $|\lambda| \le 4$, (for $\vec{P} \ne \vec{0}$) into lattice irreps, $\Lambda$~\cite{Moore:2005dw}.  Here $\tilde{\eta} \equiv P(-1)^J$, $\vec{P}$ is given in units of $\tfrac{2\pi}{L}$ and $n,m$ are non-zero integers with $n \ne m$.  We show the double-cover groups but only give the irreps relevant for integer spin.}
\label{table:latticeirreps}
\end{table}

When we use the variational method to find the optimal linear combination of operators to interpolate a pion, we will include in the basis operators of other $J$ subduced into $A_1^-$ (for $\vec{k} = \vec{0}$) and other helicity $\lambda$ subduced into $A_2$ (for $\vec{k} \ne \vec{0}$).  The pattern of subductions is given in Table \ref{table:latticeirreps}; the subduction coefficients for zero momentum are given in Ref.~\cite{Dudek:2010wm} and those for non-zero momentum are given in Appendix \ref{app:operators}.  Henceforth, we will use $\pi(\vec{k})$ as a shorthand to represent $\mathcal{O}_{A_1^-}(\vec{k} = \vec{0})$  or $\mathbb{O}_{A_2}(\vec{k} \ne \vec{0})$ as appropriate.

\subsection{Multi-hadron operators}

In general, a $\pi\pi$ creation operator can be constructed from the product of two single-pion creation operators,
\begin{equation}
\label{equ:twopionop}
\left(\pi\pi\right)^{\left[\vec{k}_1,\vec{k}_2\right]\dagger}_{\vec{P},\Lambda,\mu} = \sum_{\substack{\vec{k}_1 \in \{\vec{k}_1\}^{\star} \\ \vec{k}_2 \in \{\vec{k}_2\}^{\star} \\ \vec{k}_1 + \vec{k}_2 = \vec{P}}}
\mathcal{C}(\vec{P},\Lambda,\mu;\; \vec{k}_1;\,\vec{k}_2)\; \pi^{\dagger}(\vec{k}_1)\; \pi^{\dagger}(\vec{k}_2) ~,
\end{equation}
where $\pi(\vec{k})$ is a single-pion operator and $\mathcal{C}$ is a Clebsch-Gordan coefficient for $\Lambda_1 \otimes \Lambda_2 \rightarrow \Lambda$ with $\Lambda_{1,2} = A_1^-$ of $\text{O}^{\text{D}}_h$ if $\vec{k}_{1,2} = \vec{0}$ and $\Lambda_{1,2} = A_2$ of $\LG(\vec{k}_{1,2})$ if $\vec{k}_{1,2} \neq \vec{0}$, and where $\Lambda$ is an irrep of $\LG(\vec{P})$.  For present purposes, the particular construction of $\pi(\vec{k})$ from quark and gluon fields is not important.  It is only necessary that $\pi(\vec{k})$ transforms in the appropriate lattice irrep.  

The sum over $\vec{k}_{1,2}$ is a sum over all momenta in the \emph{stars} of $\vec{k}_{1,2}$, which we denote by $\{\vec{k}_{1,2}\}^\star$, and by which we mean all momenta related to $\vec{k}_{1,2}$ by an allowed lattice rotation.  In other words, the sum is over $R\,\vec{k}_{1,2} ~ \forall ~ R \in \text{O}^{\text{D}}_h$; the restriction that $\vec{k}_1 + \vec{k}_2 = \vec{P}$ is equivalent to requiring $R \in \LG(\vec{P})$.  We will write $\vec{k}_1$, $\vec{k}_2$ and $\vec{P}$ in units of $\tfrac{2\pi}{L}$, using square braces to indicate the suppression of the dimensionful factor, i.e. $\vec{P}=[1,0,0]$ denotes a momentum of $\tfrac{2\pi}{L}(1,0,0)$.

The Clebsch-Gordan coefficients, $\mathcal{C}$, can be determined by a group theoretic construction.  When $\vec{P} = \vec{k}_1 = \vec{k}_2 = \vec{0}$, there is only one momentum direction in the sum and $\mathcal{C}$ are just the usual Clebsch-Gordan coefficients for $\text{O}^{\text{D}}_h$~\cite{Basak:2005aq}.  In the case of two pions the only relevant Clebsch-Gordan is the trivial $A_1^- \otimes A_1^- \rightarrow A_1^+$, $\mathcal{C}=1$, giving a two-pion operator in the $A_1^+$ irrep.

For the two-pion system with $\vec{k}_1 \ne \vec{0}$ but overall at rest, $\vec{P} = \vec{k}_1 + \vec{k}_2 = \vec{0}$, $\vec{k}_2=-\vec{k}_1$, the Clebsch-Gordans required are those for $A_2 (\{\vec{k}_1\}^\star) \otimes A_2 (\{\vec{k}_2\}^\star) \rightarrow \Lambda^P$ with $A_2$ of $\LG(\vec{k}_1)$ and $\Lambda^P$ of $\text{O}^{\text{D}}_h$.  The irreps, $\Lambda^P$, arising are given in Ref.~\cite{Moore:2006ng} and summarised in Table \ref{table:twopionops}.  We discuss how to calculate the corresponding explicit Clebsch-Gordan coefficients using the induced representation and give values in Appendix \ref{app:operators}.

For the remaining case, $\vec{P} \ne \vec{0}$, we require the Clebsch-Gordan coefficients for $A_2 (\{\vec{k}_1\}^\star) \otimes A_2(\{\vec{k}_2\}^\star) \rightarrow \Lambda$, or if $\vec{k}_2 = \vec{0}$, $A_2 (\{\vec{k}_1\}^\star) \otimes A_1^- (\vec{0}) \rightarrow \Lambda$ and correspondingly for $\vec{k}_1 = \vec{0}$.  Again, these are calculated using the induced representation as discussed in Appendix \ref{app:operators} and we give the irreps which arise in Table \ref{table:twopionops}.

In this work we restrict ourselves to $\vec{P} = [0,0,0]$, $[0,0,1]$, $[0,1,1]$ and $[1,1,1]$, and the various combinations of $\vec{k}_1$ and $\vec{k}_2$ used are given in Table \ref{table:twopionops}.  Because the two pions are identical bosons, Bose symmetry requires them to be symmetric under interchange and we only use operators with the correct symmetry for isospin-2.  However, for completeness, those operators with the wrong symmetry are shown in parentheses in the table.  

We want to use these operator constructions at both the source and the sink of correlation functions.  This requires us to be able to project single-pion operators onto a given momentum at arbitrary times, something that can be achieved efficiently using the \emph{distillation} methodology~\cite{Peardon:2009gh}.

\begin{table}
\begin{ruledtabular}
\begin{tabular}{c | l l l}
$\vec{P}$ & \quad$\vec{k}_1$ & \quad$\vec{k}_2$ & \multicolumn{1}{c}{$\Lambda^{(P)}$} \\
\hline \hline
\multirow{5}{*}{$\begin{matrix}[0,0,0]\\ \text{O}^{\text{D}}_h \end{matrix}$}
 & $[0,0,0]$ & $[0,0,0]$ & $A_1^+$ \\
 & $[0,0,1]$ & $[0,0,\text{-}1]$ & $A_1^+$, $E^+$, ($T_1^-$) \\
 & $[0,1,1]$ & $[0,\text{-}1,\text{-}1]$ & $A_1^+$, $T_2^+$, $E^+$, ($T_1^-$, $T_2^-$) \\
 & $[1,1,1]$ & $[\text{-}1,\text{-}1,\text{-}1]$ & $A_1^+$, $T_2^+$, ($T_1^-$, $A_2^-$) \\
 & $[0,0,2]$ & $[0,0,\text{-}2]$ & $A_1^+$, $E^+$, ($T_1^-$) \\
\hline
\multirow{7}{*}{$\begin{matrix}[0,0,1]\\ \text{Dic}_4 \end{matrix}$}
 & $[0,0,0]$ & $[0,0,1]$ & $A_1$ \\
 & $[0,\text{-}1,0]$ & $[0,1,1]$ & $A_1$, $E_2$, $B_1$ \\
 & $[\text{-}1,\text{-}1,0]$ & $[1,1,1]$ & $A_1$, $E_2$, $B_2$ \\
 & $[0,0,\text{-}1]$ & $[0,0,2]$ & $A_1$ \\
 & $[0,\text{-}1,\text{-}1]$ & $[0,1,2]$ & $A_1$, $E_2$, $B_1$ \\
 & $[\text{-}2,0,0]$ & $[2,0,1]$ & $A_1$, $E_2$, $B_1$ \\
 & $[\text{-}1,\text{-}1,\text{-}1]$ & $[1,1,2]$ & $A_1$, $E_2$, $B_2$ \\
\hline
\multirow{8}{*}{$\begin{matrix}[0,1,1]\\ \text{Dic}_2 \end{matrix}$}
 & $[0,0,0]$ & $[0,1,1]$ & $A_1$ \\
 & $[0,1,0]$ & $[0,0,1]$ & $A_1$, ($B_1$) \\
 & $[\text{-}1,0,0]$ & $[1,1,1]$ & $A_1$, $B_2$ \\
 & $[1,1,0]$ & $[\text{-}1,0,1]$ & $A_1$, $A_2$, ($B_1$, $B_2$) \\
 & $[0,1,\text{-}1]$ & $[0,0,2]$ & $A_1$, $B_1$ \\
 & $[0,\text{-}1,0]$ & $[0,2,1]$ & $A_1$, $B_1$ \\
 & $[1,\text{-}1,1]$ & $[\text{-}1,2,0]$ & $A_1$, $A_2$, $B_1$, $B_2$ \\
 & $[1,\text{-}1,0]$ & $[\text{-}1,2,1]$ & $A_1$, $A_2$, $B_1$, $B_2$ \\
\hline
\multirow{5}{*}{$\begin{matrix}[1,1,1]\\ \text{Dic}_3 \end{matrix}$}
 & $[0,0,0]$ & $[1,1,1]$ & $A_1$ \\
 & $[1,0,0]$ & $[0,1,1]$ & $A_1$, $E_2$ \\
 & $[2,0,0]$ & $[\text{-}1,1,1]$ & $A_1$, $E_2$ \\
 & $[1,\text{-}1,0]$ & $[0,2,1]$ & $A_1$, $A_2$, $2E_2$ \\
 & $[\text{-}1,0,0]$ & $[2,1,1]$ & $A_1$, $E_2$ \\
\end{tabular}
\end{ruledtabular}

\caption{The two-pion operators for each $\vec{P}$; also shown is $\LG(\vec{P})$ -- we show the double-cover groups but only give the irreps relevant for integer spin.  Example momenta $\vec{k}_1$ and $\vec{k}_2$ are shown; all momenta in $\{\vec{k}_1\}^{\star}$ and $\{\vec{k}_2\}^{\star}$ are summed over in Eq.~\ref{equ:twopionop}.  Swapping around $\vec{k}_1$ and $\vec{k}_2$ gives the same operators up to an overall phase.  The irreps given in parentheses do not occur for two identical bosons with a symmetric flavour coupling (e.g. $\pi\pi$ in $I=0$ or $2$) because of the constraints arising from Bose symmetry.}
\label{table:twopionops}
\end{table}


\section{Distillation and correlator construction}\label{sec:dist}

Within \emph{distillation}~\cite{Peardon:2009gh}, we construct operators capable of interpolating a single pion of momentum $\vec{k}$ from the vacuum as
\begin{equation}
\pi^\dag(\vec{k},t) = \sum_{\vec{x}} e^{i \vec{k}\cdot\vec{x}}\left[\bar\psi\Box_\sigma\boldsymbol{\Gamma}^\dagger_t\Box_\sigma\psi\right](\vec{x},t),
\label{eq:distop}
\end{equation}
where the $\boldsymbol{\Gamma}_t$ are, in general, operators acting in space, color and Dirac spin on a time slice, $t$, whose explicit construction is described in detail in Ref.~\cite{Thomas:2011rh}. The quark fields $\psi$ in Equation \ref{eq:distop} are acted upon by a distillation smearing operator $\Box_\sigma$ that emphasizes the low momentum quark and gluon modes that dominate low mass hadrons. This smearing operator is defined as 
\begin{equation}
\Box^{ij}_\sigma(\vec{x},\vec{y};t) = \sum_{n=1}^{N_{\mathrm{vecs}}} e^{\sigma^2\lambda_n/4}\xi^i_n(\vec{x},t)\xi^{j\dagger}_n(\vec{y},t)
\label{eq:box}
\end{equation}
where $\lambda_n$, $\xi^i_n(\vec{x},t)$ are the $n^\mathrm{th}$ eigenvalue and eigenvector (in color, $i$, and position, $\vec{x}$) of the gauge-covariant three-dimensional Laplacian operator on a time-slice, $t$. In the present study, the smearing weight $\sigma$ is set to 0 and the number of vectors used is $N_\mathrm{vecs}=64,\,128,\,162$ on the $L/a_s = 16,\,20,\,24$ lattices respectively (a shorthand $\Box$ is used to represent $\Box_{\sigma=0}$).

The outer-product nature of the distillation smearing operator is such that correlators can be factorized into products of factors containing only propagation and factors containing only operator construction. The propagation factors, $\tau$ (called ``perambulators''), and momentum projected operators, $\Phi$, are constructed as matrices in the space of the eigenvectors (the distillation space): where $\tau_{nm}(t',t) = \xi^\dagger_n(t') M^{-1}(t',t)\xi_m(t)$ and $\Phi_{nm}(t)=\xi^\dagger_n(t) \boldsymbol{\Gamma}_t\xi_m(t)$, and $M$ is the lattice representation of the Clover-Dirac operator for the light quarks used in this study. 

As outlined in Section~\ref{sec:multi}, two-hadron operators 
are constructed from sums over products of two single-hadron operators of definite momentum, as in Eq.~\ref{equ:twopionop}. The resulting correlators for the $\pi\pi$ operators are of the generic form
\begin{equation}
C_{ij}(t',t) = \langle 0|\big(\pi\pi\big)_i(t')\cdot \big(\pi\pi\big)^\dagger_j(t)|0\rangle,
\label{eq:corr}
\end{equation}
where each operator $\pi$ is of the bilinear form given in Equation \ref{eq:distop}. For isospin-2, quark integration leads to only Wick contractions featuring quark propagation from source time $t$ to sink time $t'$; there are no annihilation contributions. The resulting traces are over the set of eigenvectors used in Equation \ref{eq:box} which is much smaller than the full lattice space, allowing for the efficient computation of the correlation functions. In particular, it is the factorization of the smearing operator that allows for the projection of both the source and sink operators onto definite inter-pion momentum, something that is not possible in the traditional ``point-all'' method. This factorization allows for the construction of the full hermitian correlation matrix among source and sink operators in Eq.~\ref{eq:corr}, and hence makes possible the application of the variational method \cite{Michael:1985ne,Luscher:1990ck,Blossier:2009kd}. In this method, the manifest orthogonality among states provides the essential key for determining high lying excited states and separating nearly degenerate states. 

To increase statistics, the correlation functions in Eq.~\ref{eq:corr} are averaged over multiple time sources. The number of time sources, along with the number of eigenvectors of the Laplacian, $N_\mathrm{vecs}$, and the number of configurations for each of the three volumes used in this study are shown in Table~\ref{tab:lattices}.


\section{Optimised pion operators}\label{sec:projection}

\begin{figure}[b]
\includegraphics[width=0.5\textwidth
]{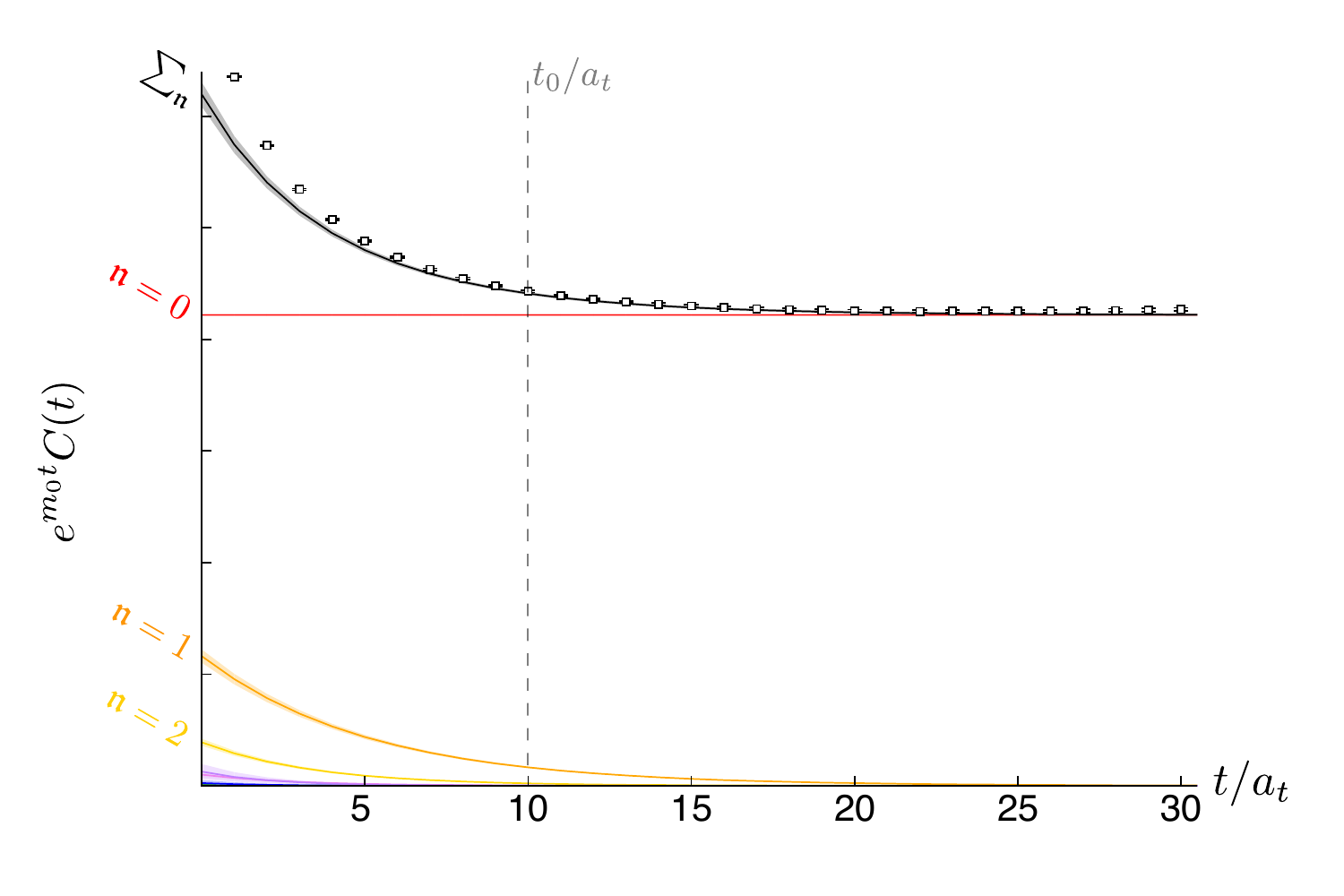}
\caption{Contributions of ground state ($\mathfrak{n}=0$) pion (red) and excited pion states (other colors) to the single pion correlator at zero momentum, $C(t) = \big\langle \big(\bar{\psi} \Box \gamma_5 \Box \psi\big)(t) \cdot \big(\bar{\psi} \Box \gamma_5 \Box \psi\big)(0)   \big\rangle$  and $N_\mathrm{vecs}=162$ on the $24^3$ lattice. Summed contribution of all states indicated by the grey curve. Excited state pions are observed to contribute significantly until $t \gtrsim 20 \,a_t$. (Excited state contributions determined from the results of variational analysis using a large operator basis, see the text) 
\label{pion_recon}}
\end{figure}

In our previous study of $\pi\pi$ isospin-2 scattering~\cite{Dudek:2010ew} we made use only of the simplest composite QCD operators capable of interpolating a pion, $\sim \bar{\psi} \Box_\sigma \gamma_5 \Box_\sigma \psi$ (Eq.~\ref{eq:distop} with $\boldsymbol{\Gamma} = \gamma_5$) where the distillation smearing operator $\Box_\sigma$ in Eq.~\ref{eq:box} took on two different values of the smearing weight $\sigma$. As well as interpolating the ground-state pion from the vacuum, this operator has significant amplitudes to interpolate various excited mesons with pion quantum numbers ($\pi^\star$). 
In correlation functions, the contribution of the excited states will die away more rapidly than the ground-state (see the decomposition in Equation \ref{spectral}), but at modest times, the excited states are present to some degree, as shown in Figure \ref{pion_recon}. For consideration of $\pi\pi$ scattering, these excited-state contributions are an unwanted pollution in our correlators that ideally we would like to be absent. Their presence forces us to analyse $\pi\pi$ correlators only at large times where effects of the finite-temporal extent of the lattice are more keenly felt.

In principal if we could find an operator which has increased overlap onto the ground-state pion and reduced overlap onto low-lying excited states, its use would lead to $\pi\pi$ correlators that are truly dominated by $\pi\pi$ at smaller times, with the contribution of unwanted $\pi \pi^\star$ being reduced. Our approach to finding such an ``optimised" single-pion operator is to variationally diagonalise a matrix of single-hadron correlators in a basis of operators, taking as our optimised operator the linear combination of basis operators having lowest energy.

The basis of operators used is as described in Section \ref{sec:singleops} and presented in detail in Refs.~\cite{Dudek:2009qf,Dudek:2010wm,Thomas:2011rh}.  It corresponds to fermion bilinears with Dirac gamma matrices and gauge-covariant derivatives\footnote{in this work we use all operators with the correct quantum numbers constructed from any possible gamma matrix and up to three derivatives (for an operator at rest) or up to one derivative (for an operator at non-zero momentum)} between them, constructed to be of definite spin or helicity in a continuum theory and then subduced into irreducible representations of the octahedral group or the appropriate little group.  For the pion this is $A_1^{-}$ for zero momentum and $A_2$ for all the non-zero momenta that we consider.

The variational analysis corresponds to solution of the generalised eigenvalue problem \cite{Michael:1985ne,Luscher:1990ck,Blossier:2009kd}
\begin{equation}
	C(t) v^{(\mathfrak{n})} = \lambda_\mathfrak{n}(t) C(t_0) v^{(\mathfrak{n})} \label{GEVP}
\end{equation}
where the state energies are obtained from fits to $\lambda_\mathfrak{n}(t) \sim e^{-E_\mathfrak{n}(t-t_0)}$. The optimal combination of operators, $\mathcal{O}_i$, to interpolate state $|\mathfrak{n}\rangle$ from the vacuum is $\Omega_\mathfrak{n}^\dag = \sum_i v^{(\mathfrak{n})}_i \mathcal{O}_i^\dag$. Our implementation of the variational method is described in Ref.~\cite{Dudek:2010wm}.

\begin{figure}
\includegraphics[width=0.5\textwidth
]{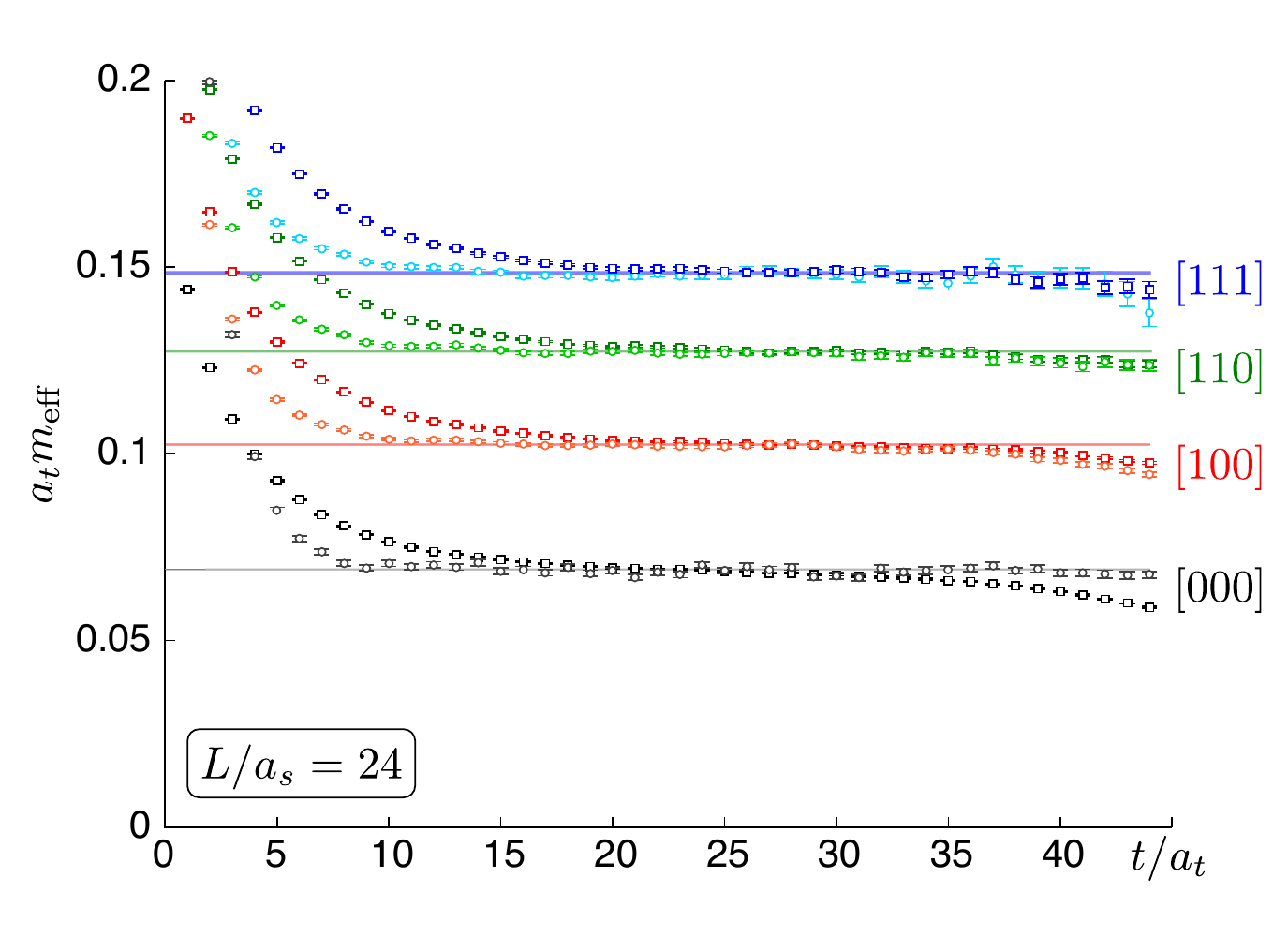}
\caption{
Effective masses
\footnote{throughout this paper we define the effective mass of a correlator $C(t)$ to be $m_\mathrm{eff} = \tfrac{1}{3a_t}\log \left[ \frac{C(t)}{C(t+3a_t)} \right]  $} of single-pion correlators computed using $\bar{\psi}\Box \gamma_5 \Box \psi$ (darker shades, squares) and ``optimised" operators, $\Omega_{\mathfrak{n}=0}$ (lighter shades, circles). Shown for a range of momenta on the $L/a_s =24$ lattice.
\label{pion_projection}}
\end{figure}

In Figure \ref{pion_projection} we show, for a range of momenta, the improvement obtained using the ``optimised" pion operators alongside the simple $\bar{\psi}\Box \gamma_5 \Box \psi$ operators, where clearly the correlators computed with the optimised operators relax to the ground state more rapidly that the simpler operators, typically at or before $10 a_t$ from the source (a time comparable with the values of $t_0$ found to be optimal in solution of equation \ref{GEVP}).

Use of these optimised operators will lead to some confidence when dealing with $\pi\pi$ correlators where for times $\gtrsim 10 a_t$ away from the source, we will be able to largely neglect the contribution of $\pi \pi^\star$ states.

\section{Pion mass and dispersion relation}\label{sec:dispersion}

\begin{figure}[b]
\includegraphics[width=0.45\textwidth
]{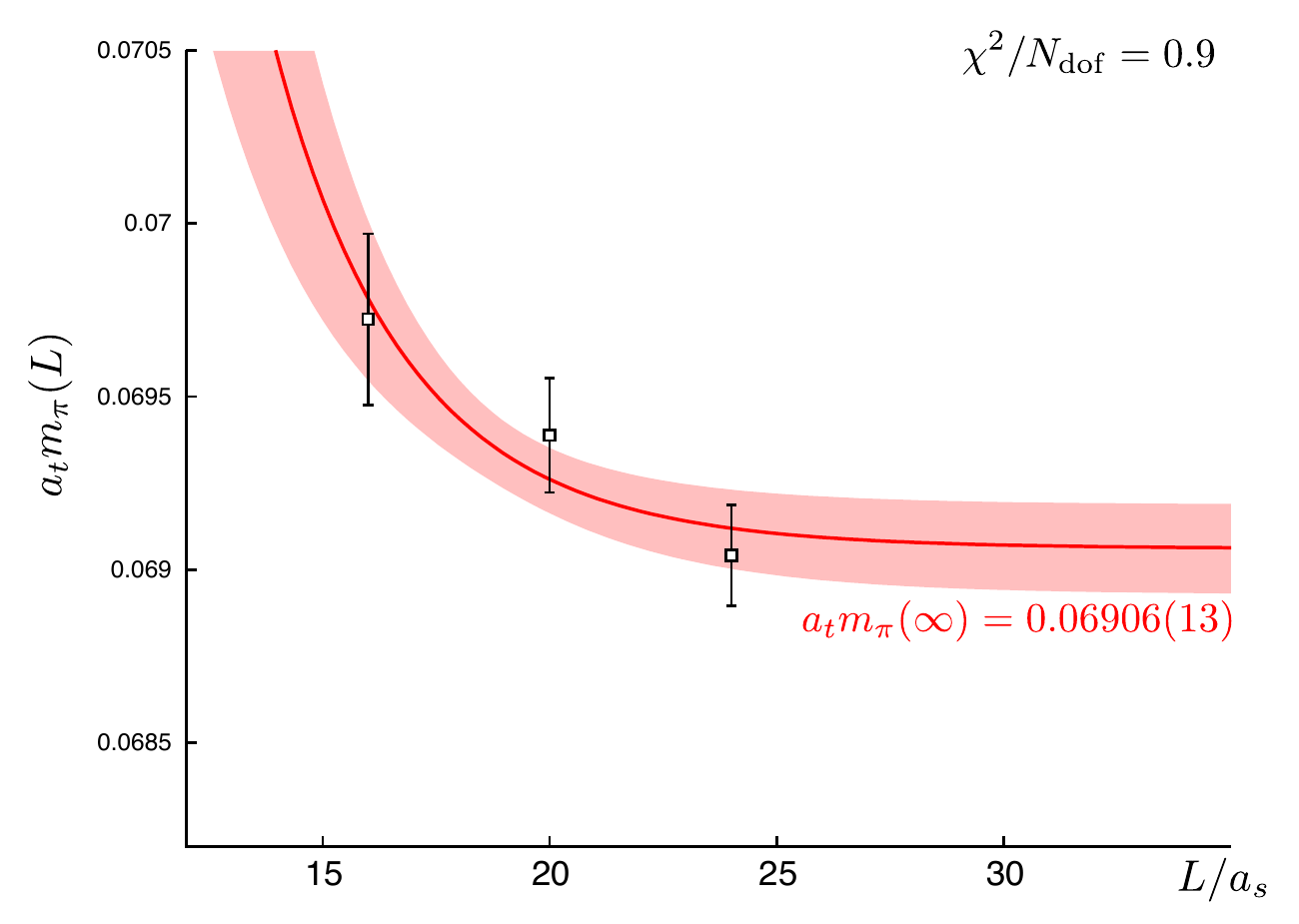}
\caption{Pion mass as a function of lattice spatial volume. Volume dependence fitted with Equation (\ref{voldep}).
\label{pion_mass}}
\end{figure}

As well as the volume dependence of energies of multi-hadron states owing to hadron interactions suggested by the L\"uscher formalism, there can also be exponential dependence of single-hadron energies on $L$. We can attempt to determine any such behavior for the pion by computing its mass on the three volumes at our disposal. In Figure \ref{pion_mass} we show the pion mass extracted on our three lattice volumes where there is seen to be very little volume dependence $\left(\frac{m_\pi(L/a_s =24)}{m_\pi(L/a_s=16)} = 0.990(4)\right)$. In \cite{Beane:2011sc}, NPLQCD suggest a $\chi$PT motivated form for the $L$ dependence,
\begin{equation}
m_\pi(L) = m_\pi + c \frac{e^{-m_\pi L}}{\left(m_\pi L \right)^{3/2}}. \label{voldep}
\end{equation}
Fitting to this form we find $a_t m_\pi = 0.06906(13)$ and $a_t c = 0.24(10)$ in good agreement with NPLQCD's $0.069073(63)(62)$, $0.23(12)(7)$ respectively. We use $a_t m_\pi = 0.06906(13)$ as our best estimate for the pion mass in all subsequent calculations\footnote{fitting the same data to a constant leads to $a_t m_\pi = 0.06928(18)$ with a somewhat poorer fit, $\chi^2/N_\mathrm{dof} = 3.0$.}.

\begin{figure}
\includegraphics[width=0.5\textwidth
]{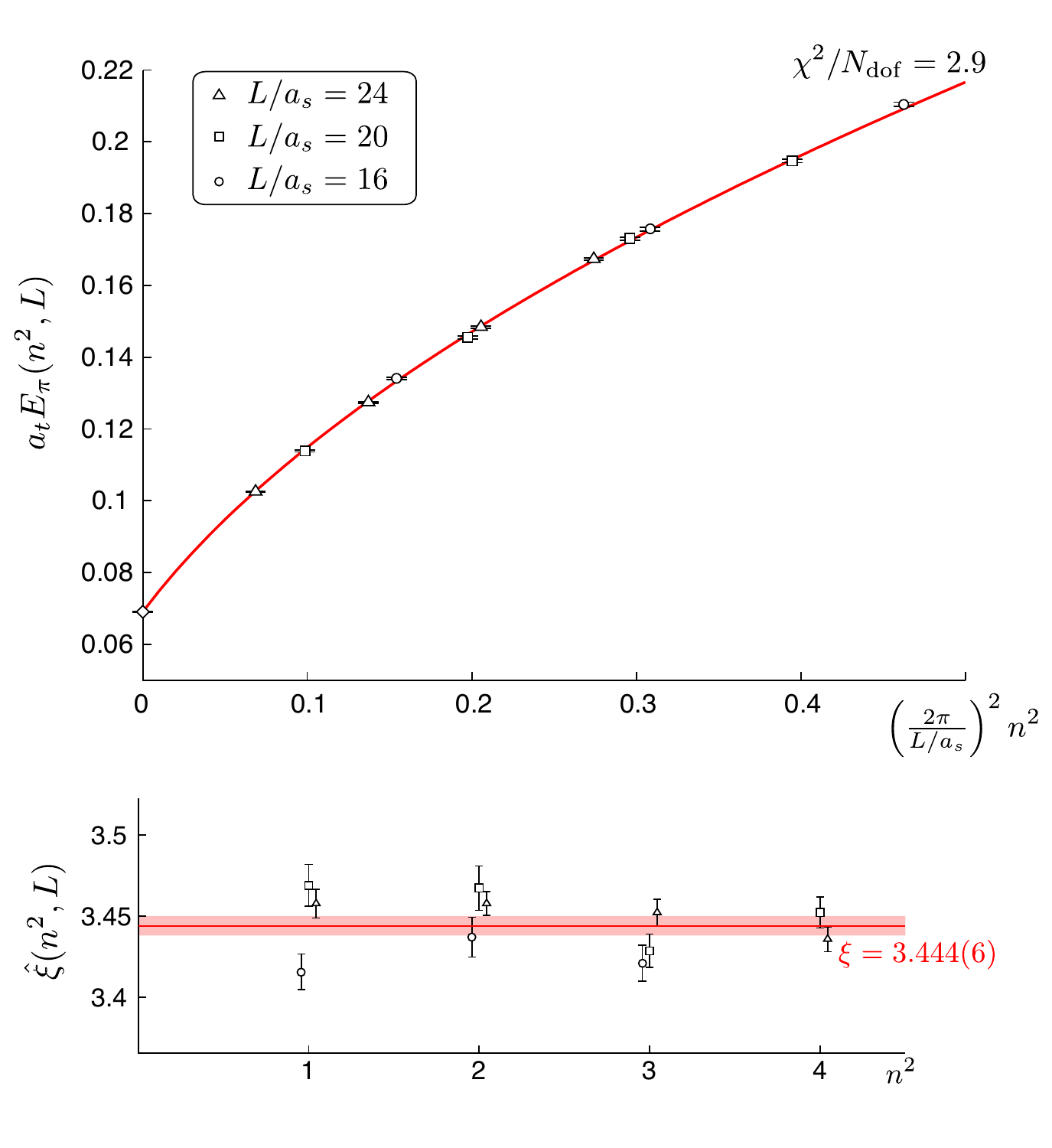}
\caption{
Pion dispersion relation. Fit as described in the text. Lower plot shows $\hat{\xi}(n^2,L) \equiv \frac{ \tfrac{2\pi}{L/a_s} \sqrt{n^2} }{\sqrt{(a_t E_{n^2})^2 - (a_t m)^2 }}$ and the fitted value of $\xi=3.444(6)$.
\label{pion_dispersion}}
\end{figure}

A complication which arises from our use of an anisotropic lattice is the need to determine the precise value of the anisotropy, $\xi$, which relates the spatial lattice spacing $a_s$ to the temporal lattice spacing $a_t = a_s / \xi$. The anisotropy appears in the dispersion relation of a free-particle, where the periodic boundary conditions in space lead to allowed momenta $\vec{p} = \frac{2\pi}{L}\big(n_x, n_y, n_z\big)$ for integer $n_x,\,n_y,\,n_z$, so that
\begin{equation}
\big(a_t E_{n^2} )^2 = \big( a_t m \big)^2 + \frac{1}{\xi^2} \left( \frac{2\pi}{L/a_s} \right)^2 n^2,  \label{disp}
\end{equation}
if we assume that mesons on the lattice have a continuum-like dispersion relation. Whether this is a good description will be determined by explicit fits to extracted pion energies at various momenta. In Figure \ref{pion_dispersion} we show pion energies on the three volumes along with a fit to Equation \ref{disp} with $\xi$ as a free parameter. The fit is acceptable leading to $\xi = 3.444(6)$. Using other parameterisations of the dispersion relation (adding a $p^4$ term, using cosh/sinh etc...), lead to fits which are indistinguishable within the thickness of the line in Figure \ref{pion_dispersion} and to compatible values of $\xi$. In the remainder of the paper we use $\xi = 3.444(6)$ as our best estimate\footnote{in correlated fitting to obtain $a_t m_\pi$ and $\xi$ simultaneously we find a relatively small correlation between the parameters and for error propagations in the remainder of the calculation we treat them as independent variables.}.

\section{Effects of finite temporal extent}\label{sec:finiteT}

Our extractions of finite-volume $\pi\pi$ energy spectra follow from analysis of the time-dependence of correlation functions and the form of these time-dependencies is affected by the finite temporal extent of the lattice. The size of finite-$T$ effects are generically determined by the size of $e^{-m_\pi T}$, which while small on these lattices, is large enough for its effects to be visible, particularly in the $\pi\pi$ sector.

\begin{figure}
\includegraphics[width=0.5\textwidth
]{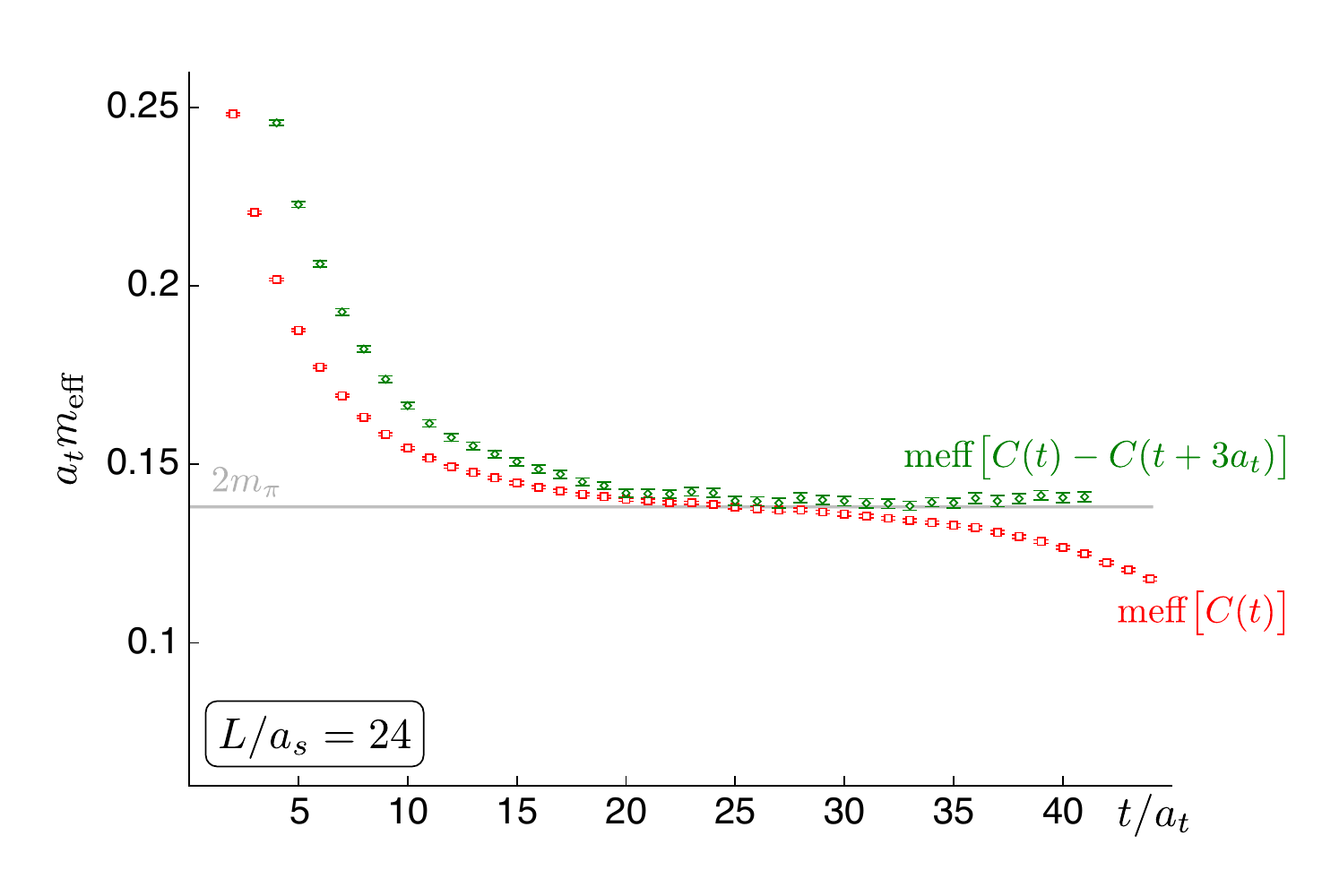}
\caption{Effective masses of a ``$\pi\pi$" correlator as described in the text. Raw correlator (red squares) and shifted correlator (green diamonds).
\label{corr_rest_finiteT_shifting}}
\end{figure}

As an explicit example of a systematic effect whose origin will turn out to be the finite temporal extent of the lattice, we show in Figure \ref{corr_rest_finiteT_shifting} the effective mass of a very simple ``$\pi\pi$" correlator. The same ``$\pi\pi$" operator, $\sum_{\vec{x}} \big[\bar{\psi}\Box \gamma_5 \Box \psi\big](\vec{x}) \cdot \sum_{\vec{y}} \big[\bar{\psi}\Box \gamma_5 \Box \psi\big](\vec{y})$ appears at source ($t_\mathrm{src}=0$) and sink ($t$). The effective mass of the raw correlator is observed to continue falling after appearing to briefly plateau near an energy equal to twice the pion mass. This behavior can occur if the correlator features, as well as a sum of exponentially decaying time-dependencies corresponding to discrete energy eigenstates, as in Equation \ref{spectral}, also a contribution that is \emph{constant in time}. Such a term can be eliminated by considering the \emph{shifted} correlator, $\widehat{C}_{\delta t}(t) \equiv C(t) - C(t+\delta t)$. An effective mass of this construction with $\delta t = 3 a_t$ is also shown in Figure \ref{corr_rest_finiteT_shifting} where it is observed to plateau to an energy slightly above twice the pion mass\footnote{the very slow relaxation to the plateau is mainly due to not using optimised pion operators in this construction.}. A direct estimate of the size of the constant term comes from fitting $C(t)$ to the form
\begin{equation}
\sum_\mathfrak{n} A_\mathfrak{n} e^{- E_\mathfrak{n} t} + c, \label{exp_and_const}
\end{equation} 
where $\{ A_\mathfrak{n}\},\, \{E_\mathfrak{n} \}$ and $c$ are the fit parameters. A fit to the raw correlator over a time region $15 \to 43$ with two exponentials and a constant gives a $\chi^2/N_\mathrm{dof} = 0.7$ and $a_t E_0 = 0.13966(4)$, $A_0 = 8.76(8) \times 10^4$ and $c = 26.4(13)$, indicating a statistically significant constant term. We propose that the origin of the constant term is the finite temporal extent of the lattice and notice that $2 A_0 e^{- m_\pi T} = 25.4(2)$ is very close to the fitted value of $c$. In the remainder of this section we will attempt to describe the effect of finite-$T$ on our computed $\pi\pi$ correlators at rest and in-flight.

\subsection{Finite-$T$ effects for correlators at rest}

Let us begin by considering a correlator constructed using pion interpolating fields of definite momentum,
\begin{equation}
C^{\vec{k}_1',\vec{k}_2'}_{\vec{k}_1,\vec{k}_2}(t) = \big\langle \pi^-_{\vec{k}_1'}(t)\pi^-_{\vec{k}_2'}(t) \cdot  \pi^+_{\vec{k}_1}(0)\pi^+_{\vec{k}_2}(0) \big\rangle, \nonumber
\end{equation}
where in this section the operator $\pi^+$ interpolates a positively charged pion from the vacuum. In practice we will always project these products into definite little group irreps, $\Lambda$, for a given $\vec{P}=\vec{k}_1+\vec{k}_2 = \vec{k}_1'+\vec{k}_2'$ as described in Section \ref{sec:multi}. With anti-periodic boundary conditions in the finite time direction, two-point correlators have the decomposition\footnote{see \cite{Beane:2009kya} for a  discussion of these finite-$T$ effects on the spectrum of single particle systems and \cite{Detmold:2011kw} for discussion of many-hadron states.}
\begin{align}
	C(t) &= \big\langle \mathcal{O}'(t) \mathcal{O}(0) \big\rangle \nonumber \\
	&= \mathrm{tr}\big[ e^{-H T} \mathcal{O}'(t) \mathcal{O}^\dag(0) \big] / \mathrm{tr}\big[ e^{-H T} \big]  \nonumber \\
		&\propto \sum_{\mathfrak{n}, \mathfrak{m}} e^{-E_\mathfrak{n} T} e^{ (E_\mathfrak{n} - E_\mathfrak{m})t} \big\langle \mathfrak{n} \big| \mathcal{O}'(0) \big| \mathfrak{m} \big\rangle \big\langle \mathfrak{m} \big| \mathcal{O}^\dag(0) \big| \mathfrak{n} \big\rangle , \label{finT}
\end{align}
in terms of eigenstates of the Hamiltonian, $H|\mathfrak{n}\rangle = E_\mathfrak{n} |\mathfrak{n}\rangle$, which will be discrete in a finite spatial volume. The contribution to this sum we are interested in is the only one to survive in the limit $T\to \infty$ and is of the form
\begin{eqnarray}
\sum_\mathfrak{n} \big\langle 0 \big| \pi^- \pi^- \big| (\pi^+\pi^+)_\mathfrak{n} \big\rangle \big\langle (\pi^+\pi^+)_\mathfrak{n} \big| \pi^+ \pi^+ \big|0  \big\rangle e^{-E^\mathfrak{n}_{\pi\pi} t} \nonumber\\
= \sum_\mathfrak{n}\left(Z^\mathfrak{n}_{\pi\pi}\right)^2 e^{-E^\mathfrak{n}_{\pi\pi} t}\label{eq:wanted}
\end{eqnarray}
where $\big|(\pi^+\pi^+)_\mathfrak{n}\big\rangle$ are $I=2$ eigenstates. 

At finite $T$ there are other terms in the sum in Equation \ref{finT}, the largest being of form
\begin{equation}
\sum_{\vec{p},\vec{q}} e^{-E_\pi(\vec{p}) \,T} \big\langle \pi^-_{\vec{p}} \big | \pi^-_{\vec{k}_1'}(t) \pi^-_{\vec{k}_2'}(t) \big| \pi^+_{\vec{q}} \big\rangle \big\langle \pi^+_{\vec{q}} \big| \pi^+_{\vec{k}_1}(0) \pi^+_{\vec{k}_2}(0) \big| \pi^-_{\vec{p}} \big\rangle \nonumber
\end{equation}
which has a time-dependence of 
\begin{align}
	z_{\vec{k}_1}^2 & z_{\vec{k}_2}^2 \big[\delta_{\vec{k}_1',\vec{k}_1} \delta_{\vec{k}_2', \vec{k}_2} + \delta_{\vec{k}_1',\vec{k}_2} \delta_{\vec{k}_2', \vec{k}_1} \big] \nonumber\\
	&\times \big[ e^{-E_\pi(\vec{k}_1')\, T} e^{- \left(E_\pi(\vec{k}_2') - E_\pi(\vec{k}_1') \right) t} \nonumber \\
	&\quad\quad + e^{-E_\pi(\vec{k}_2')\, T} e^{- \left(E_\pi(\vec{k}_1') - E_\pi(\vec{k}_2') \right) t}   \big]
\end{align}
where $z_{\vec{k}} \equiv \big\langle \pi^+_{\vec{k}} \big| \pi^+_{\vec{k}} \big| 0 \big\rangle$. As a first example, consider the case of correlators in the $\pi\pi$ rest frame, $\vec{P}=\vec{0}$, $C^{\vec{k},-\vec{k}}_{\vec{k},-\vec{k}}(t)$, where this term becomes
\begin{equation}
2\, \left(z_{\vec{k}}\right)^4\, e^{-E_\pi(\vec{k})\, T}   \label{constant}
\end{equation}
which is simply a constant in time. 

\begin{figure}[h]
\includegraphics[width=0.4\textwidth
]{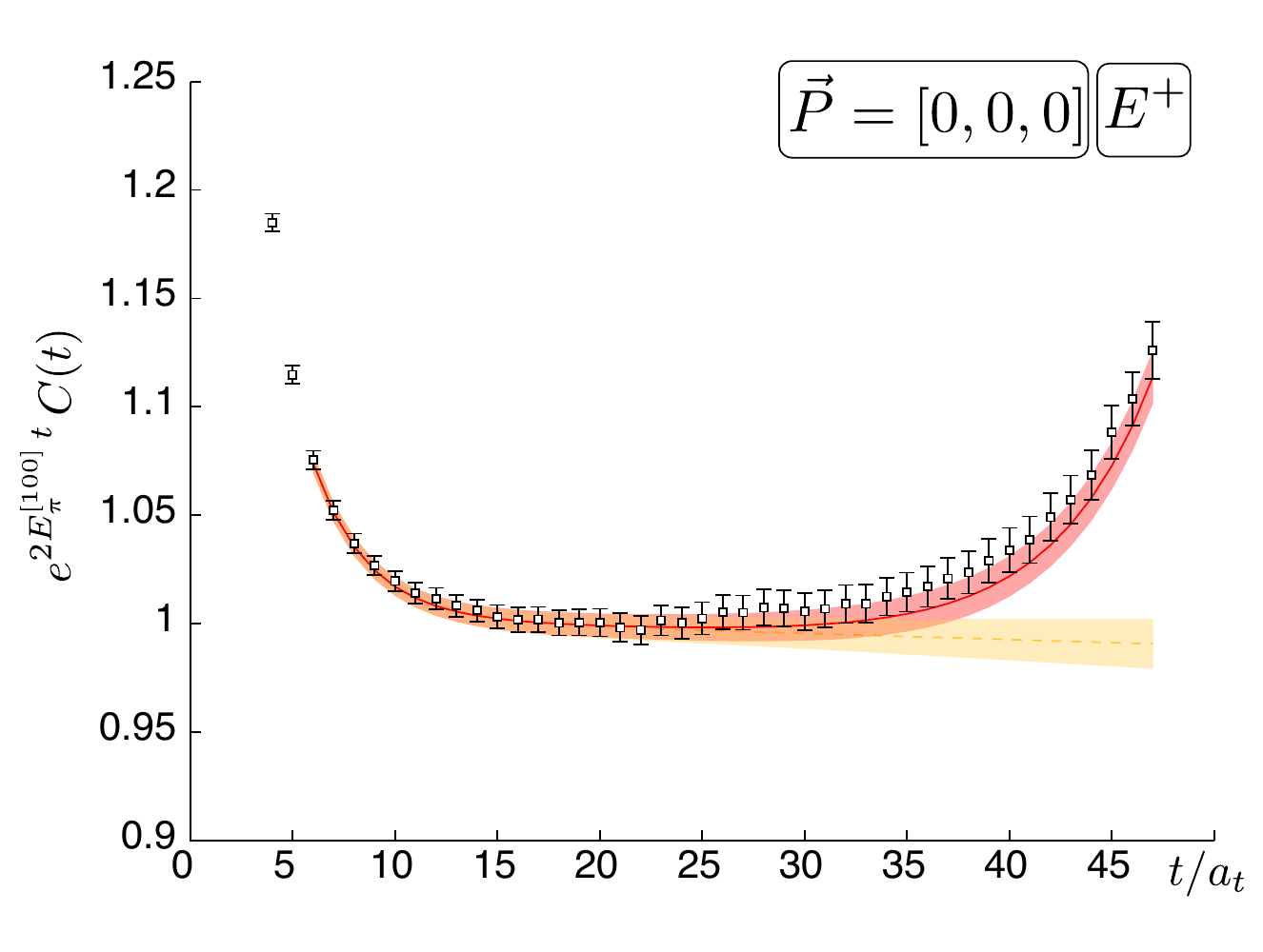}
\includegraphics[width=0.4\textwidth
]{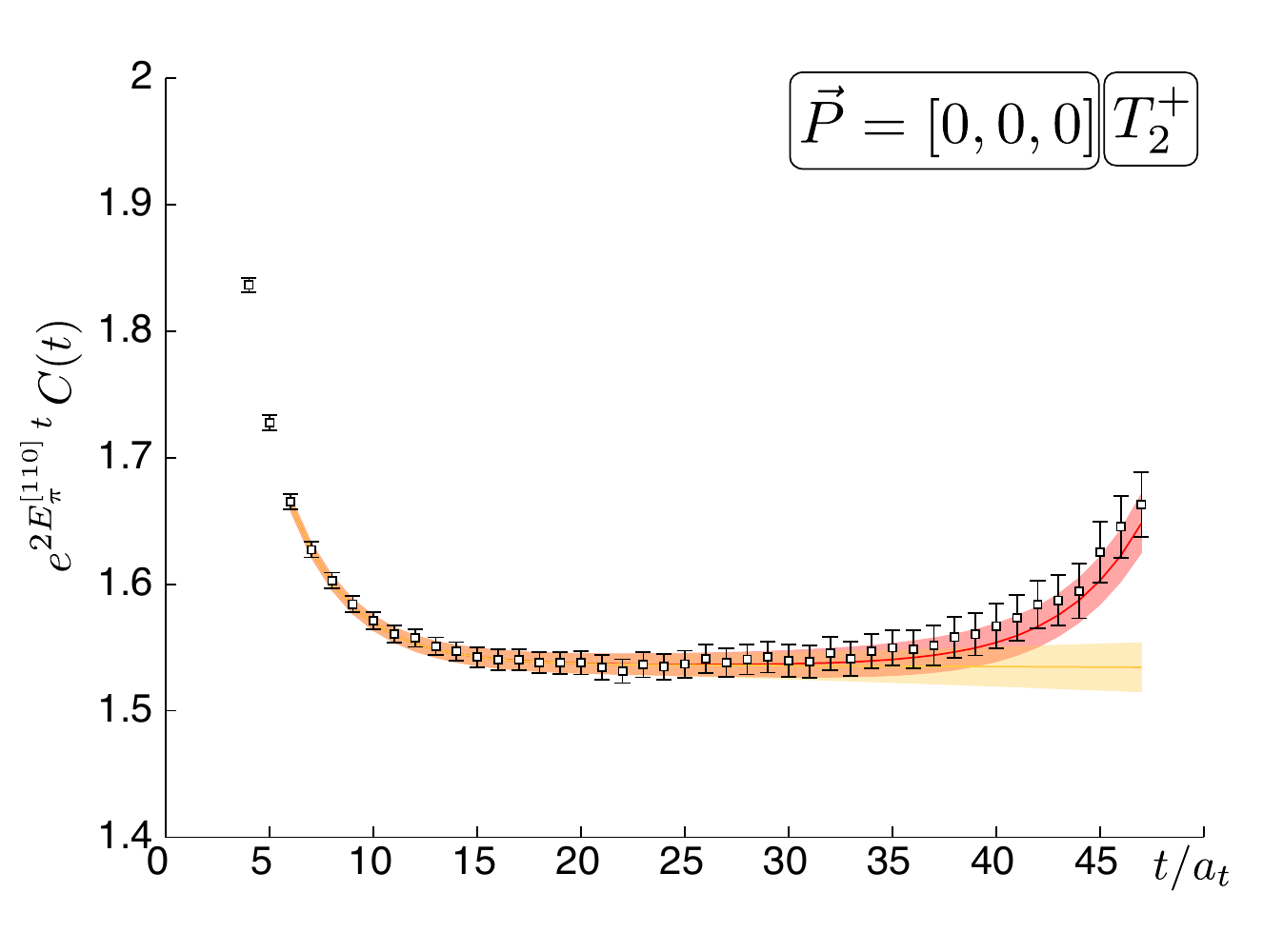}
\caption{
Fits to diagonal $\pi\pi$ correlators with $\vec{P}=[0,0,0]$ using the lowest allowed $|\vec{k}|$ that gives rise to irrep $\Lambda^P$. Correlator is plotted via $e^{2 E_\pi(\vec{k}) t} \, C(t)$ such that in the limit of non-interacting pions and $T\to \infty$ we would have a horizontal line. The solid red line shows the result of the fit using equation (\ref{exp_and_const}) while the orange dashed line shows the result of excluding the constant contribution, which should correspond to the $T\to \infty$ behavior. Fit parameters given in Table \ref{tab:corr_rest_finiteT}.
\label{corr_rest_finiteT}}
\end{figure}

We may now address the observation made at the start of this section that the correlator constructed with $\vec{k}=\vec{k}'=[0,0,0]$ has a clear constant term. Our analysis above suggests that its magnitude would be $2\, \left(z_{[000]}\right)^4 \,e^{-m_\pi T}$, while the leading $T$-independent term is of form $\left(Z^{(0)}_{\pi\pi}\right)^2 e^{-E^{(0)}_{\pi \pi} t }$. In the limit of weakly-interacting pions we would expect $Z^{(0)}_{\pi\pi} \to \left(z_{[000]}\right)^2$ and as such $c \to 2 A_0 e^{-m_\pi T}$. This appears to hold true to a rather good approximation in the data. 

We expect other finite-$T$ terms to be negligibly small in practice; in particular a term often considered in single-particle analysis, $Z^2 e^{-E T} e^{E t}$, which turns exponentially decaying time-dependence into cosh-like time-dependence can be ignored here. It is suppressed by at least $e^{- 2 m_\pi T}$ and only becomes relevant close to $t = T/2$ while we consider correlators only at earlier times\footnote{In practical terms, while the constant term could contribute $\sim 10\%$ of the correlator at $t=48$, the extra term ``in the cosh"  would only be at the $1\%$ level. Other contributions featuring $\langle \pi^+ \pi^+ \pi^\pm | \pi^+ \pi^+ |\pi^\pm\rangle$ formally appear at $\mathcal{O}(e^{-m_\pi T})$, but their $t$-dependence ensures that they provide negligible contributions to the correlators.}.

In Figure \ref{corr_rest_finiteT} we show further evidence for the presence of the constant term in $\pi\pi$ correlators
. These correlators, evaluated on the $L/a_s=24$ lattice, using optimised pion operators, have $\vec{P}=[0,0,0]$ and use $\pi_{\vec{k}_1}\pi_{\vec{k}_2}$ products projected into definite irreps $\Lambda^P$ constructed from the lowest allowed $|\vec{k}_1|,\,|\vec{k}_2|$ as detailed in Section \ref{sec:multi} (
$E^+ \to \vec{k} = [1,0,0]$, $T_2^+ \to \vec{k} = [1,1,0]$). The results of the correlator fits (of the form given in equation (\ref{exp_and_const})) are presented in Table \ref{tab:corr_rest_finiteT}, where we see that the size of the constant term is in rather good agreement with $4 A_0 e^{-E_\pi(\vec{k}) T}$, the value in a non-interacting theory (see the Clebsch-Gordan coefficients in Appendix \ref{app:operators} for the appropriate combination of pion momenta). Clearly the ``polluting" constant term plays a significant role in the correlator as early as $t \sim 25 a_t$ and if we want to use timeslices beyond this point in variational analysis, we will need to take some account of its presence.

\begin{table}[t]
\begin{tabular}{cc|c| ccc}
irrep. & $\vec{k}$ & $\chi^2/N_\mathrm{dof}$ & $E_0/2E_\pi(\vec{k})$ & $c$ & $4 A_0 e^{-E_\pi(\vec{k}) T}$ \\
\hline
$E^+$ & $[1,0,0]$ & $0.9$ & $1.0014(17)$ & $7.9(5)\times 10^{-6}$ & $ 7.8 \times 10^{-6} $ \\
$T_2^+$ & $[1,1,0]$ & $0.9$ & $1.0002(16)$ & $6.9(11)\times 10^{-7}$ & $ 4.9 \times 10^{-7} $
\end{tabular}
\caption{
Fits, using two exponentials in equation (\ref{exp_and_const}), to diagonal $\pi\pi$ correlators with $\vec{P}=[0,0,0]$ using the lowest allowed $|\vec{k}|$ that gives rise to irrep $\Lambda^P$. 
\label{tab:corr_rest_finiteT}}
\end{table}

Our solution is to completely remove the effect of all time-independent terms from correlators, by instead of considering $C(t)$, using \emph{shifted} correlators,
\begin{equation}
\widehat{C}_{\delta t}(t) = C(t) - C(t+\delta t), \label{shift}
\end{equation}
which exactly cancel contributions constant in time for any choice of $\delta t \neq 0$. The desired $\pi\pi$ contributions, equation (\ref{eq:wanted}), are changed only by a rescaling of the $Z^\mathfrak{n}_{\pi\pi}$ to
\begin{equation}
\widehat{Z}^\mathfrak{n}_{\pi\pi} = Z^\mathfrak{n}_{\pi\pi} \left[1 - e^{-E_{\pi\pi}^\mathfrak{n} \delta t} \right]^{1/2}. \label{Zscale}
\end{equation}
This is just a change in scale of overlaps that for a given state, $\mathfrak{n}$, is common to all operators. 
Shifting then does not violate any of the conditions for carrying out a variational analysis and we can proceed with use of $\widehat{C}_{\delta t}(t)$ in equation (\ref{GEVP}) to yield the finite-volume energy spectrum $E^\mathfrak{n}_{\pi\pi}$. 

\subsection{Finite-$T$ effects for correlators in-flight}

The unwanted contributions to correlators ``in-flight" ($\vec{P} \neq \vec{0}$) are not time-independent and cannot be removed by simply shifting in time. Following equation (\ref{finT}), they take the generic form 
\begin{align}
(z_{\vec{k}_1})^2 \,& (z_{-\vec{k}_1+\vec{P}})^2 \,\big[\delta_{\vec{k}_1',\vec{k}_1} + \delta_{\vec{k}_1',-\vec{k}_1+\vec{P}} \big] \nonumber \\
&\times \big[  e^{-E_\pi(\vec{k}_1')\, T} e^{-\left( E_\pi(-\vec{k}_1'+\vec{P}) - E_\pi(\vec{k}_1') \right)t} \nonumber \\
&\quad\quad + e^{-E_\pi(-\vec{k}_1'+\vec{P})\, T} e^{-\left( E_\pi(\vec{k}_1') - E_\pi(-\vec{k}_1'+\vec{P}) \right)t} \big],\nonumber
\end{align}
where the contributions of largest magnitude occur if either $\vec{k}_1'$ or $-\vec{k}_1'+\vec{P}$ are equal to zero as then the finite-$T$ suppression factor is only $e^{-m_\pi T}$. The largest ``polluting" term in this case would not be a constant but rather have a time dependence $\sim e^{- \Delta E_\pi \, t}$ where $\Delta E_\pi $ is the energy gap between a single pion of momentum $\vec{k}$ and one with momentum $\vec{P}-\vec{k}$. In the case $\vec{P}=\vec{0}$ this reverts to a constant in time as expected.

\begin{figure}
\includegraphics[width=0.5\textwidth
]{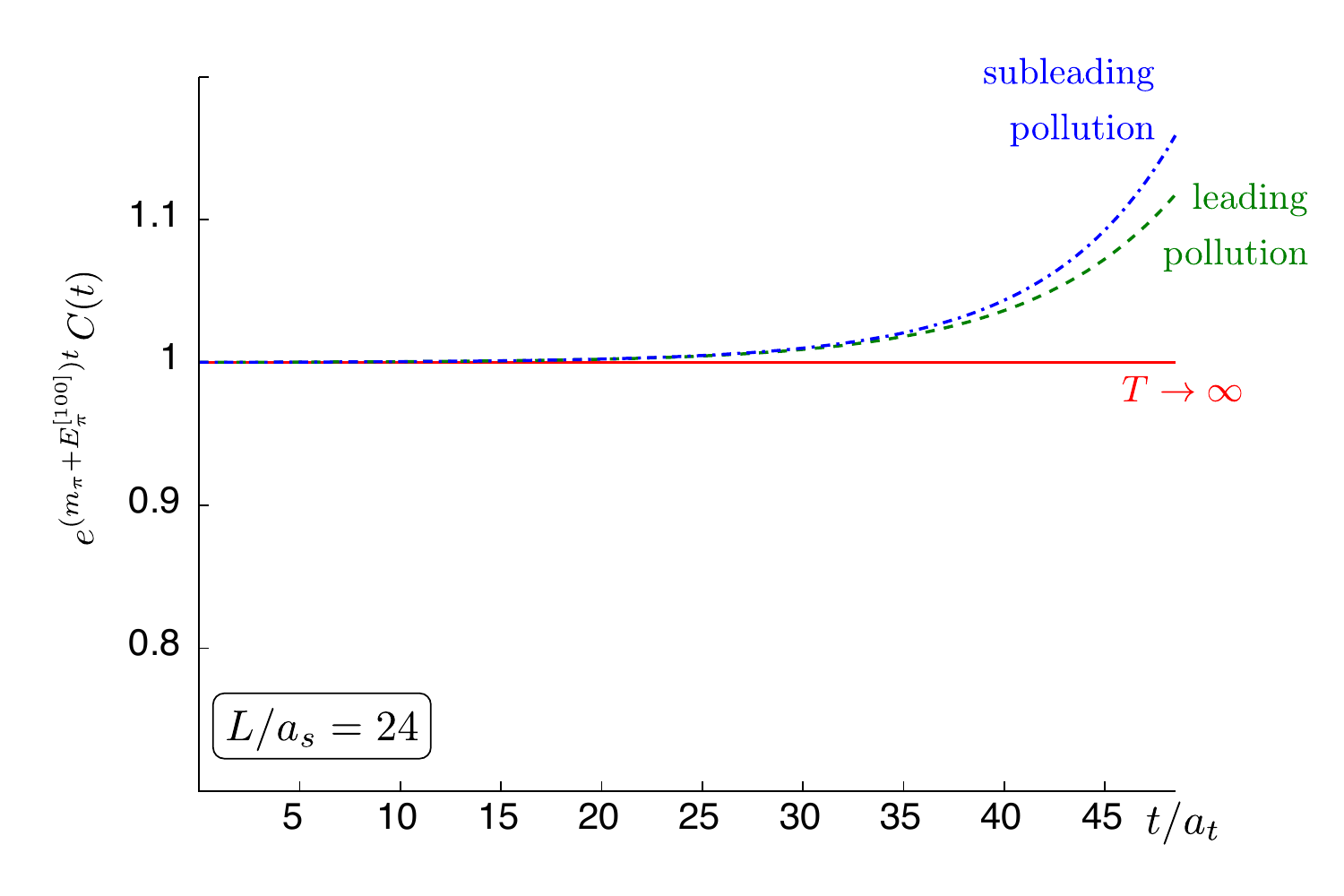}
\caption{Simulated contributions to a correlator ($\vec{k}_1=[0,0,0],\, \vec{k}_2 = [1,0,0]$) of the desired ($T \to \infty$, red) term, equation \ref{flight_good}, and two ``polluting" (finite-$T$) terms from equation \ref{flight_bad} - the first term (leading, green dashed) and the sum of the two terms (leading plus subleading, blue, dot-dashed). Observe that in the time region we will consider, the leading term dominates over the subleading term. 
\label{corr_flight_finiteT_model}}
\end{figure}

\begin{figure}
\includegraphics[width=0.4\textwidth
]{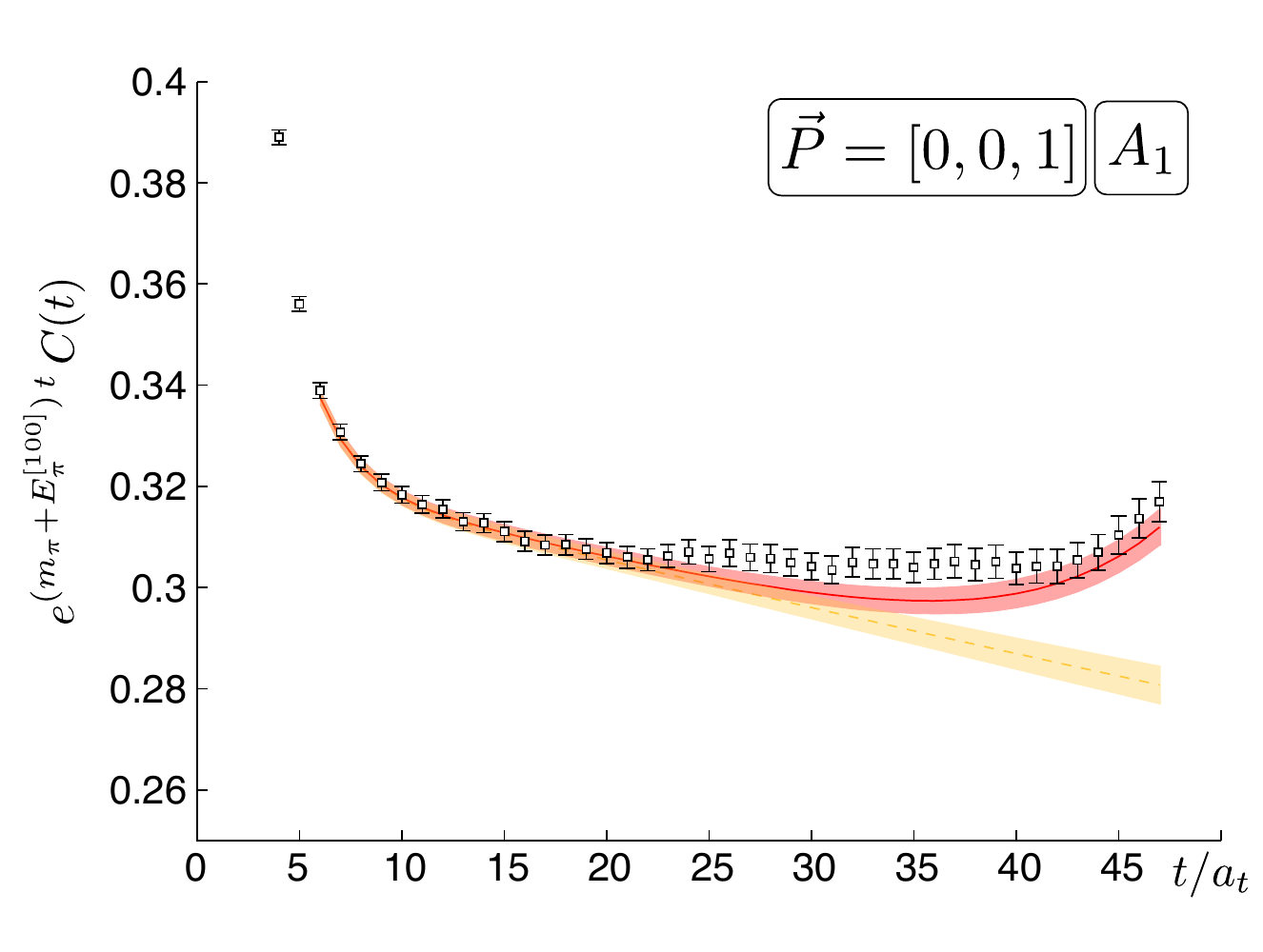}
\includegraphics[width=0.4\textwidth
]{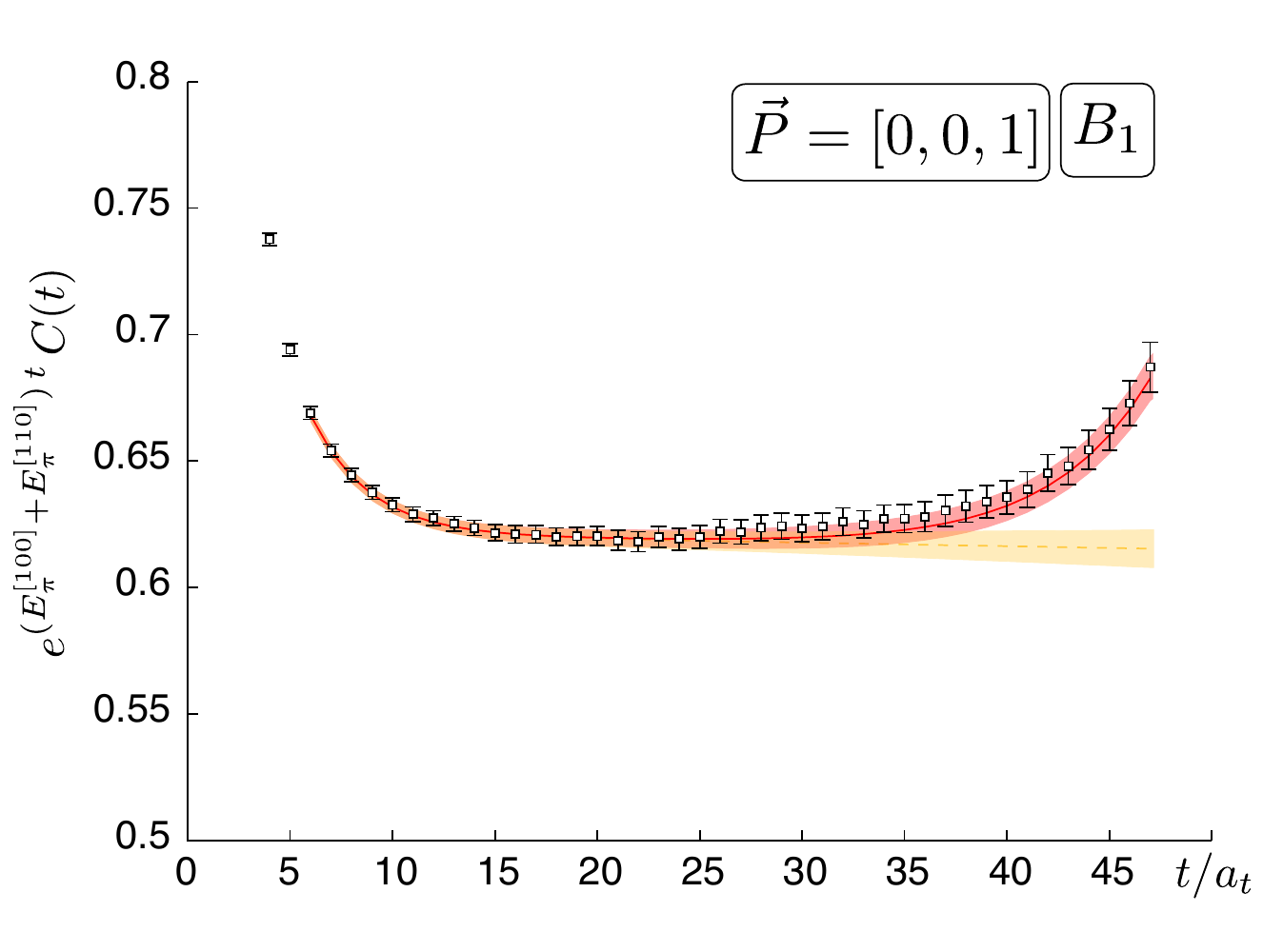}
\includegraphics[width=0.4\textwidth
]{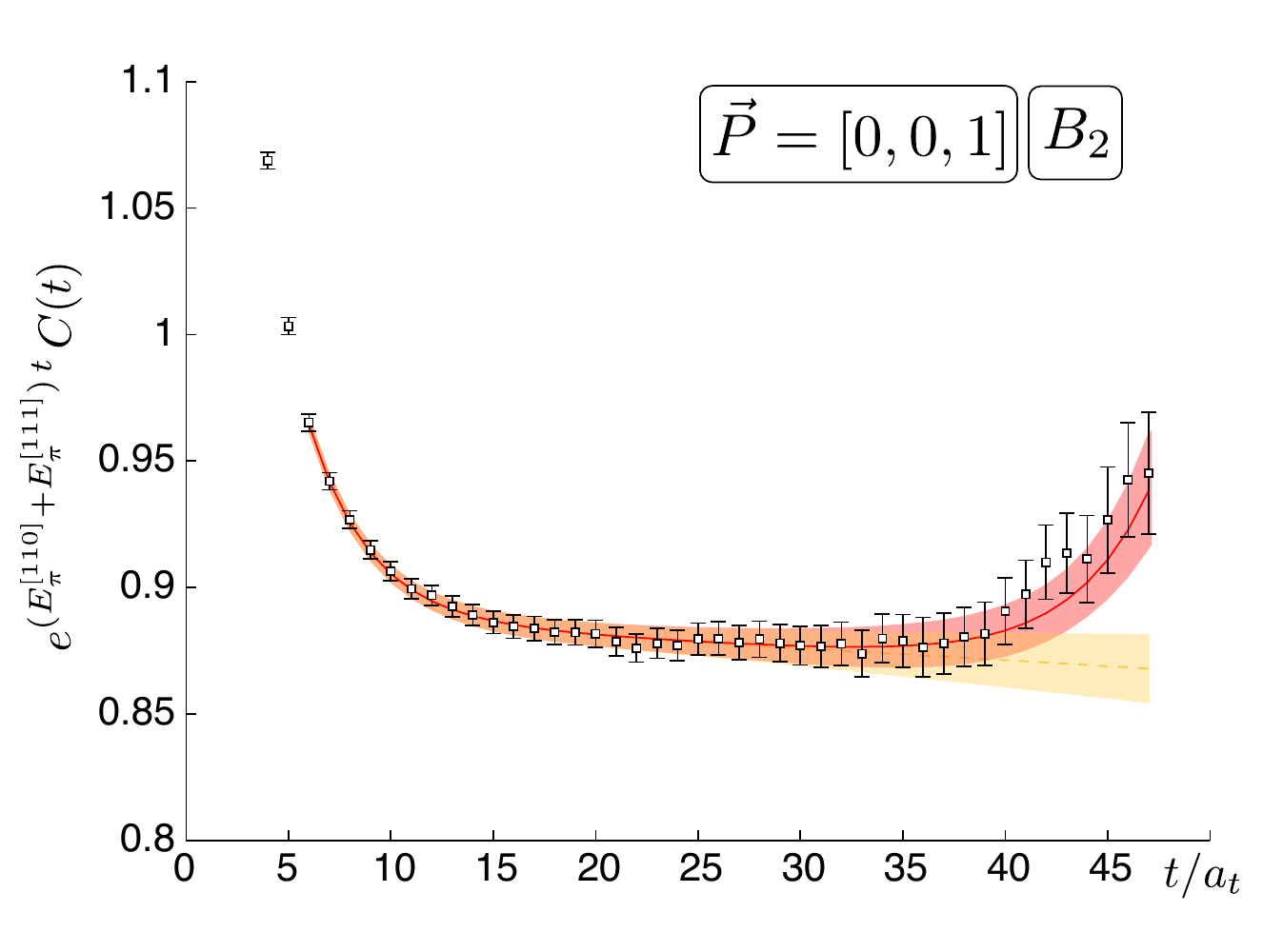}
\caption{
Fits to diagonal $\pi\pi$ correlators with $\vec{P}=[0,0,1]$ using the lowest allowed $|\vec{k}_1|,\, |\vec{k}_2|$ that gives rise to irrep $\Lambda$. Correlator is plotted via $e^{(E_\pi(\vec{k}_1) + E_\pi(\vec{k}_2)  ) t} \, C(t)$ such that in the limit of non-interacting pions and $T\to \infty$  we would have a horizontal line. The solid red line shows the result of the fit using equation (\ref{exp_flight}) while the orange dashed line shows the result of excluding the contribution proportional to $c$, which should correspond to the $T\to \infty$ behavior. 
\label{corr_flight_finiteT}}
\end{figure}

Consider the concrete example of a correlator with $\pi^+_{[000]} \pi^+_{[100]}$ at the source and $\pi^-_{[000]} \pi^-_{[100]}$ at the sink. In this case, as well as the desired term which is approximately\footnote{this would be exact for non-interacting pions, in $\pi\pi$ $I=2$ scattering the interaction is weak so the approximation should be a reasonable guide.}
\begin{equation}
\sim \left(z_{[000]}\right)^2 \left(z_{[100]}\right)^2 e^{-(m_\pi + E_\pi^{[100]}) t}, \label{flight_good}
\end{equation}
we would have ``polluting" terms
\begin{align}
\sim \left(z_{[000]}\right)^2 \left(z_{[100]}\right)^2 \Big( &e^{-m_\pi T} e^{-(E_\pi^{[100]} - m_\pi )t} \nonumber \\
 &+  e^{-E_\pi^{[100]} T} e^{-(m_\pi - E_\pi^{[100]} )t} \Big),\label{flight_bad}
\end{align}
where as shown in Figure \ref{corr_flight_finiteT_model}, the first of these polluting terms is expected to  dominate the pollution for the time regions we consider. We can observe the effect of this leading pollution term in fits to correlators having $\vec{P}=[1,0,0]$ computed on the $L/a_s = 24$ lattice using optimised pion operators - in Figure \ref{corr_flight_finiteT} we show the irreps $\Lambda = A_1,\, B_1,\, B_2$, constructed using the smallest allowed magnitudes of pion momentum. The fit form (which neglects the subleading pollution) is 
\begin{equation}
	\sum_\mathfrak{n} A_\mathfrak{n} e^{-E_\mathfrak{n} t} + c\, e^{- \big(E_\pi(\vec{k}_\mathrm{max}) - E_\pi(\vec{k}_\mathrm{min}) \big) t} \label{exp_flight}
\end{equation}
with $\{A_\mathfrak{n}\}$, $\{E_\mathfrak{n}\}$ and $c$ as fit variables, using fixed $E_\pi(\vec{k})$ obtained from the dispersion relation (equation \ref{disp}). 

It would appear that these diagonal correlators can be reasonably well described by the fit form proposed indicating a small but statistically significant impact of finite-$T$ effects on the correlators. We will need to address these terms in any variational extraction of the in-flight $\pi\pi$ spectrum. Our approach is to remove the worst of the pollution exactly and settle for approximate reduction of less acute terms. The largest polluting term has a time-dependence $\propto e^{-E_\pi(\vec{k}_\mathrm{min}) T} e^{-\Delta E_\mathrm{min} \, t}$ where $\vec{k}_\mathrm{min}$ is the lowest momentum that appears in \emph{any} of the correlators making up our correlator matrix, and $\Delta E_\mathrm{min}$ is whatever positive energy gap appears in the corresponding time-dependence. This term can be converted into a constant by forming the following \emph{weighted} correlator,
\begin{align}
	\widetilde{C}(t) &= e^{\Delta E_\mathrm{min} \, t} C(t) ,
\end{align}		
and the constant term can be removed by then \emph{shifting} the weighted correlator,
\begin{align}
\widehat{\widetilde{C}}_{\delta t}(t) &= \widetilde{C}(t) - \widetilde{C}(t+ \delta t).
\end{align}

We refer to these as \emph{weighted-shifted} correlators. The exact same weighting and shifting procedure is applied to every element of the matrix of correlators such that the effect of the weighting is to shift \emph{all} energies down by a common $\Delta E_\mathrm{min}$. This can be corrected for by adding $\Delta E_\mathrm{min}$ to the variationally obtained spectrum.

In summary, while finite-$T$ effects are modest in our two-pion correlators, precision extraction of a $\pi\pi$ energy spectrum requires that we account for them in our analysis. Through appropriate weighting and shifting of correlators before applying the variational method, we believe that we are able to remove the leading systematic effects leaving only sub-leading effects that we find to be smaller than our level of statistical uncertainty.


\section{Finite-volume spectrum}\label{sec:spectrum}

We compute correlator matrices in each irrep $\vec{P},\,\Lambda$ using the basis of operators defined in Section \ref{sec:multi}. After modifying the correlator matrix with the appropriate weighting and/or shifting as described in the previous section, the spectrum is obtained by solution of the generalised eigenvalue problem, equation \ref{GEVP}. Each irrep is considered independently and the entire procedure is repeated on each of the three lattice volumes. The two-pion operators used are given in Table \ref{table:opsused} and the number of operators for each $\vec{P}$ and irrep are given in Table \ref{table:numopsirreps}.  We illustrate the method here with the example of the $\vec{P}=[1,0,0]$, $\Lambda = A_1$ irrep on the $L/a_s = 24$ lattice.

\begin{table}[b]
\begin{ruledtabular}
\begin{tabular}{c | c | l l l}
 $\vec{P}$ & Volumes & \quad $\vec{k}_1$ & \quad$\vec{k}_2$ & \multicolumn{1}{c}{$\Lambda^{(P)}$} \\
\hline \hline
\multirow{5}{*}{$\begin{matrix}[0,0,0] \\ \text{O}^{\text{D}}_h\end{matrix}$}
& \multirow{5}{*}{$16^3, 20^3, 24^3$}
   & $[0,0,0]$ & $[0,0,0]$ & $A_1^+$ \\
 & & $[0,0,1]$ & $[0,0,\text{-}1]$ & $A_1^+, E^+$ \\
 & & $[0,1,1]$ & $[0,\text{-}1,\text{-}1]$ & $A_1^+, T_2^+, E^+$ \\
 & & $[1,1,1]$ & $[\text{-}1,\text{-}1,\text{-}1]$ & $A_1^+, T_2^+$ \\
 & & $[0,0,2]$ & $[0,0,\text{-}2]$ & $A_1^+, E^+$ \\
\hline
\multirow{7}{*}{$\begin{matrix}[0,0,1] \\ \text{Dic}_4 \end{matrix}$}
& \multirow{4}{*}{$16^3, 20^3, 24^3$}
   & $[0,0,0]$ & $[0,0,1]$ & $A_1$ \\
 & & $[0,\text{-}1,0]$ & $[0,1,1]$ & $A_1, E_2, B_1$ \\
 & & $[\text{-}1,\text{-}1,0]$ & $[1,1,1]$ & $A_1, E_2, B_2$ \\
 & & $[0,0,\text{-}1]$ & $[0,0,2]$ & $A_1$ \\
\cline{2-5}
 & \multirow{3}{*}{$20^3, 24^3$}
   & $[0,\text{-}1,\text{-}1]$ & $[0,1,2]$ & $A_1, E_2, B_1$ \\
 & & $[\text{-}2,0,0]$ & $[2,0,1]$ & $A_1, E_2, B_1$ \\
 & & $[\text{-}1,\text{-}1,\text{-}1]$ & $[1,1,2]$ & $A_1, E_2, B_2$ \\
\hline
\multirow{8}{*}{$\begin{matrix}[0,1,1] \\ \text{Dic}_2 \end{matrix}$}
& \multirow{5}{*}{$16^3, 20^3, 24^3$}
 & $[0,0,0]$ & $[0,1,1]$ & $A_1$ \\
 & & $[0,1,0]$ & $[0,0,1]$ & $A_1$ \\
 & & $[\text{-}1,0,0]$ & $[1,1,1]$ & $A_1, B_2$ \\
 & & $[1,1,0]$ & $[\text{-}1,0,1]$ & $A_1, A_2$ \\
 & & $[0,1,\text{-}1]$ & $[0,0,2]$ & $A_1, B_1$ \\
\cline{2-5}
 & \multirow{3}{*}{$20^3, 24^3$}
   & $[0,\text{-}1,0]$ & $[0,2,1]$ & $A_1, B_1$ \\
 & & $[1,\text{-}1,1]$ & $[\text{-}1,2,0]$ & $A_1, A_2, B_1, B_2$ \\
 & & $[1,\text{-}1,0]$ & $[\text{-}1,2,1]$ & $A_1, A_2, B_1, B_2$ \\
\hline
\multirow{5}{*}{$\begin{matrix}[1,1,1] \\ \text{Dic}_3 \end{matrix}$}
& \multirow{3}{*}{$16^3, 20^3, 24^3$} 
 & $[0,0,0]$ & $[1,1,1]$ & $A_1$ \\
 & & $[1,0,0]$ & $[0,1,1]$ & $A_1$ \\
 & & $[2,0,0]$ & $[\text{-}1,1,1]$ & $A_1$ \\
\cline{2-5}
 & \multirow{2}{*}{$20^3, 24^3$}
   & $[1,\text{-}1,0]$ & $[0,2,1]$ & $A_1$ \\
 & & $[\text{-}1,0,0]$ & $[2,1,1]$ & $A_1$ \\
\end{tabular}
\end{ruledtabular}
\caption{The two-pion operators used presented for each $\vec{P}$ on various volumes; also shown is $\LG(\vec{P})$.  We give only the irreps that we considered in this work.  Example momenta $\vec{k}_1$ and $\vec{k}_2$ are shown; all momenta in $\{\vec{k}_1\}^{\star}$ and $\{\vec{k}_2\}^{\star}$ are summed over in Eq.~\ref{equ:twopionop}.}
\label{table:opsused}
\end{table}

\begin{table}[t]
\begin{ruledtabular}
\begin{tabular}{c | c c c}
 $\vec{P}$ & $\Lambda^{(P)}$ & $16^3$ & $20^3,24^3$ \\
\hline \hline
\multirow{3}{*}{$[0,0,0]$}
 & $A_1^+$ & 5 & 5 \\
 & $E^+$ & 3 & 3 \\
 & $T_2^+$ & 2 & 2 \\
\hline
\multirow{4}{*}{$[0,0,1]$}
 & $A_1$ & 4 & 7 \\
 & $E_2$ & 2 & 5 \\
 & $B_1$ & 1 & 3 \\
 & $B_2$ & 1 & 2 \\
\hline
\multirow{4}{*}{$[0,1,1]$}
 & $A_1$ & 5 & 8 \\
 & $A_2$ & 1 & 3 \\
 & $B_1$ & 1 & 4 \\
 & $B_2$ & 1 & 3 \\
\hline
\multirow{1}{*}{$[1,1,1]$}
 & $A_1$ & 3 & 5 \\
\end{tabular}
\end{ruledtabular}
\caption{The number of two-pion operators used for each $\vec{P}$ and irrep on the various lattice volumes.}
\label{table:numopsirreps}
\end{table}

\subsection{Example of $\vec{P}=[0,0,1]$, $\Lambda = A_1$}

\begin{figure*}
\includegraphics[width=0.9\textwidth
]{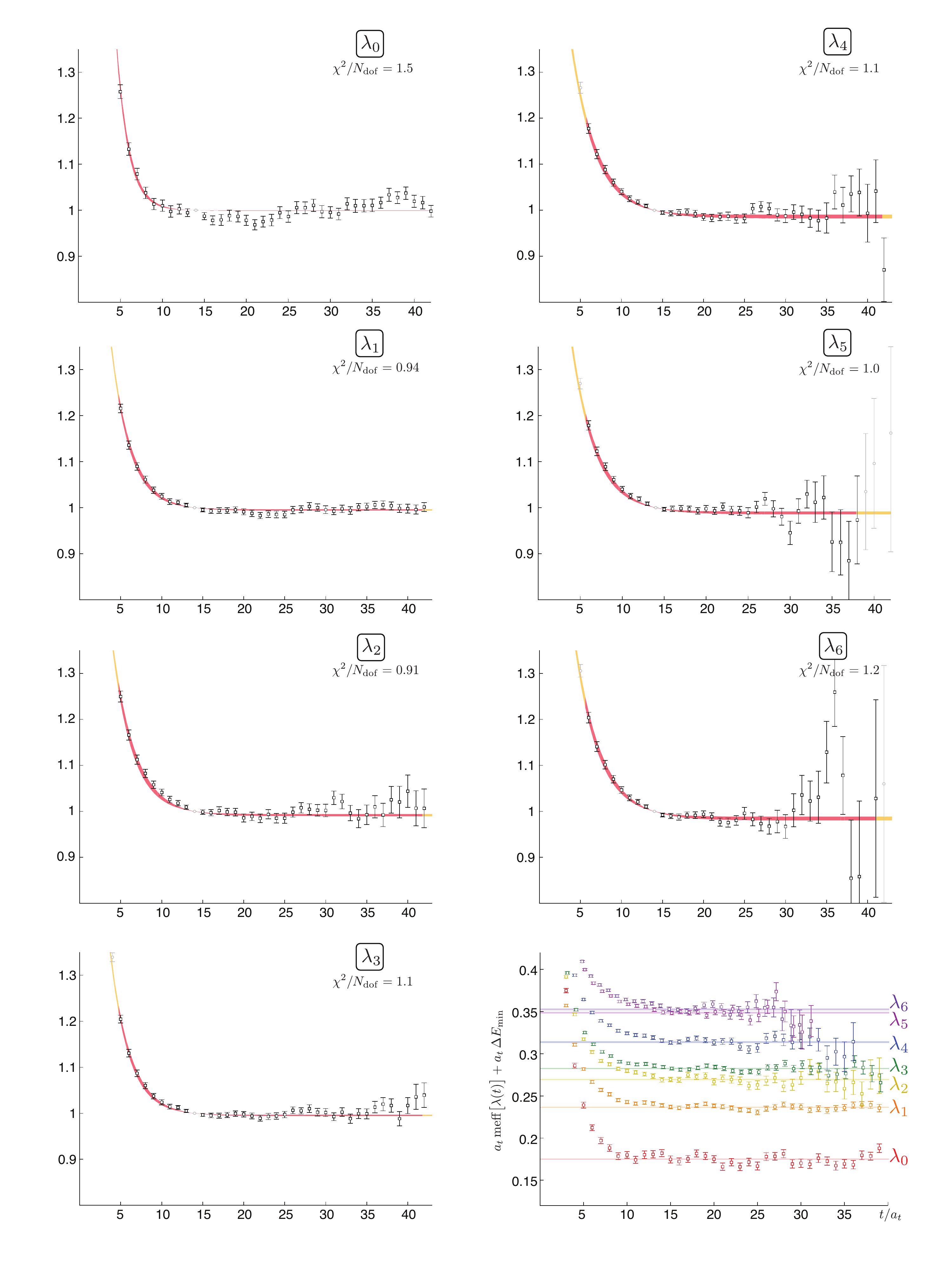}
\caption{
Principal correlators from solution of equation \ref{GEVP} applied to the weighted-shifted correlator matrix $\widehat{\widetilde{C}}(t)$ for $\vec{P}=[0,0,1]$, $\Lambda = A_1$ with $t_0 = 14 a_t$. Plotted is $e^{E (t-t_0)}\lambda(t)$ against $t/a_t$ along with fits to the time-dependence according to equation \ref{lambda_fit}. Also plotted in the bottom-right are the effective masses of the principal correlators (with the energy weighting $\Delta E_\mathrm{min}$ corrected) and the fit values $E$ superimposed as horizontal bands. All energies are those in the frame in which the lattice is at rest.
\label{P100_A1_prin_corr}}
\end{figure*}

\begin{figure*}
\includegraphics[width=0.9\textwidth
]{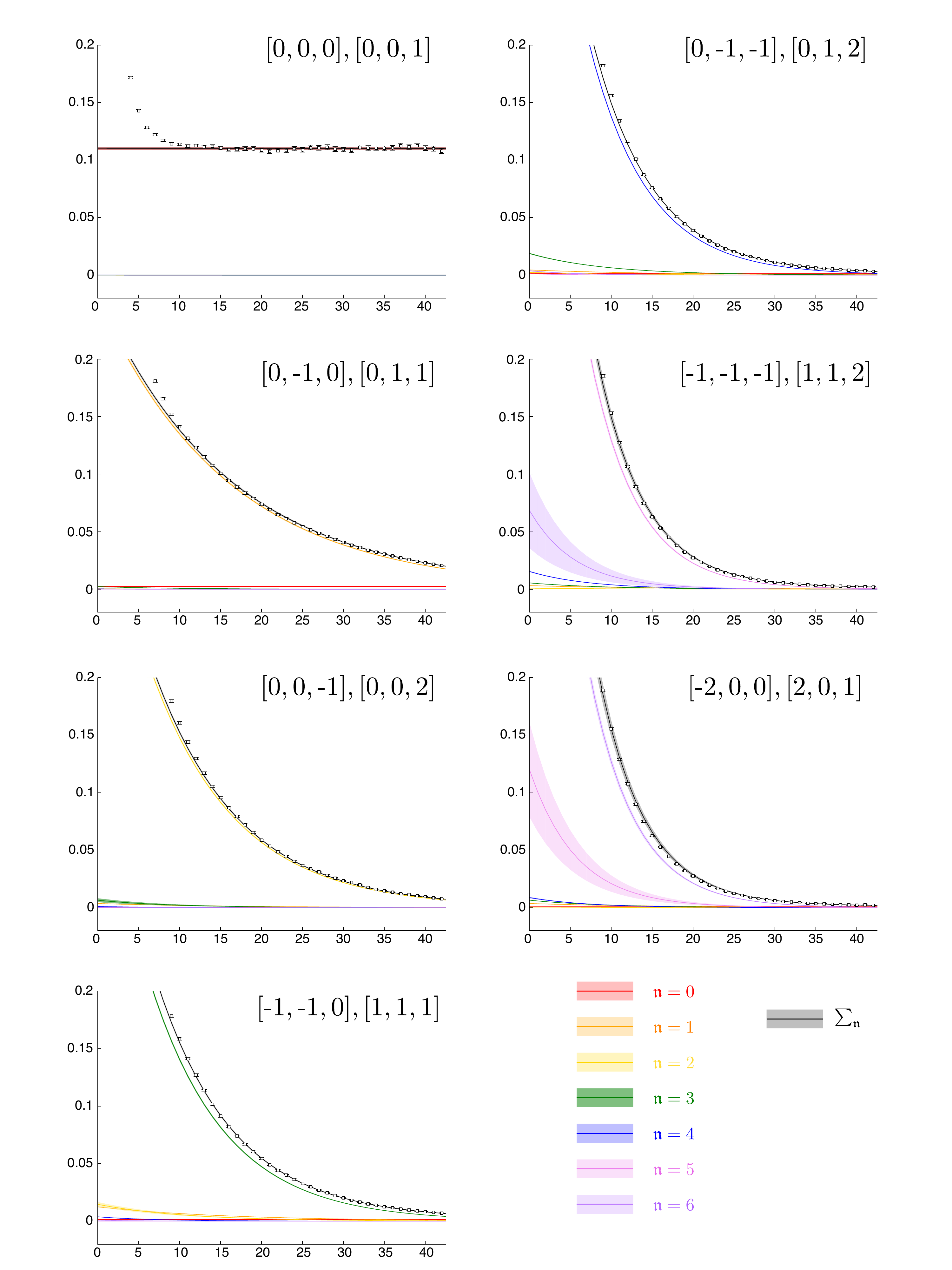}
\caption{
Diagonal elements of the weighted-shifted correlation matrix for $\vec{P}=[0,0,1]$, $\Lambda = A_1$: $\widehat{\widetilde{C}}_{[\vec{k}_1, \vec{k}_2]}^{[\vec{k}_1, \vec{k}_2]}(t)$ and their reconstruction using terms in the sum over states in equation \ref{eq:wanted}. Plotted is $e^{\widetilde{E}_{\pi\pi}^{(0)} t} \, \widehat{\widetilde{C}}(t)$. 
\label{P100_A1_recon}}
\end{figure*}

\begin{figure}
\includegraphics[width=0.45\textwidth
]{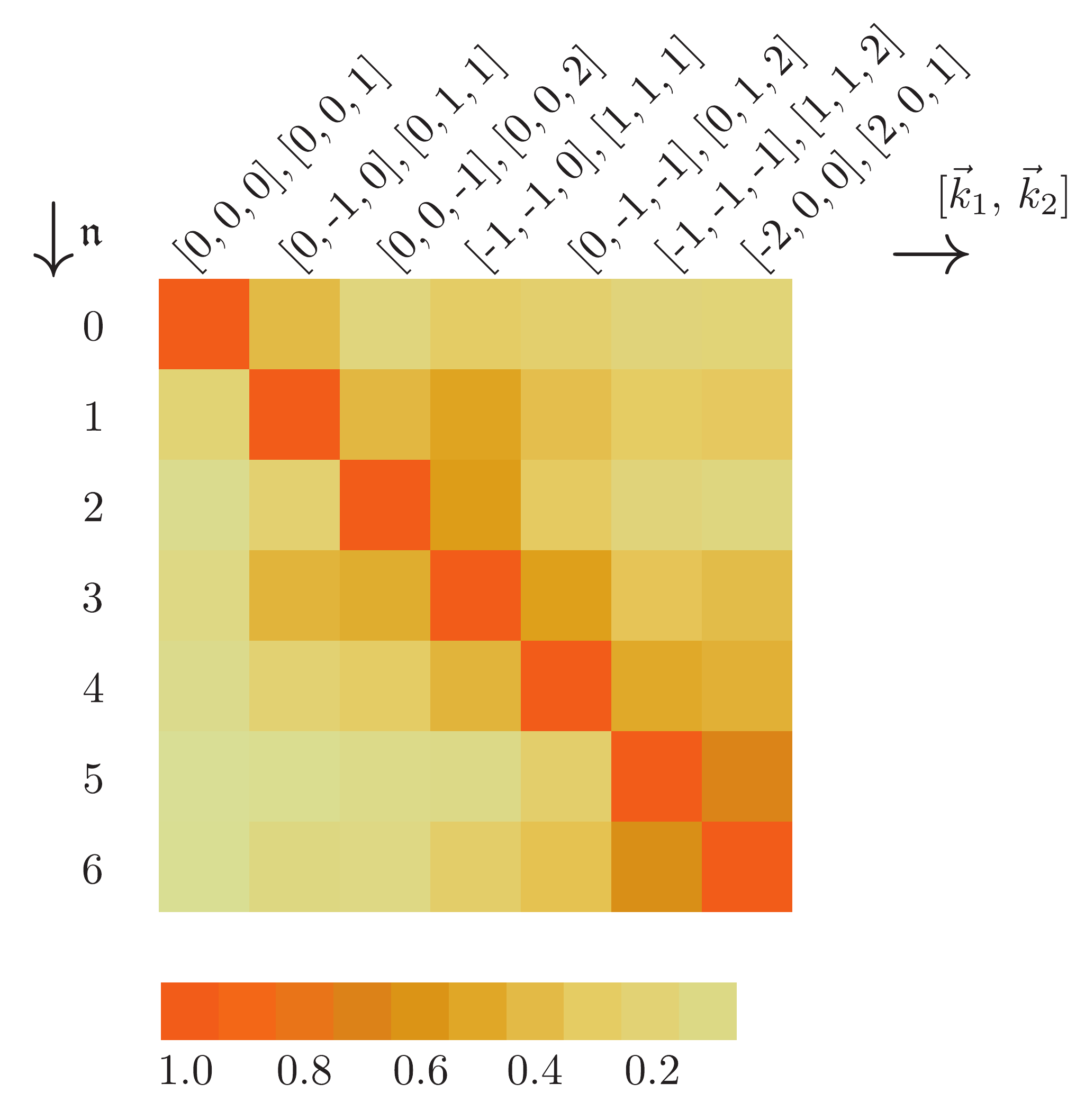}
\caption{
``Matrix" plot of values of $Z^{(\mathfrak{n})}_{[\vec{k}_1, \vec{k}_2]}$ normalised according to $\frac{Z^{(\mathfrak{n})}_{[\vec{k}_1, \vec{k}_2]} }{ \mathrm{max}_\mathfrak{n}\left[ Z^{(\mathfrak{n})}_{[\vec{k}_1, \vec{k}_2]} \right]}$ so that the largest overlap across all states for a given operator $[\vec{k}_1, \vec{k}_2]$ is unity.
\label{Z_P100}}
\end{figure}

As an explicit example of our variational fitting procedure consider the $\vec{P}=[0,0,1]$, $A_1$ irrep evaluated on the $L/a_s = 24$ lattice. Our basis of operators here is obtained by applying equation \ref{equ:twopionop} and is thus of the form 
\begin{equation}
\big( \pi\pi \big)_{[001], A_1}^{[\vec{k}_1, \vec{k}_2]\dag} = \sum_{\substack{\vec{k}_1 \in \{\vec{k}_1\}^\star \\ \vec{k}_2 \in \{\vec{k}_2\}^\star \\\vec{k}_1+\vec{k}_2 = [001] }}  \mathcal{C}([0,0,1],A_1; \vec{k}_1; \vec{k}_2 )\; \pi^\dag(\vec{k}_1)\, \pi^\dag(\vec{k}_2) \label{p100ops}
\end{equation}
with constructions using the pion momenta given in Table \ref{table:opsused}. This gives a correlation matrix, $C(t)$, of dimension 7. In order to remove the largest finite-$T$ effects, as discussed in the previous section, the \emph{weighted-shifted} correlation matrix, $\widehat{\widetilde{C}}(t)$ is formed, using $\Delta E_\mathrm{min} = E_\pi([0,0,1]) - m_\pi$. This matrix is analysed using equation \ref{GEVP} - for $t_0 = 14 a_t$, the obtained principal correlators, $\lambda_\mathfrak{n}(t)$ are shown in Figure \ref{P100_A1_prin_corr} along with fits of the form,
\begin{equation}
\lambda(t) = (1 - A) e^{- E (t -t _0)} + A e^{-E' (t-t_0)} \label{lambda_fit}
\end{equation}
where $E$, $E'>E$ and $A \ll 1$ are the fit parameters. The second exponential allows for the excited state\footnote{by ``excited states" here we might have several types, including $\pi\pi$ with large relative momenta, $\pi \pi^\star$ and other inelastic contributions.} pollution expected to be present for $t \lesssim t_0$ (our reported spectra are just the values of $E$, $E'$ is discarded).  The fits are very good and the absence of any significant upward curvature at larger $t$ (as in Figure \ref{corr_flight_finiteT_model}) suggests that our weighting-shifting procedure has removed the bulk of the finite-$T$ pollution\footnote{such upward curvature \emph{is} seen in variational analysis of the raw correlator matrix, $C(t)$.}.

The solution of equation \ref{GEVP} also provides eigenvectors $v^{(\mathfrak{n})}$ which can be converted into overlaps, $Z^{(\mathfrak{n})}_{[\vec{k}_1, \vec{k}_2]} \equiv \big\langle (\pi\pi)_\mathfrak{n};\,[0,0,1], \, A_1 \big| \big( \pi\pi \big)_{[001], A_1}^{\dag[\vec{k}_1, \vec{k}_2]} \big| 0 \big\rangle$ using $\widehat{Z}^{(\mathfrak{n})}_{[\vec{k}_1, \vec{k}_2]} = \big(\widehat{v}^{(\mathfrak{n})\dag} \widehat{\widetilde{C}}(t_0)\big)_{[\vec{k}_1, \vec{k}_2]} \,e^{\widetilde{E}_\mathfrak{n} t_0 / 2}$. Our method of solution of the generalised eigenvalue problem treats each timeslice independently such that we actually obtain $v^{(\mathfrak{n})}(t)$ and thus $\widehat{Z}(t)$. This time-dependence is fitted to a constant (or a constant plus an exponential if that is required to get a good fit) and the resulting constant is rescaled to undo the effect of the shifting of the correlators in the manner prescribed by equation \ref{Zscale}.

The overall quality of description of the correlators by the variational solution can be seen in Figure \ref{P100_A1_recon} along with an indication of how much each $\big|(\pi\pi)_\mathfrak{n}\big\rangle$ state contributes to each of the diagonal correlators. These contributions are reconstructed from the results of the variational analysis by building the sum in equation \ref{eq:wanted} state-by-state. The description, as one would expect, is excellent for $t > t_0$; indeed the ability to get a good description of the correlators using only the number of states equal to the basis size is our condition to determine an appropriate value of $t_0$ \cite{Dudek:2007wv}. That we are able to countenance a value as low as $t_0 = 14 a_t$ is due to our use of optimised pion operators so that $\pi\pi^\star$ contributions to the correlators are much reduced.

It is apparent in Figure \ref{P100_A1_recon} that the basis of operators, defined by equation \ref{p100ops}, is rather close to a diagonalising basis and this can be clearly seen in Figure \ref{Z_P100} which shows the $Z$ values for each state $\mathfrak{n}$ and each operator $[\vec{k}_1, \vec{k}_2]$. This indicates that the finite-volume $\pi\pi$ eigenstates are close to being states of definite pion momentum which agrees with the expectation that the $I=2$ interpion interaction strength is weak and the observation of only small shifts from non-interacting $\pi\pi$ energies. It is interesting to note that the largest deviations from diagonal behaviour, i.e. the largest mixing of the non-interacting state basis, occurs for levels which are very close in energy. This is precisely what we would expect from perturbation theory, where small energy denominators enhance mixing of near-degenerate states. That we are able to resolve this mixing with a high degree of confidence is an advantage of our use of a variational approach.

\subsection{Volume dependence of $\pi\pi$ spectra}

\begin{figure}
\includegraphics[width=0.5\textwidth
]{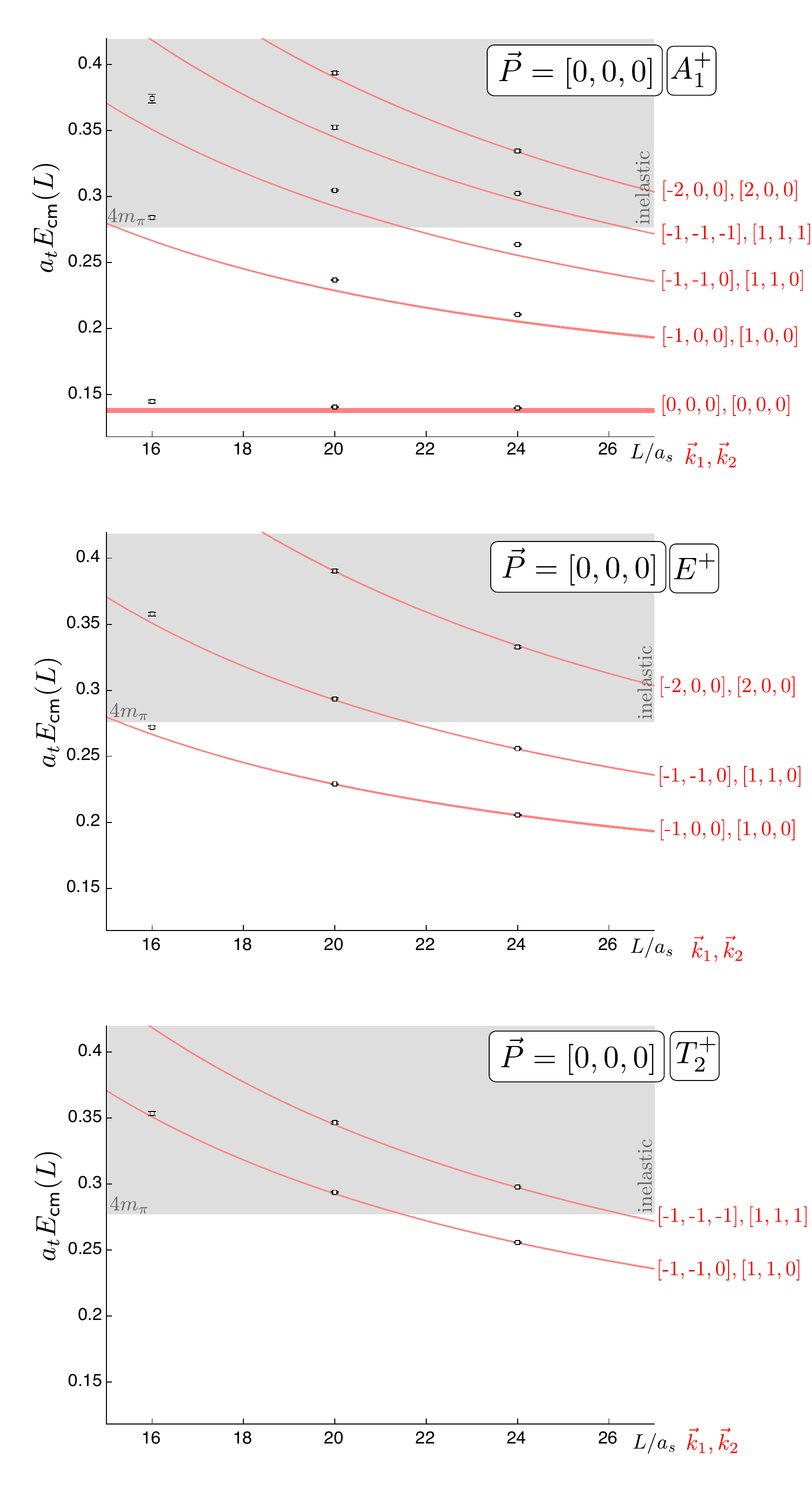}
\caption{
Extracted center-of-momentum frame energy spectra for $\vec{P}=[0,0,0]$ irreps $A_1^+,E^+,T_2^+$. Also shown (in red) are non-interacting pion pair energies, $\sqrt{m_\pi^2 + |\vec{k}_1|^2} + \sqrt{m_\pi^2 + |\vec{k}_2|^2}$ whose uncertainty is determined by the uncertainty on $a_t m_\pi$ and $\xi$ determined in Section \ref{sec:dispersion}. Grey area represents opening of inelastic ($4\pi$) threshold.
\label{P000}}
\end{figure}

\begin{figure}
\includegraphics[width=0.5\textwidth
]{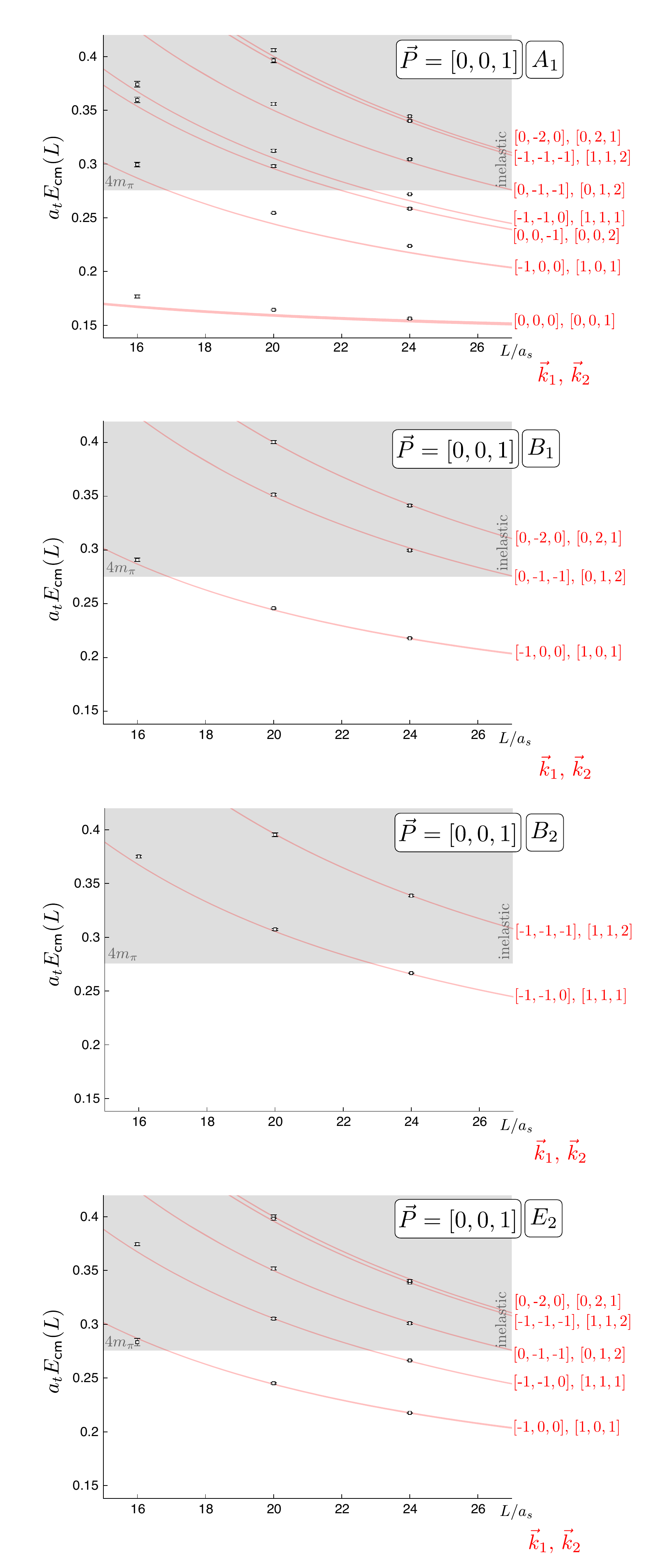}
\caption{
As Figure \ref{P000} for $\vec{P}=[0,0,1]$.
\label{P100}}
\end{figure}

\begin{figure}[h]
\includegraphics[width=0.5\textwidth
]{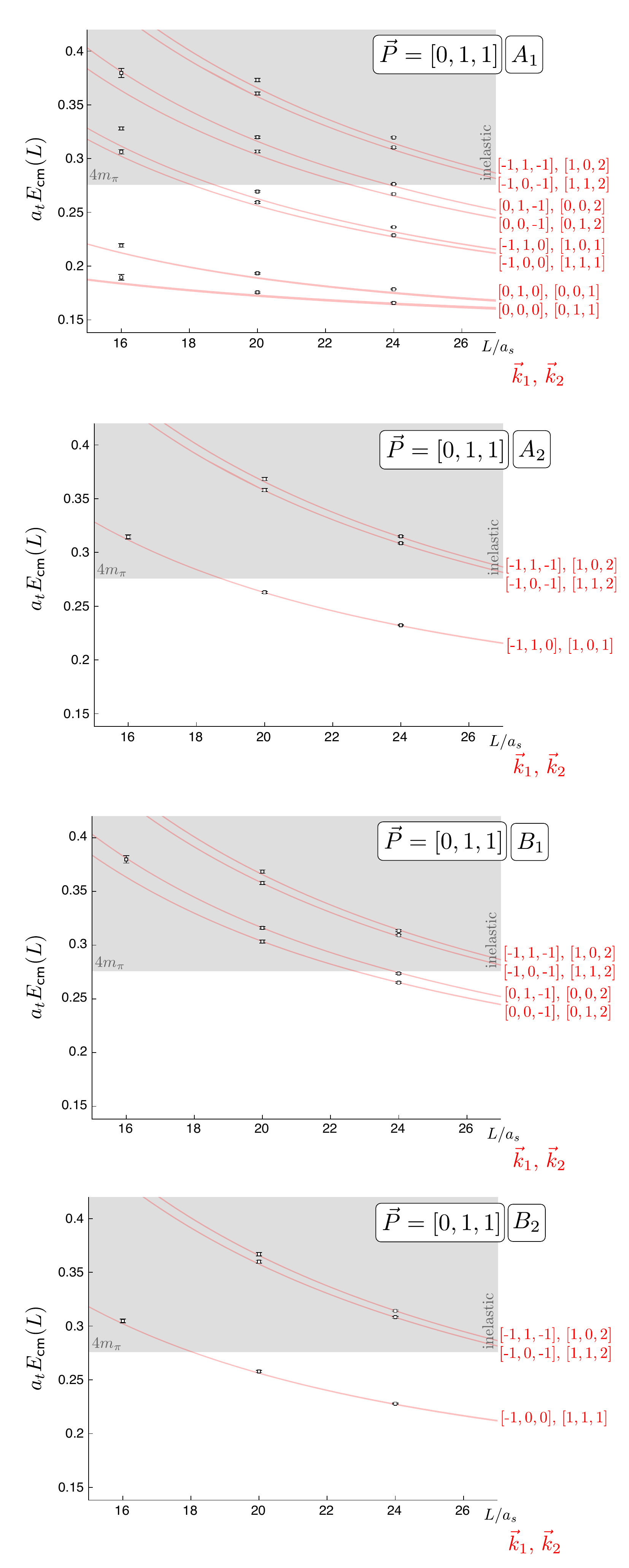}
\caption{
As Figure \ref{P000} for $\vec{P}=[0,1,1]$.
\label{P110}}
\end{figure}

\begin{figure}[h]
\includegraphics[width=0.5\textwidth
]{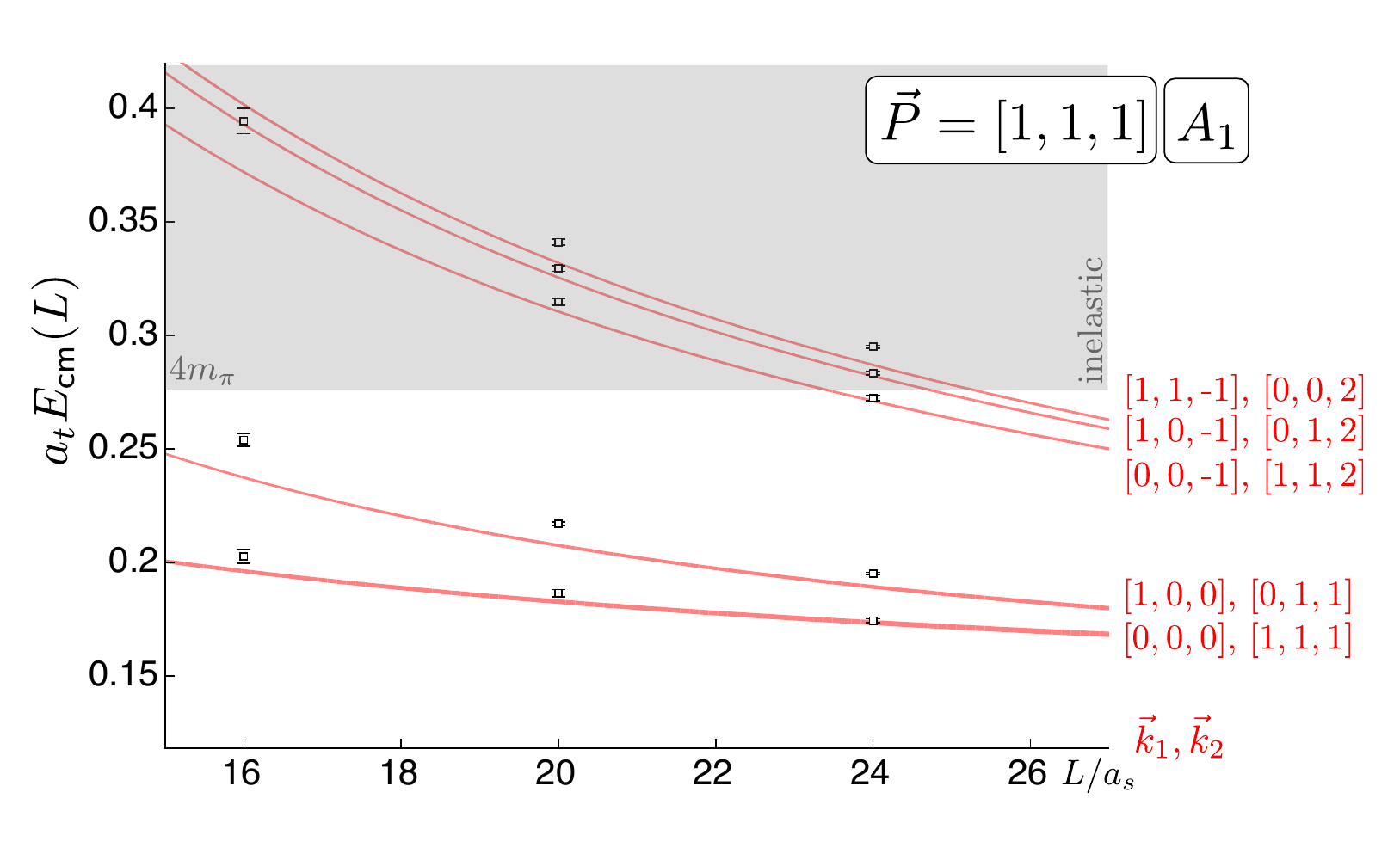}
\caption{
As Figure \ref{P000} for $\vec{P}=[1,1,1]$.
\label{P111}}
\end{figure}

We perform this analysis procedure independently for each $\vec{P},\Lambda$ on each volume. The energies obtained are in the frame in which the lattice is at rest, and can be more usefully expressed in the $\pi\pi$ center-of-momentum frame, 
\begin{align}
E_\mathsf{cm} &= \sqrt{ E_\mathsf{lat}^2 - |\vec{P}|^2} \nonumber \\
\big( a_t E_\mathsf{cm} \big) &= \bigg[ \big(a_t E_\mathsf{lat}\big)^2 - \tfrac{1}{\xi^2} \left(\tfrac{2\pi}{L/a_s} \right)^2 n^2_{\vec{P}}  \bigg]^{1/2} \label{kinematics}
\end{align}
where we use the anisotropy, $\xi$, determined from the pion dispersion relation in Section \ref{sec:dispersion}. In Figures \ref{P000},\ref{P100},\ref{P110},\ref{P111} we show the volume dependence of the extracted center-of-momentum frame energy spectrum along with the energies of pairs of non-interacting pions carrying various allowed lattice momenta. 

In all cases we observe small energy shifts, with the largest shifts in $A_1$ irreps, reflecting the expected strongest interaction in $S$-wave scattering.


\clearpage
\section{Phase-shifts from finite-volume spectra}\label{sec:luescher}

The formalism to relate the amplitude for two-particle elastic scattering in partial waves labelled by angular momentum $\ell$, to the spectrum of states in a finite cubic spatial volume, is laid down in \cite{Luscher:1991cf}, with extensions to the case of a moving frame presented in \cite{Rummukainen:1995vs,Kim:2005gf,Christ:2005gi}.  Because we are considering $\pi\pi$ scattering in isospin-2 where only even $\ell$ occur, there is a one-to-one correspondence between the irreps of the symmetry group relevant for the L\"uscher formalism in a moving frame and the little group irreps\footnote{We note that the symmetry group relevant for the L\"uscher formalism here is the subgroup of $\text{O}^{\text{D}}_h$ under which $\vec{P} \rightarrow \pm\vec{P}$ rather than the constraint for little groups that $\vec{P} \rightarrow \vec{P}$.  The irreps are similar to those of the little groups but have an additional ``parity'' label.}
 and so we will refer to the little group irreps.

The formalism can be compactly expressed in a single equation,
\begin{equation}
	\det\left[ \mathbf{E}(p_\mathsf{cm}) - \mathbf{U}^{(\vec{P},\Lambda)}\Big( \big(\tfrac{p_\mathsf{cm} L}{2\pi} \big)^2 \Big) \right] = 0.\label{luescher}
\end{equation}
$\mathbf{U}$ is a formally infinite-dimensional matrix of known functions whose rows and columns are each labelled by the pair $(\ell, n)$, $U_{\ell n;\ell' n'}$. $\{\ell\}$ are the angular momenta which subduce into the irrep, $\Lambda$, and $n$ is an index indicating the $n^\mathrm{th}$ embedding of that $\ell$ into this irrep; the pattern of these subductions is given in Table \ref{table:pipiirreps}.  $\mathbf{U}$ is a function of the dimensionless variable $q^2 = \big(\tfrac{p_\mathsf{cm} L}{2\pi} \big)^2$, featuring the center-of-momentum frame scattering momentum and the spatial length of the cubic lattice, $L$. 

$\mathbf{E}$ is a diagonal matrix, independent of $L$, which encodes the scattering amplitude through the elastic scattering phase-shifts, $\delta_\ell(p_\mathsf{cm})$, as $E_{\ell n;\ell' n'} = e^{2i \delta_\ell(p_\mathsf{cm})} \delta_{\ell' \ell} \delta_{n'n}$. 

$\mathbf{U}$ is conveniently expressed in terms of a matrix $\mathbf{M}$ as $\mathbf{U} = \big( \mathbf{M} +i \mathbf{1} \big) \big( \mathbf{M} -i \mathbf{1} \big)^{-1}$ where we can obtain the elements of $\mathbf{M}$ using
\begin{widetext}
\begin{equation}
	\mathcal{M}^{(\vec{P},\Lambda, \mu)}_{\ell n; \ell' n'}(q^2) \;\delta_{\Lambda,\Lambda'} \delta_{\mu,\mu'} = 
	\sum_{\substack{\hat{\lambda}=\pm |\lambda| \\ m=-\ell \ldots \ell }} \;
	\sum_{\substack{\hat{\lambda}'=\pm |\lambda'| \\ m'=-\ell' \ldots \ell' }} \;
	\mathcal{S}_{\vec{P},\Lambda,\mu}^{\tilde{\eta},\lambda*} \, D^{(\ell)*}_{m \lambda}(R) \cdot \mathcal{M}^{(\vec{P})}_{\ell m; \ell' m'}(q^2) \cdot \mathcal{S}_{\vec{P},\Lambda,\mu'}^{\tilde{\eta},\lambda'} \, D^{(\ell')}_{m' \lambda'}(R).
\end{equation}
\end{widetext}
In this equation, $R$ is a rotation carrying the $J_z$ quantisation axis $(0,0,P)$ into $\vec{P}$, with $D^{(\ell)}_{m \lambda}(R)$ relating $J_z$ values, $m$, to helicities, $\lambda$. A convention for constructing $R$ is given in \cite{Thomas:2011rh}. $\mathcal{S}_{\vec{P},\Lambda,\mu}^{\tilde{\eta},\lambda}$ is the subduction from helicity $\lambda$ to the $\mu^\mathrm{th}$ row of the lattice irrep $\Lambda$ (see Appendix \ref{app:operators}). Different magnitudes of helicity, $|\lambda|,|\lambda'|$ give rise to the different embeddings $n,n'$. The ``reflection parity", $\tilde{\eta} \equiv P(-1)^\ell = +$ for a system of two pseudoscalars.  
$\mathcal{M}^{(\vec{P})}_{\ell m; \ell' m'}$ is the same object defined in equation (89) of \cite{Rummukainen:1995vs} where it is expressed in terms of a known linear combination of generalised zeta functions of argument $q^2$.

One potential use of equation \ref{luescher} is to take a scattering problem where the amplitudes are known and find the corresponding spectrum of states in a certain finite-volume box. For a known set of scattering phase-shifts, $\{\delta_\ell(p_\mathsf{cm})\}$, the finite-volume spectrum on an $L\times L \times L$ spatial lattice can be obtained by solving equation \ref{luescher} for discrete values of $p_\mathsf{cm}$ which give discrete values of $E_\mathsf{lat}$. Of course in practice, for any given lattice irrep, $\Lambda$, we need to truncate the infinite $(\ell, n)$ basis to the set of phase-shifts $\{\delta_\ell(p_\mathsf{cm})\}$ known to us. Fortunately, at low scattering momentum there is a hierarchy in $\delta_\ell(p_\mathsf{cm})$ which follows from angular momentum conservation, $\delta_\ell(p_\mathsf{cm}) \sim p_\mathsf{cm}^{2\ell +1}$, such that $\delta_0 \gg \delta_2 \gg \delta_4 \ldots$, and we may be justified in making a finite truncation in $\ell$.

\begin{table}[!h]
\begin{ruledtabular}
\begin{tabular}{c c | c  l }
$\vec{P}$ & $\LG(\vec{P})$ & $\Lambda^{(P)}$ & $\pi\pi ~ \ell^N$\\
\hline \hline
\multirow{5}{*}{$[0,0,0]$} & \multirow{5}{*}{$\text{O}^{\text{D}}_h$} 
   & $A_1^+$ & $0^1,~4^1$\\
 & & $T_1^+$ & $4^1$\\
 & & $T_2^+$ & $2^1,~4^1$\\
 & & $E^+$   & $2^1,~4^1$\\
\hline
\multirow{5}{*}{$[0,0,n]$} & \multirow{5}{*}{$\text{Dic}_4$} 
   & $A_1$ & $0^1,~2^1,~4^2$\\
 & & $A_2$ & $4^1$\\
 & & $E_2$ & $2^1,~4^2$\\
 & & $B_1$ & $2^1,~4^1$\\
 & & $B_2$ & $2^1,~4^1$\\
\hline
\multirow{4}{*}{$[0,n,n]$} & \multirow{4}{*}{$\text{Dic}_2$} 
   & $A_1$ & $0^1,~2^2,~4^3$\\
 & & $A_2$ & $2^1,~4^2$ \\
 & & $B_1$ & $2^1,~4^2$\\
 & & $B_2$ & $2^1,~4^2$\\
\hline
\multirow{3}{*}{$[n,n,n]$} & \multirow{3}{*}{$\text{Dic}_3$} 
   & $A_1$ & $0^1,~2^1,~4^2$\\
 & & $A_2$ & $4^1$\\
 & & $E_2$ & $2^2,~4^3$\\
\hline
$[n,m,0]$ & \multirow{2}{*}{$\text{C}_4$}
   & $A_1$ & $0^1,~2^3,~4^5$\\
 $[n,n,m]$ & & $A_2$ & $2^2,~4^4$\\
\end{tabular}
\end{ruledtabular}
\caption{The pattern of subductions of $I=2$ $\pi\pi$ partial waves, $\ell \leq 4$, into lattice irreps, $\Lambda$, where $N$ is the number of embeddings of this $\ell$ in this irrep.  This table is derived from Table~\ref{table:latticeirreps} by considering the subductions of the $\ell$ for $\vec{P}=\vec{0}$ and the various helicity components for each $\ell$ for $\vec{P}\neq\vec{0}$.  Here $\vec{P}$ is given in units of $\tfrac{2\pi}{L}$ and $n,m$ are non-zero integers with $n \ne m$.  We show the double-cover groups but only give the irreps relevant for integer spin.  
}
\label{table:pipiirreps}
\end{table}

\clearpage
\subsection{A toy model of $\pi\pi$ scattering}

\begin{figure*}[!t]
\includegraphics[width=0.9\textwidth
]{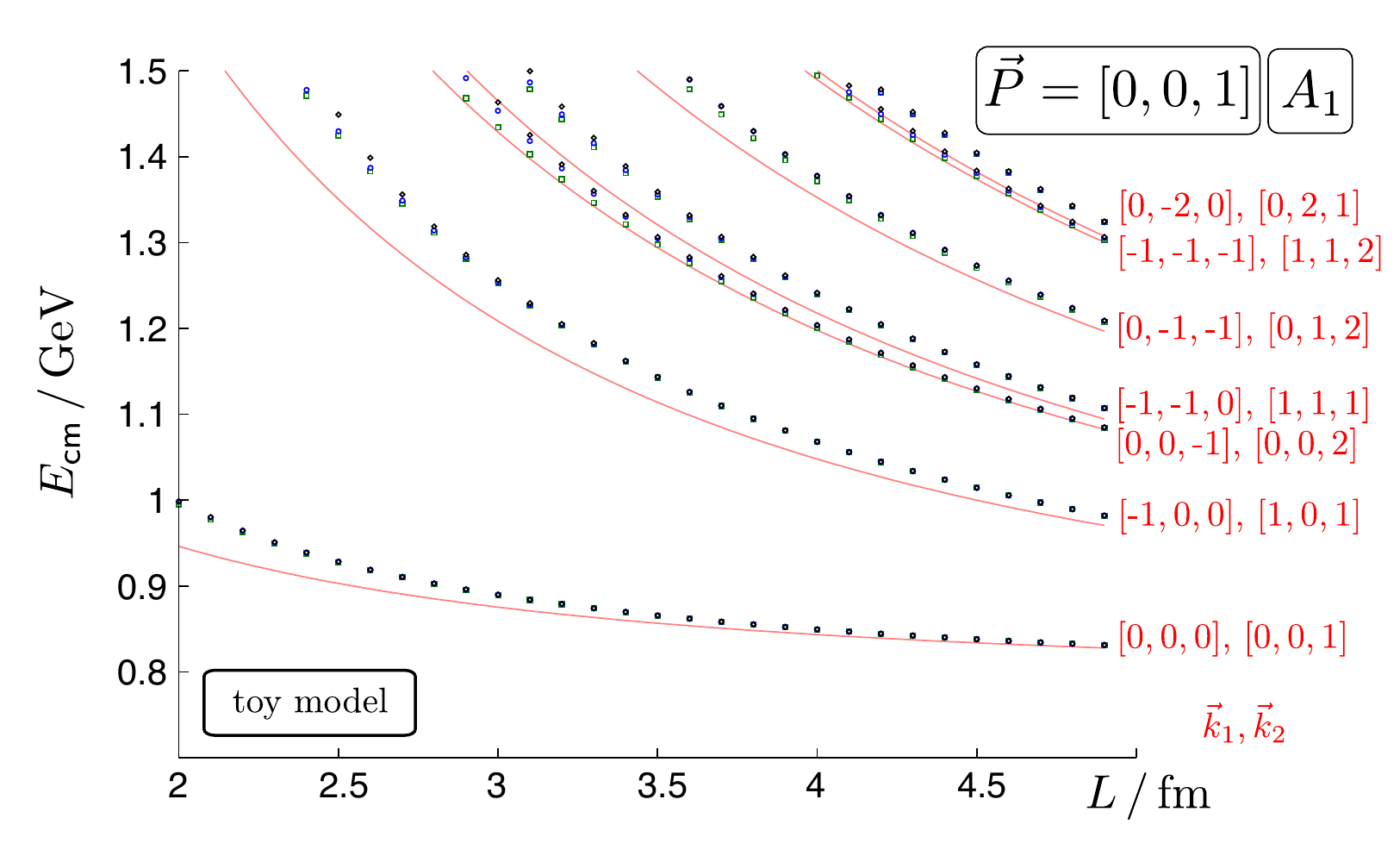}
\caption{
Finite-volume spectrum for the toy model of effective-range parameterisations in the irrep $\vec{P}=[0,0,1]$, $A_1$. Green squares indicate the spectrum including only $\ell=0$ scattering, blue circles include $\ell=0,2$ and black diamonds include $\ell=0,2,4$. Note that for many of the energy levels the squares, circles and diamonds lie on top of each other. Red curves show non-interacting energies of pion pairs with momenta $\vec{k}_1, \vec{k}_2$.
\label{P100_A1_rev_eng}}
\end{figure*}

In order to demonstrate the formalism, we will briefly break away from analysis of lattice QCD obtained finite-volume spectra to consider a simple toy-model of $\pi\pi$ scattering in which the scattering amplitudes are known to us.
The toy model is built from an effective range parameterisation of elastic scattering in $\ell = 0,2,4$ partial waves. We have 
\begin{equation}
p_\mathsf{cm}^{2\ell +1} \cot \delta(p_\mathsf{cm}) = \frac{1}{a_\ell} + \frac{1}{2} r_\ell \,p_\mathsf{cm}^2, \label{effrange}
\end{equation}
with parameters
\begin{align}
a_0 &= -0.8\, \mathrm{GeV}^{-1}, \, r_0 = +2.5 \, \mathrm{GeV}^{-1}, \nonumber \\
a_2 &= -2.4 \, \mathrm{GeV}^{-5},\, r_2 \equiv 0, \nonumber \\
a_4 &= -5.0 \, \mathrm{GeV}^{-9}, \, r_4 \equiv 0, \nonumber \label{effrangevalues}
\end{align}
which happens to reasonably describe the experimental $\pi\pi$ $I=2$ scattering data up to a momentum $p_\mathsf{cm} \sim 0.6 \, \mathrm{GeV}$ \cite{Hoogland:1977kt,Cohen:1973yx,Zieminski:1974ex,Durusoy:1973aj}. 

Given this parameterisation and the choice $m_\pi = 0.396 \,\mathrm{GeV}$ we solve equation \ref{luescher} for the finite-volume spectrum in several irreps, $(\vec{P},\Lambda)$, over a range of volumes, $L = 2.0\to 5.0 \, \mathrm{fm}$. In Figure \ref{P100_A1_rev_eng} we show the center-of-momentum frame finite-volume energy spectrum for one example irrep $\vec{P}=[0,0,1],\, \Lambda = A_1$. 
At each volume we show the spectrum obtained from three different scattering parameterisations: 
the green squares show the spectrum with only $S$-wave scattering ($\delta_2 = \delta_4 \equiv 0$), the blue circles include also $D$-wave scattering ($\delta_4 \equiv 0$), and the black diamonds correspond to all of $\delta_{0,2,4}$ being described by the effective range parameterisations given above. We observe that the contribution of higher partial waves to determining the finite-volume energy varies with excitation level.

The problem to be solved in lattice QCD is actually the inverse of that just described - we start with the finite-volume spectrum determined through analysis of correlation functions and want to find the phase-shifts as a function of scattering momentum. If a given irrep received contributions from only a single $\ell$ this would be relatively simple - we would solve equation \ref{luescher} for unknown $\delta_\ell(p_\mathsf{cm})$ by inputting the determined value of $p_\mathsf{cm}$ extracted from $E_\mathsf{lat}$ using equation \ref{kinematics}. The toy model construction indicates the potential difficulty with such a naive approach - equation \ref{luescher} depends on the value of many $\delta_\ell$ simultaneously and on the face of it this is an underconstrained problem. 

\begin{figure*}
\includegraphics[width=0.47\textwidth
]{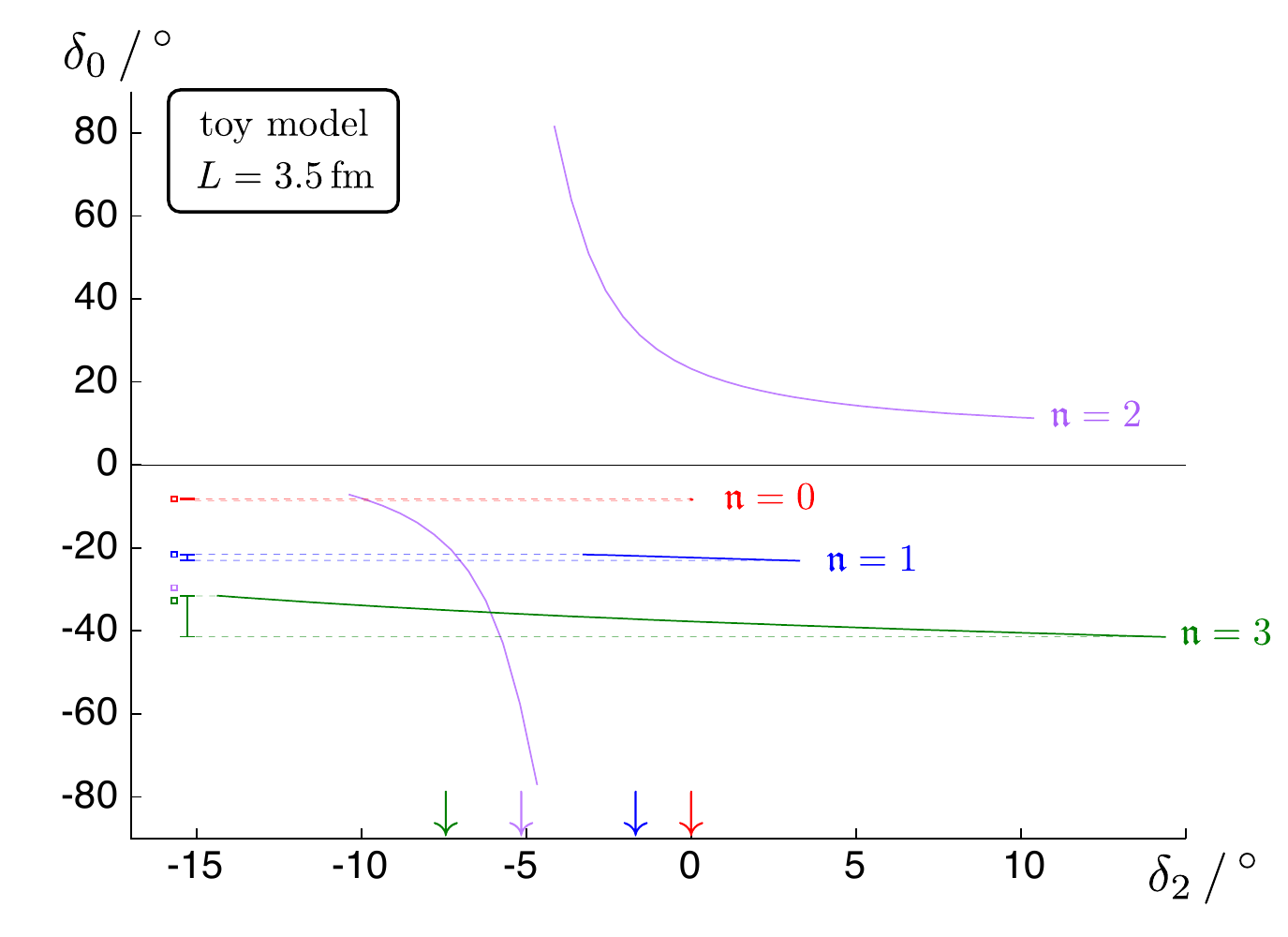}
\includegraphics[width=0.47\textwidth
]{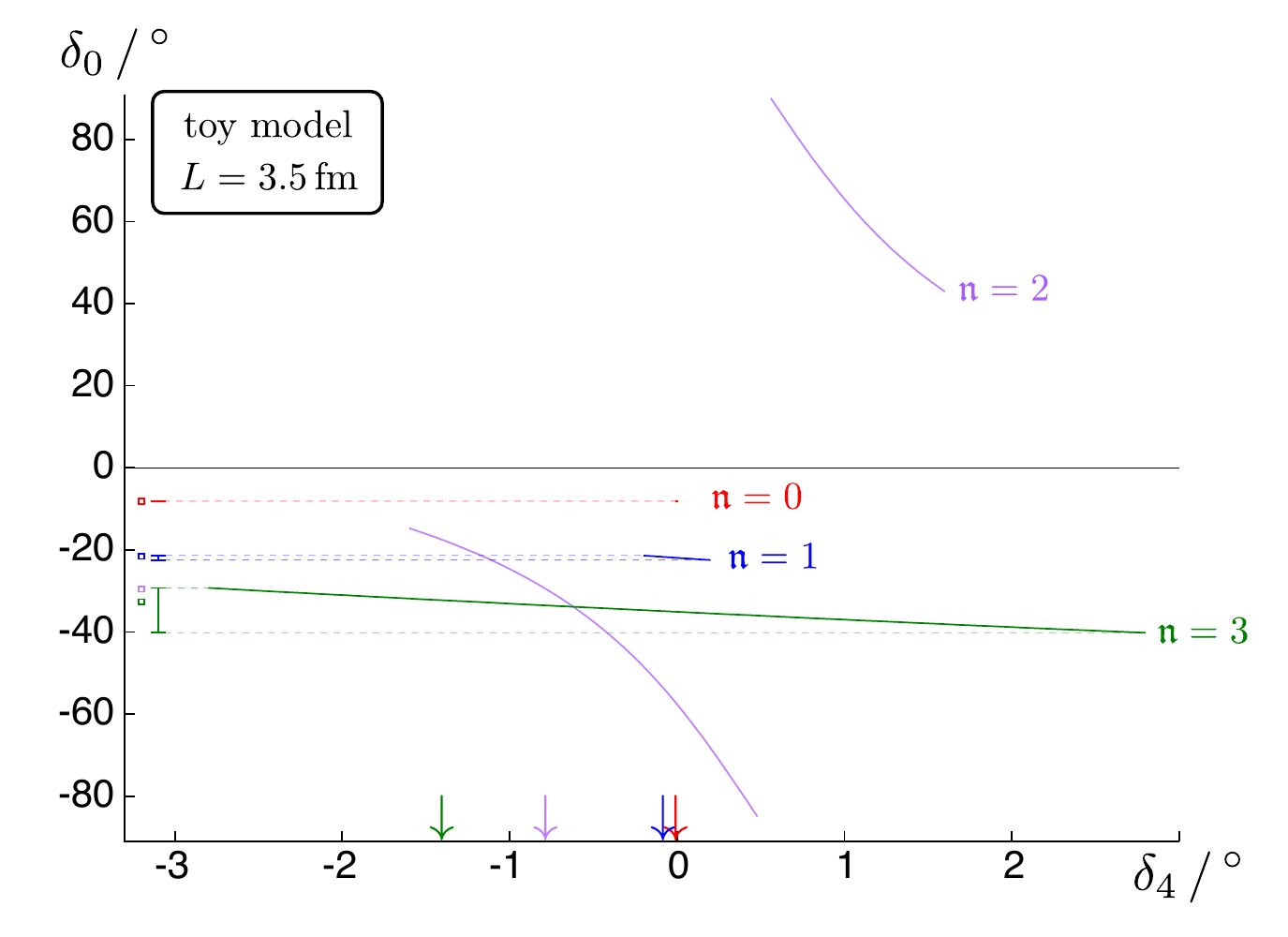}
\caption{Lowest four energy levels ($\mathfrak{n} = 0,1,2,3$) in toy model with volume $L=3.5\,\mathrm{fm}$ in irrep $\vec{P}=[0,0,1]$, $A_1$.\\
\hspace{0.5cm}(a) Sensitivity of $\delta_0$ extracted from equation \ref{luescher} as a function of assumed values of $\delta_2$ in range $\pm 2|\delta^\mathrm{exact}_2|$ with $\delta_4=0$\\
\hspace{0.5cm}(b) Sensitivity of $\delta_0$ extracted from equation \ref{luescher} as a function of assumed values of $\delta_4$ in range $\pm 2|\delta^\mathrm{exact}_4|$ with $\delta_2=\delta_2^\mathrm{exact}$\\
Boxes on far left indicate exact values of $\delta_0$ at the corresponding scattering momenta. Arrows on $x$-axis indicate exact values of $\delta_{2,4}$
\label{delta2_sensitivity}}
\end{figure*}

Within the toy model we can explore the effect of the simplest possible assumption that higher partial waves contribute only negligibly - consider the spectrum in $\vec{P}=[0,0,1],\, \Lambda=A_1$ for $L = 3.5\,\mathrm{fm}$. In Figure \ref{delta2_sensitivity}(a) we show the extracted $\delta_0$ for the lowest four energy levels as a function of a supplied value\footnote{included in equation \ref{luescher} in $\mathbf{E}$ as a fixed parameter} of $\delta_2$ (and with $\delta_4 = 0$). The naive assumption of $\delta_2 = 0$ is seen to give reasonable estimates of $\delta_0$ for the lowest two levels, but to be significantly discrepant for the next two levels. Varying $\delta_2$ between $\pm2|\delta_2^\mathrm{exact}|$ (which we know because this is a toy model) gives the curves shown. Figure \ref{delta2_sensitivity}(b) shows the sensitivity to $\delta_4$ assuming that $\delta_2$ is known exactly.

In this exercise we explicitly see that the influence of higher partial waves can vary significantly between levels; for $\mathfrak{n}=0,1,3$ the influence of $\delta_{2,4}$ is modest and given a ``reasonable" estimate of their magnitude we could assign a systematic error on $\delta_0$ that would encompass the exact result. On the other hand, no information can be obtained from level $\mathfrak{n}=2$ without very precise knowledge of both $\delta_2$ and $\delta_4$ at the corresponding scattering momentum. This will not be possible in any practical calculation and we must be careful to identify those cases where an energy level shows such extreme sensitivity.

\begin{figure*}
\includegraphics[width=0.95\textwidth
]{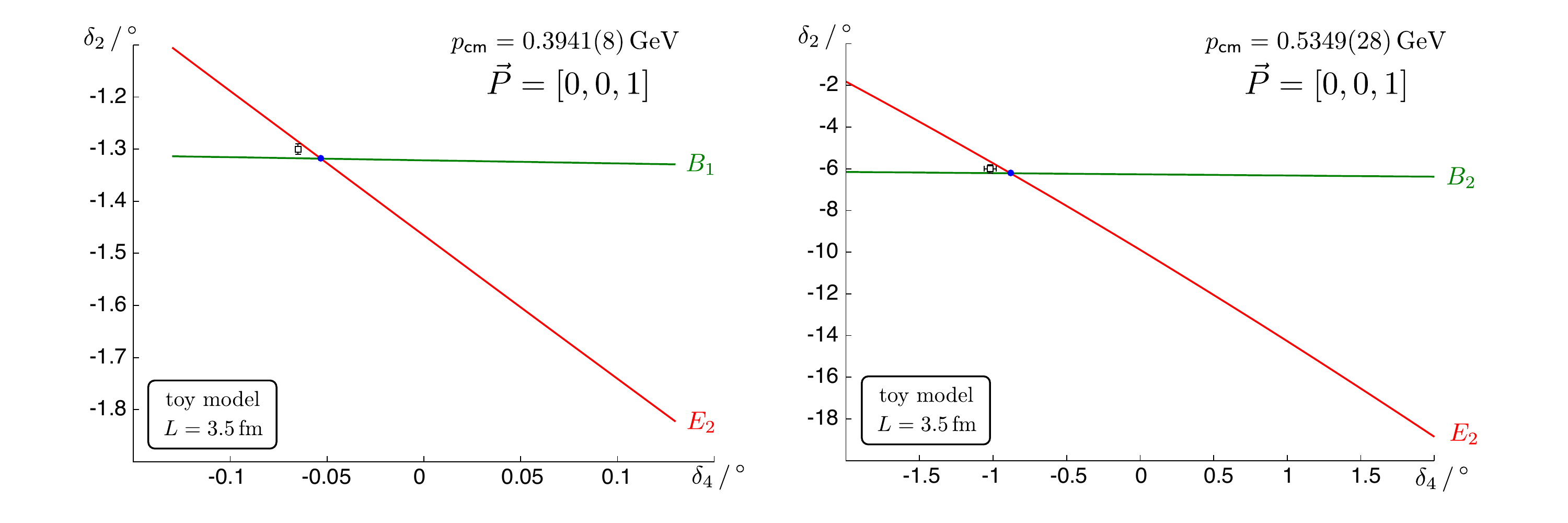}
\caption{Simultaneous solution of two equations \ref{luescher} for $\delta_{2,4}$. Open black square shows exact values with uncertainty indicating the variation in $\delta_{2,4}^\mathrm{exact}$ over the momentum region between the two determined $p_\mathsf{cm}$.\\
(a) $\vec{P}=[0,0,1]$, $\mathfrak{n}=0$ in $E_2$ and $\mathfrak{n}=0$ in $B_1$. (b) $\vec{P}=[0,0,1]$, $\mathfrak{n}=1$ in $E_2$ and $\mathfrak{n}=0$ in $B_2$.
\label{coupled}}
\end{figure*}

Thus even if our main aim is only to determine $\delta_0(p_\mathsf{cm})$ we see that it is incumbent upon us to also estimate $\delta_{\ell > 0}$. The easiest way to do this is to analyse the finite-volume spectra of irreps which receive no contribution from $\ell=0$, see Table \ref{table:pipiirreps}. Typically any irrep that features $\ell = 2$ will also feature $\ell = 4$ so we have a similar problem of estimating $\delta_2$ given no knowledge of $\delta_4$. Fortunately in the case under consideration where the interactions are weak we encounter situations in which energy levels in two different irreps have very similar energy values. For example with $\vec{P}=[0,0,1]$, the lowest level in $E_2$ and the lowest level in $B_1$ are both very close to the non-interacting $\big( \vec{k}_1= [0,\text{-}1,0],\,\vec{k}_2= [0,1,1] \big)$ level and correspond to $p_\mathsf{cm}$ values of $0.03934, 0.03950\, \mathrm{GeV}$ respectively. In this case, to the extent that $\delta_{2,4}(p_\mathsf{cm})$ do not change significantly over the small difference in $p_\mathsf{cm}$, and the functions in $\mathbf{U}$ are not rapidly varying over the corresponding range in $q^2$, we can solve the coupled system of two equations \ref{luescher} (for $E_2$ and $B_1$) for the two unknowns $\delta_2, \delta_4$. This is demonstrated in Figure \ref{coupled}(a) where the simultaneous solution of the two equations is seen to be reasonably close to the exact values. A similar extraction for two levels in $E_2, B_2$ is shown in Figure \ref{coupled}(b).

\begin{figure*}
\includegraphics[width=0.85\textwidth
]{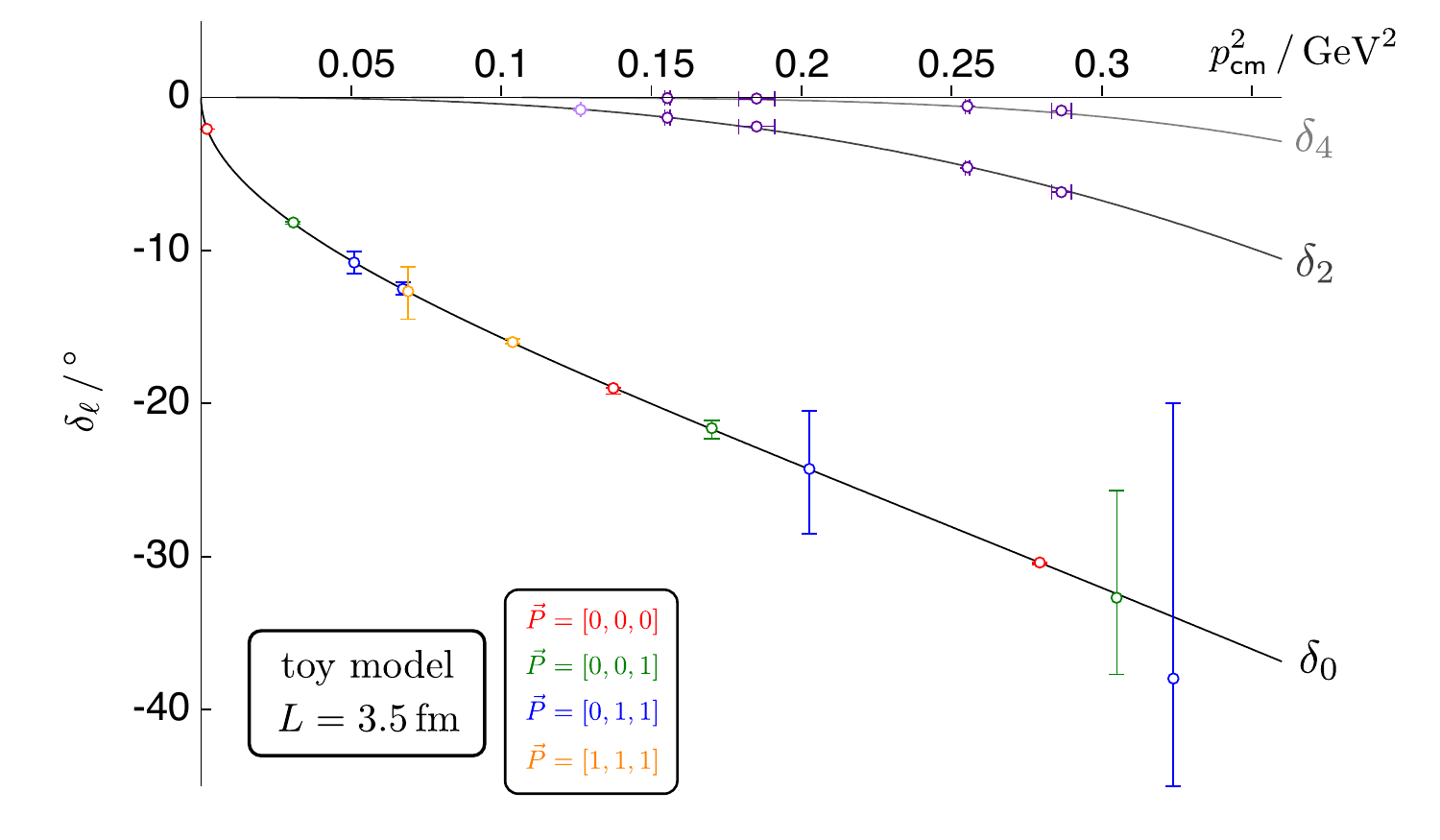}
\caption{Phase shifts, $\delta_{0,2,4}(p_\mathsf{cm})$ extracted from $L=3.5\,\mathrm{fm}$ spectrum using the method described in the text. Uncertainty in $\delta_0$ indicates the effect of a conservative assumed uncertainty on $\delta_{2,4}$. Some points with very large uncertainty not shown. Toy model input phase shifts shown by the curves.  
\label{toy_result}}
\end{figure*}

Several level pairings of this type can be identified and an estimate of a few discrete values of $\delta_2, \delta_4$ can be made as shown by the purple points in Figure \ref{toy_result}. Fitting these points with an effective-range parameterisation, or using some other method to interpolate between the discrete points, we obtain our desired estimates of $\delta_2, \delta_4$ for use in determination of $\delta_0$.

The extracted values of $\delta_0$ shown in Figure \ref{toy_result} correspond to solving equation \ref{luescher} for each energy level in the $A_1$ representation for $\vec{P}=[0,0,0],\,[0,0,1],\,[0,1,1],\,[1,1,1]$ with $\delta_2, \delta_4$ fixed at our best estimate from interpolation between the determined $\delta_2, \delta_4$ points. The error bars indicate the uncertainty in $\delta_0$ obtained by varying $\delta_2, \delta_4$ within an assumed $100\%$ uncertainty. 
 We observe that following such a procedure leads to a reasonable reproduction of the originally input toy-model phase-shifts. Note that we used only a single volume to obtain this result - using multiple volumes will further improve the determination.

An alternative approach to dealing with the contribution of higher partial waves is to parameterise all $\delta_\ell(p_\mathsf{cm})$ one expects to contribute significantly in terms of a relatively small number of variable parameters. By performing a global fit to all energy levels simultaneously, varying the parameters, one can attempt to find a description of the finite-volume spectrum that is best in a least-squares sense.
Clearly in the case of this toy model, one could use the parameterisation given in equation \ref{effrange} and by varying parameters $a_0, r_0, a_2, a_4$ come to an \emph{exact} description of the spectrum. We do not present this trivial result here, but we will return to this ``global fitting" method in the next section when we consider the finite-volume spectrum obtained from lattice QCD computations.

\clearpage
\subsection{Lattice QCD data}

\begin{table*}
\begin{ruledtabular}
\begin{tabular}{cl c c c}
$L/a_s$ & \multicolumn{1}{c}{levels} & \multicolumn{1}{c}{$a_t p_\mathsf{cm}$} & $\delta_2\,/\,^\circ$ & $\delta_4\,/\,^\circ$\\
\hline \hline
\multirow{2}{*}{$24$} & $[0,0,0],\, E^+,\, \mathfrak{n}=1$ & $0.10766(23)(8)$ & \multirow{2}{*}{$-0.39(82)(67)$} & \multirow{2}{*}{$-0.17(32)(22)$} \\
                      & $[0,0,0],\, T_2^+,\, \mathfrak{n}=0$ & $0.10764(23)(8)$ & & \\
\hline
\multirow{2}{*}{$24$} & $[0,0,1],\, B_1,\, \mathfrak{n}=0$ & $0.08427(25)(11)$ & \multirow{2}{*}{$-0.40(47)(39)$} & \multirow{2}{*}{$-0.05(26)(16)$} \\
                      & $[0,0,1],\, E_2,\, \mathfrak{n}=0$ & $0.08418(25)(11)$ & & \\
\hline
\multirow{2}{*}{$24$} & $[0,0,1],\, B_2,\, \mathfrak{n}=0$ & $0.11412(29)(8)$ & \multirow{2}{*}{$-1.60(80)(64)$} & \multirow{2}{*}{$-0.78(69)(55)$} \\
                      & $[0,0,1],\, E_2,\, \mathfrak{n}=1$ & $0.11393(28)(8)$ & & \\
\hline
\multirow{2}{*}{$20$} & $[0,0,1],\, B_1,\, \mathfrak{n}=0$ & $0.10174(35)(9)$ & \multirow{2}{*}{$-1.59(54)(36)$} & \multirow{2}{*}{$-0.018(36)(17)$} \\
                      & $[0,0,1],\, E_2,\, \mathfrak{n}=0$ & $0.10131(37)(9)$ & & \\            
\end{tabular}
\end{ruledtabular}

\caption{Levels with very similar $p_\mathsf{cm}$ values used in simultaneous solution of equations \ref{luescher}.
\label{tab:simult}}
\end{table*}

\begin{figure}[h]
\includegraphics[width=.5\textwidth
]{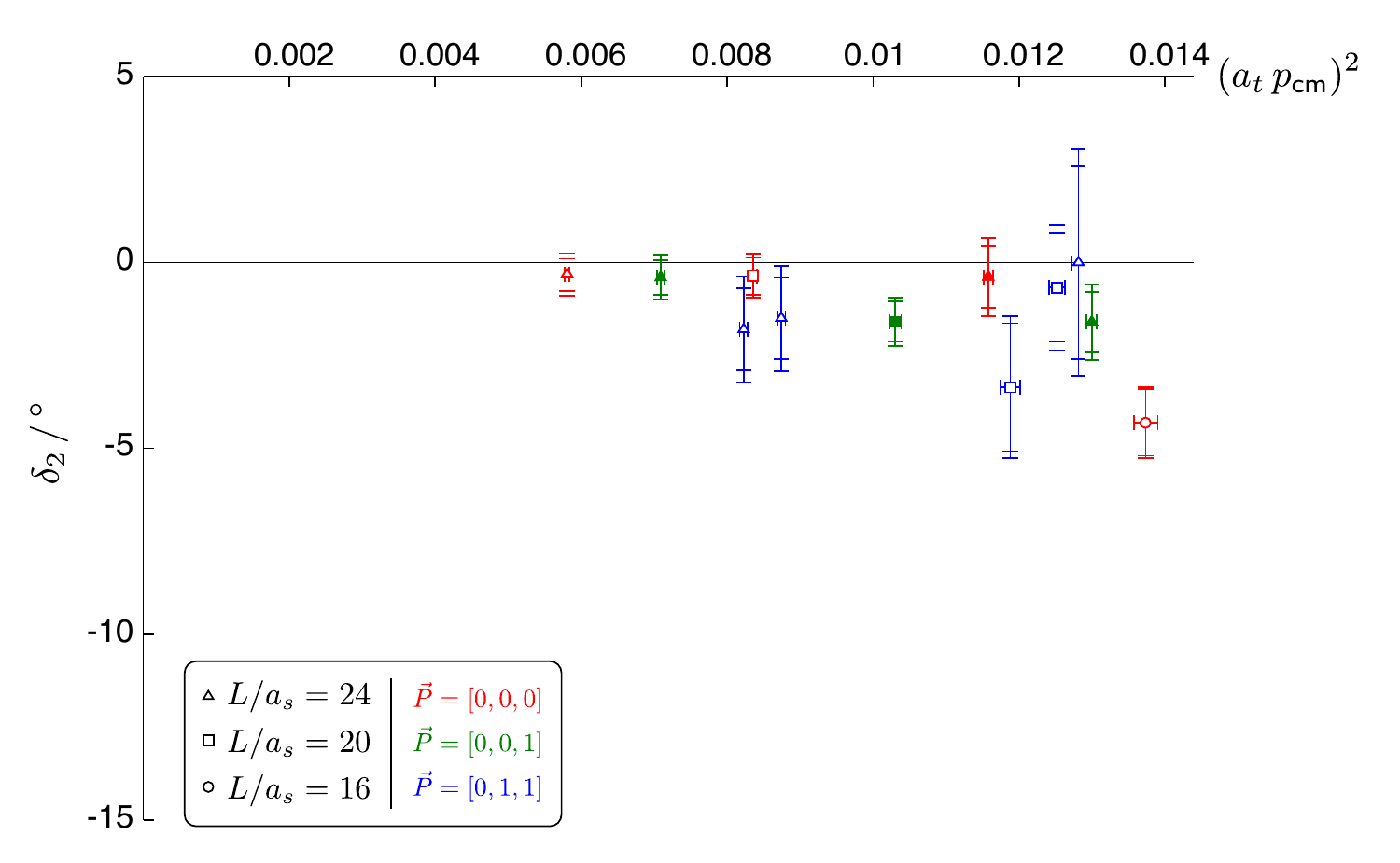}
\caption{$\delta_2$ values in elastic scattering region determined from finite-volume spectra. Filled points determined by simultaneous solution of equations \ref{luescher} (innermost errorbar statistical uncertainty, outermost errorbar reflects combined statistical uncertainty and uncertainty in $a_t m_\pi$, $\xi$ with all errors added in quadrature). Open points determined from single levels, with effect of $\delta_4$ estimated (innermost errorbar statistical uncertainty, outermost errorbar reflects combined statistical uncertainty and uncertainty in $a_t m_\pi$, $\xi$ and $\delta_4$).
\label{delta2pointwise}}
\end{figure}

\begin{figure}[h]
\includegraphics[width=.5\textwidth
]{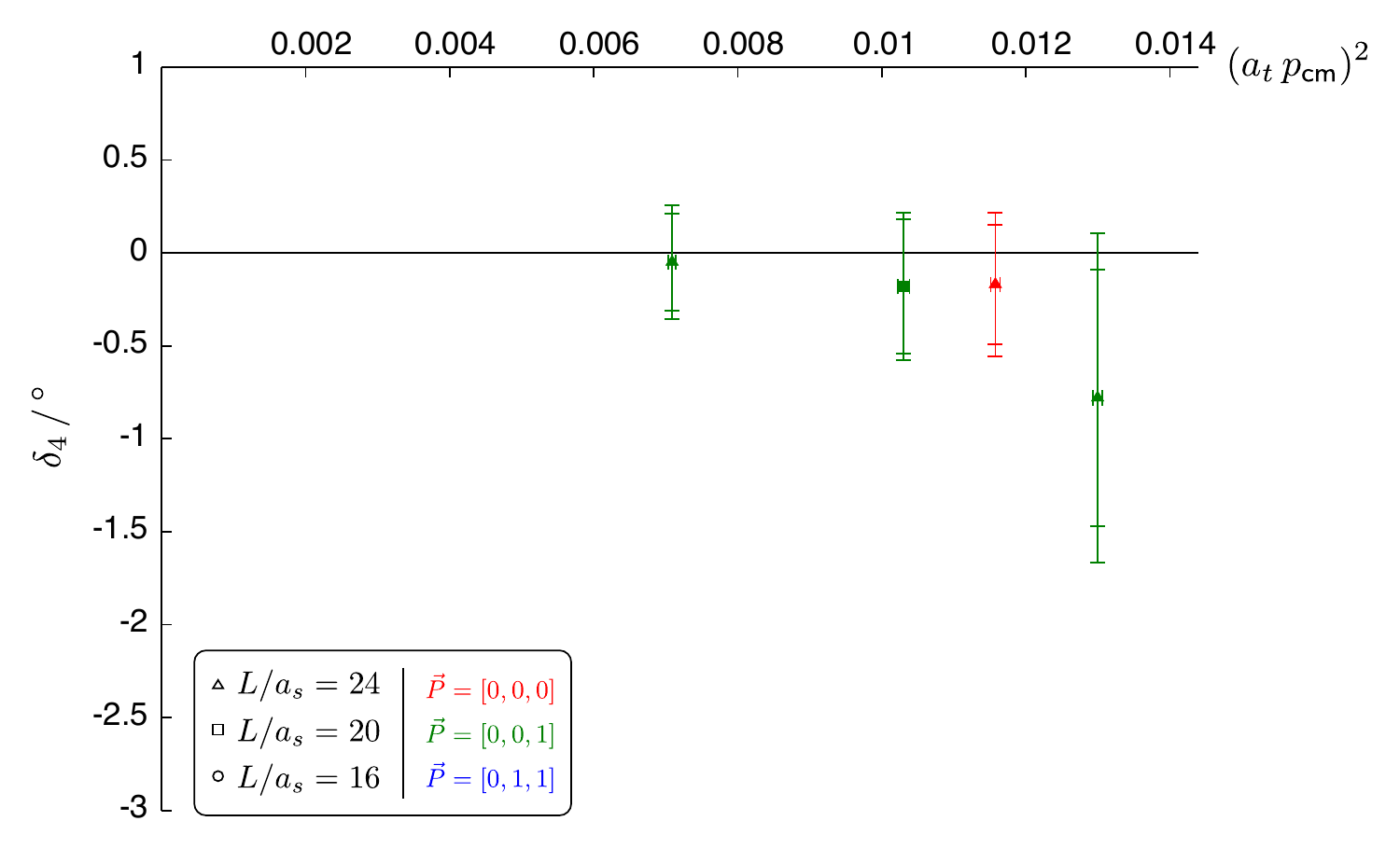}
\caption{$\delta_4$ values in elastic scattering region determined from finite-volume spectra. Filled points determined by simultaneous solution of equations \ref{luescher} (innermost errorbar statistical uncertainty, outermost errorbar reflects combined statistical uncertainty and uncertainty in $a_t m_\pi$, $\xi$ with errors added in quadrature).
\label{delta4pointwise}}
\end{figure}

We now return to consideration of the finite-volume spectrum presented in Section \ref{sec:spectrum}. The first step in our ``level-by-level" approach is to solve for $\delta_{2,4}$ using pairs of simultaneous equations \ref{luescher}. Pairs of levels below inelastic threshold that can be used to yield estimates for $\delta_{2,4}$ are presented in Table \ref{tab:simult} and are displayed by the filled points in Figures \ref{delta2pointwise}, \ref{delta4pointwise}. $\delta_4$ is observed to be statistically compatible with zero throughout the elastic region. There are also levels in irreps whose leading contribution is from $\ell=2$ that do not pair and cannot be analysed using a simultaneous solution - these are considered in isolation, where the (small) role of $\delta_4$ is estimated and included as a systematic error, they are shown by the open points in Figure \ref{delta2pointwise}.

\begin{figure}[h]
\includegraphics[width=.5\textwidth
]{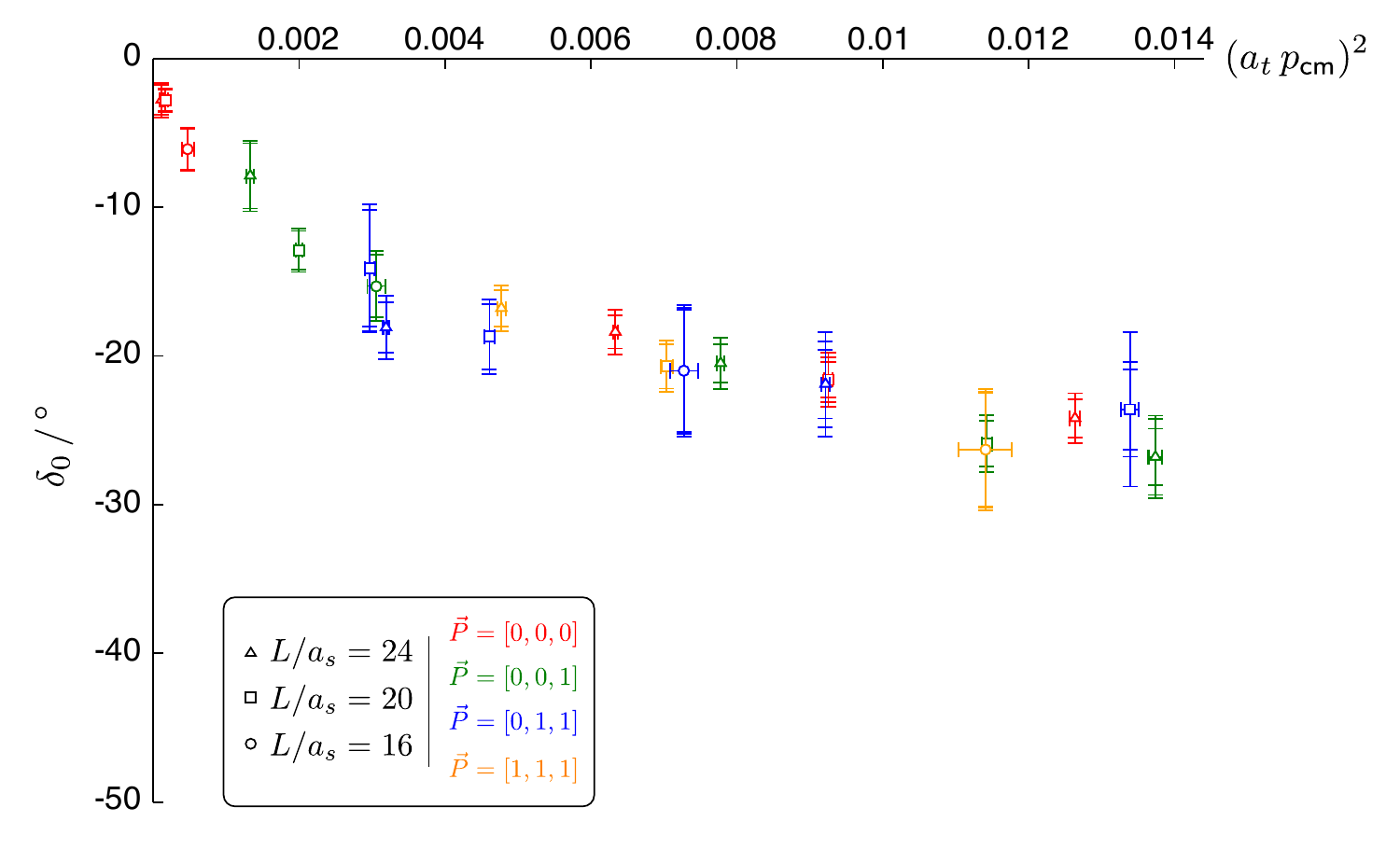}
\caption{$\delta_0$ values in elastic scattering region determined from finite-volume spectra. Innermost errorbar is the statistical uncertainty, middle errorbar combined statistical uncertainty and uncertainty in ($a_t m_\pi$, $\xi$), outermost errorbar reflects total uncertainty including imperfect knowledge of $\delta_{2,4}$ (all errors added in quadrature). Some points with very large uncertainty not shown.
\label{delta0pointwise}}
\end{figure}

Each of these $\delta_2$, $\delta_4$ data sets can be described well by a scattering length fit, $p_\mathsf{cm}^{2\ell + 1} \cot \delta_\ell(p_\mathsf{cm}) = 1/a_\ell$, and the resulting fit function is used to estimate the size of $\delta_{2,4}$ at any $p_\mathsf{cm}$ in the elastic region when determining $\delta_0$ values from $A_1$ irreps. As indicated in the previous subsection, a systematic error on $\delta_0$ due to imperfect knowledge of $\delta_{2,4}$ is assigned by 
assuming a $100\%$ error on the estimated values of $\delta_{2,4}$. 
The resulting $\delta_0$ points are displayed in Figure \ref{delta0pointwise} where it is observed that the uncertainty from imperfect knowledge of $\delta_{2,4}$ is typically much smaller that the statistical uncertainty.

\pagebreak
We now consider the second approach described above where the $\delta_\ell(p_\mathsf{cm})$ are parameterised and by varying a small number of parameters a best description of all the finite volume spectra is obtained in a ``global fit". Our procedure is to minimise a $\chi^2$ with respect to the variable parameters in the parameterisation, which we denote collectively by $\{a_i\}$. The $\chi^2$ describes the similarity between the extracted finite-volume spectrum and the spectrum predicted by the parameterisation on the appropriate volumes,
\begin{widetext}
\begin{equation}
\chi^2(\{a_i\}) = 
\sum_L \sum_{\substack{\vec{P} \Lambda \mathfrak{n} \\ \vec{P}' \Lambda'\mathfrak{n}'}}
\left[ p_\mathsf{cm}(L;\vec{P}\Lambda\mathfrak{n}) - p^{\det}_\mathsf{cm}(L; \vec{P}\Lambda\mathfrak{n}; \{a_i \}) \right] \mathbb{C}^{-1}(L; \vec{P} \Lambda \mathfrak{n}; \vec{P}' \Lambda' \mathfrak{n}') \left[ p_\mathsf{cm}(L;\vec{P}'\Lambda'\mathfrak{n}') - p^{\det}_\mathsf{cm}(L; \vec{P}'\Lambda'\mathfrak{n}'; \{a_i \}) \right]. \label{chisq}
\end{equation}
\end{widetext}
Here we have $p^{\det}_\mathsf{cm}(L; \vec{P}\Lambda\mathfrak{n}; \{a_i \})$ which is the particular solution of equation \ref{luescher} 
which is nearest to $p_\mathsf{cm}(L;\vec{P}\Lambda\mathfrak{n})$ (with the parameters set to the particular values $\{a_i\}$). The data covariance, $\mathbb{C}$, accounts for the correlation between determined energies computed on the same lattice configurations - different volumes correspond to independently generated lattice ensembles and hence are not correlated.

Statistical errors on the parameters, $\{a_i\}$, are determined by $\Delta \chi^2 = 1$. Errors from the imperfect knowledge of $a_t m_\pi$ and $\xi$ are estimated by repeating the $\chi^2$ minimisation varying the mass and anisotropy within their respective errors. We treat these as independent systematic errors, although they would naturally be reduced with increased numbers of gauge-field configurations at each lattice volume.

Fits with effective range and scattering length parameterisations (equation \ref{effrange}) were attempted. These fits never indicated the need to include significant strength in the $\ell=4$ wave. A successful fit to all energy levels with an effective range parameterisation of $\ell=0$ and scattering length in $\ell=2$ gives the following parameter values and correlations,
\nopagebreak
\vspace{0.25cm}
\begin{tabular}{rlr}
$a_{\ell=0}$ & $= (-4.45\pm0.18\pm0.06 )\, \cdot \,a_t$ & \multirow{3}{*}{ \;\;\;$\begin{bmatrix} 1 & 0.9 & 0.4 \\ & 1 & 0.2 \\ & & 1 \end{bmatrix}$  }\\ 
$r_{\ell=0}$ & $= (-3.7\pm1.8\pm0.7) \, \cdot \, a_t$ & \\
$a_{\ell=2}$ & $= (-1.20\pm0.29\pm0.17) \times 10^3 \, \cdot \, a_t^5$ & \\
\\
& $\chi^2/N_\mathrm{dof} = 116/46$, & \\
\end{tabular}
\vspace{0.25cm}
\\*
where the second set of uncertainties reflects variation of $a_t m_\pi$ and $\xi$ within their uncertainties. We see that the effective range in $\ell=0$ is barely significant and is strongly correlated with the scattering length. The degree of correlation between $\ell=0$ and $\ell=2$ is mild. Given the lack of significance for $r_0$, a fit with just a scattering length was attempted, yielding

\begin{figure}
\includegraphics[width=.5\textwidth
]{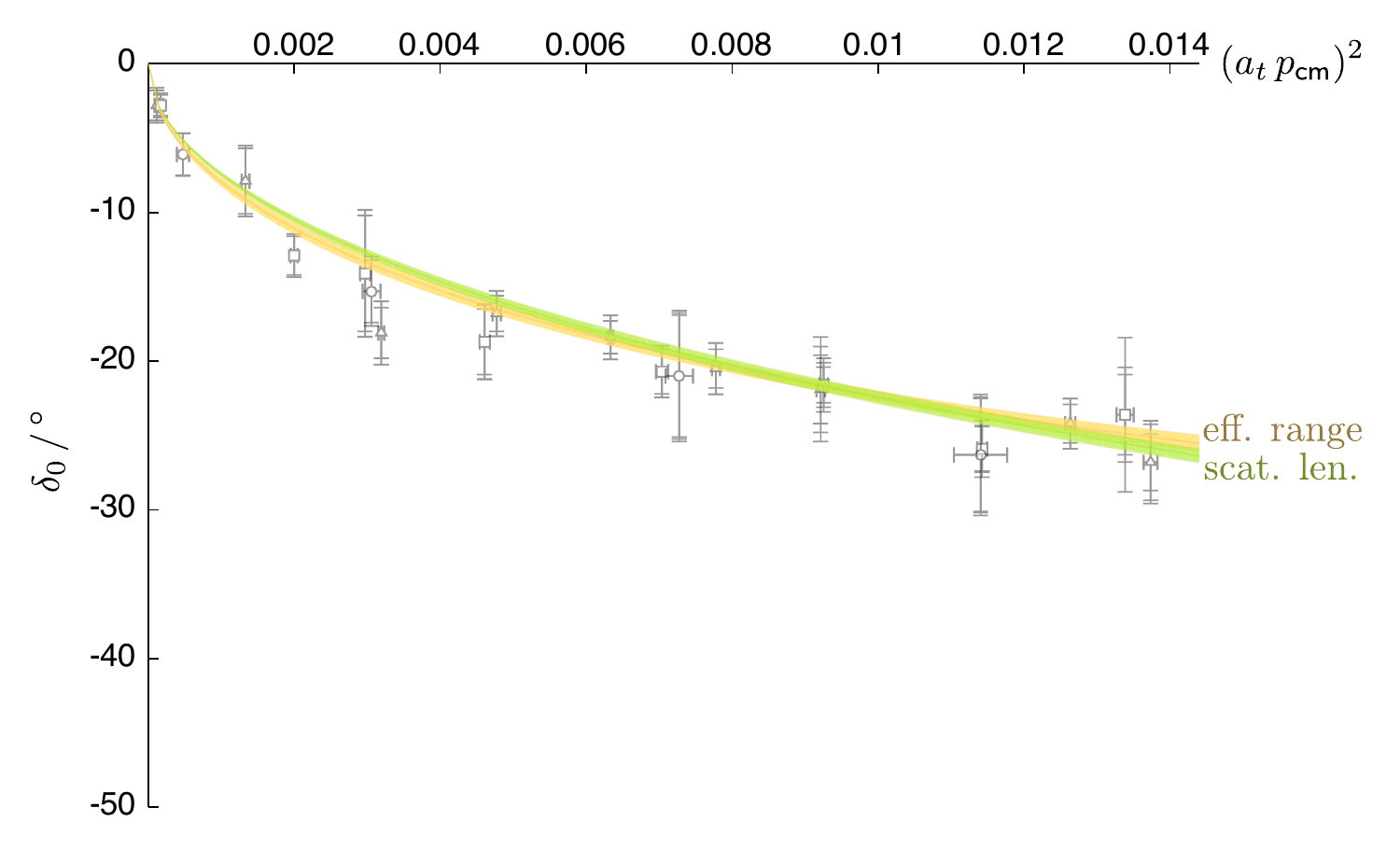}
\includegraphics[width=.5\textwidth
]{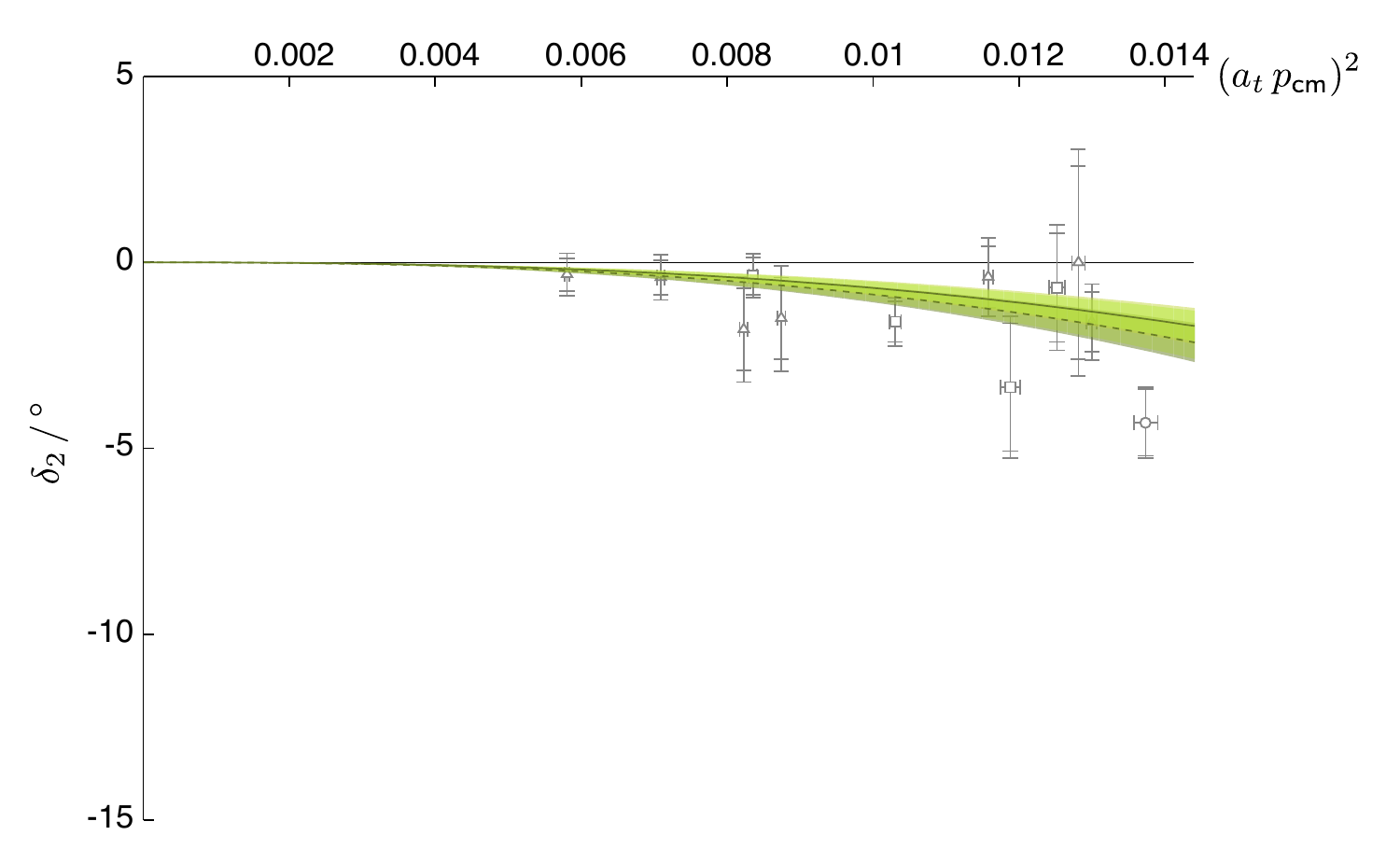}
\caption{Upper: $\delta_0(p_\mathsf{cm})$ obtained through ``global fits" to finite-volume spectra using effective range and scattering length parameterisations.
Lower: $\delta_2(p_\mathsf{cm})$ obtained through ``global fits" to finite-volume spectra using a scattering length parameterisations.
 Also shown for comparison, the ``level-by-level" analysis previously presented in Figures \ref{delta0pointwise}, \ref{delta2pointwise}.
\label{delta0global}}
\end{figure}

\vspace{0.25cm} 
\begin{tabular}{rlr}
$a_{\ell=0}$ & $= (-4.13\pm0.07\pm0.06 )\, \cdot \,a_t$ & \multirow{2}{*}{ \;\;\;$\begin{bmatrix} 1 & 0.5 \\ & 1 \end{bmatrix}$  }\\ 
$a_{\ell=2}$ & $= (-1.08\pm0.28\pm0.19) \times 10^3 \, \cdot \, a_t^5$ & \\
\\
& $\chi^2/N_\mathrm{dof} = 121/47$, & \\
\end{tabular}
\vspace{0.25cm} 
\\*
where the quality of fit is insignificantly degraded. Clearly there is no need to invoke higher terms in the effective range expansion to describe the data. A fit to only those irreps where $\ell=2$ is leading yields $a_{\ell=2} = (-1.51\pm0.31\pm0.21) \times 10^3 \, \cdot \, a_t^5$ with $\chi^2/N_\mathrm{dof} = 31/16$, in reasonable agreement with the values obtained above. These various fits are shown in Figure \ref{delta0global} 
along with the points determined using the ``level-by-level" approach described earlier where good agreement between the two methods is observed.

\pagebreak
\section{Results}\label{sec:results}
\begin{figure*}
\includegraphics[width=0.8\textwidth
]{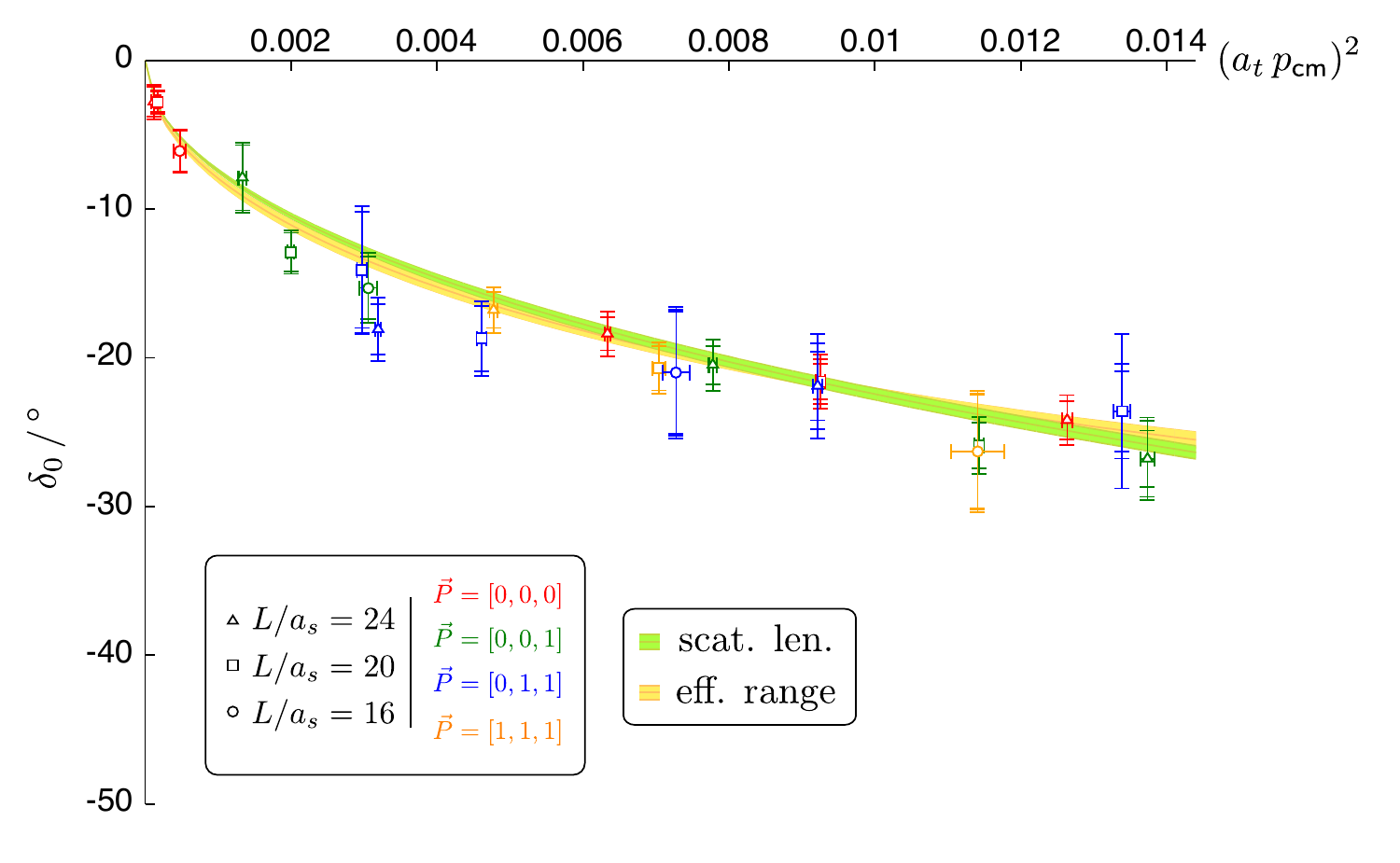}
\caption{Extracted $I=2$ $\pi\pi$ elastic scattering phase-shift in $S$-wave, $\delta_0(p_\mathsf{cm})$, as obtained from analysis of finite-volume spectra with $m_\pi = 396\,\mathrm{MeV}$. Center-of-momentum frame scattering momentum expressed in units of the temporal lattice spacing. The momentum region plotted is entirely elastic, with the $4\pi$ threshold opening at $(a_t p_\mathsf{cm})^2 = 0.014$. Colored points correspond to an analysis treating each energy level independently. The innermost errorbar is the statistical uncertainty, the middle errorbar reflects combined statistical uncertainty and uncertainty in $(a_t m_\pi$, $\xi)$ and the outermost errorbar shows the total uncertainty including imperfect knowledge of $\delta_{2,4}$ (all errors added in quadrature). Curves indicate a global analysis of all energy levels describing the phase-shift by a scattering length or an effective range parameterisation.
\label{delta0summary}}
\end{figure*}

In Figure \ref{delta0summary} we show the $\pi\pi$ $\ell=0$ elastic scattering phase shift for $m_\pi = 396 \,\mathrm{MeV}$ as a function of center-of-momentum frame scattering momentum as extracted from finite-volume spectra. Discrete points correspond to a ``level-by-level" analysis in which the L\"uscher equation is solved for $\delta_0(p_\mathsf{cm})$ at each obtained $p_\mathsf{cm}$ with some justified assumptions made about the size of $\delta_{2,4}$ at this scattering momentum, and with the degree of uncertainty about the higher $\ell$ partial waves reflected in a systematic error. The curves are the result of ``global fits" to all the finite-volume energy levels assuming either an effective range parameterisation or just a scattering length, either of which are able to describe the energy spectrum well. The best estimates for the scattering length and effective range expressed in units of the pion mass on this lattice are
\begin{align}
	m_\pi \cdot a_{\ell=0} &= -0.307 \pm 0.013 \nonumber \\
	m_\pi \cdot r_{\ell=0} &= -0.26 \pm 0.13, \nonumber
\end{align}
but there is a very high degree of correlation ($0.9$) between these values, and a pure scattering length of $m_\pi \cdot a_{\ell=0} = -0.285 \pm 0.006$ can describe the data just as well.

\begin{figure*}
\includegraphics[width=0.8\textwidth
]{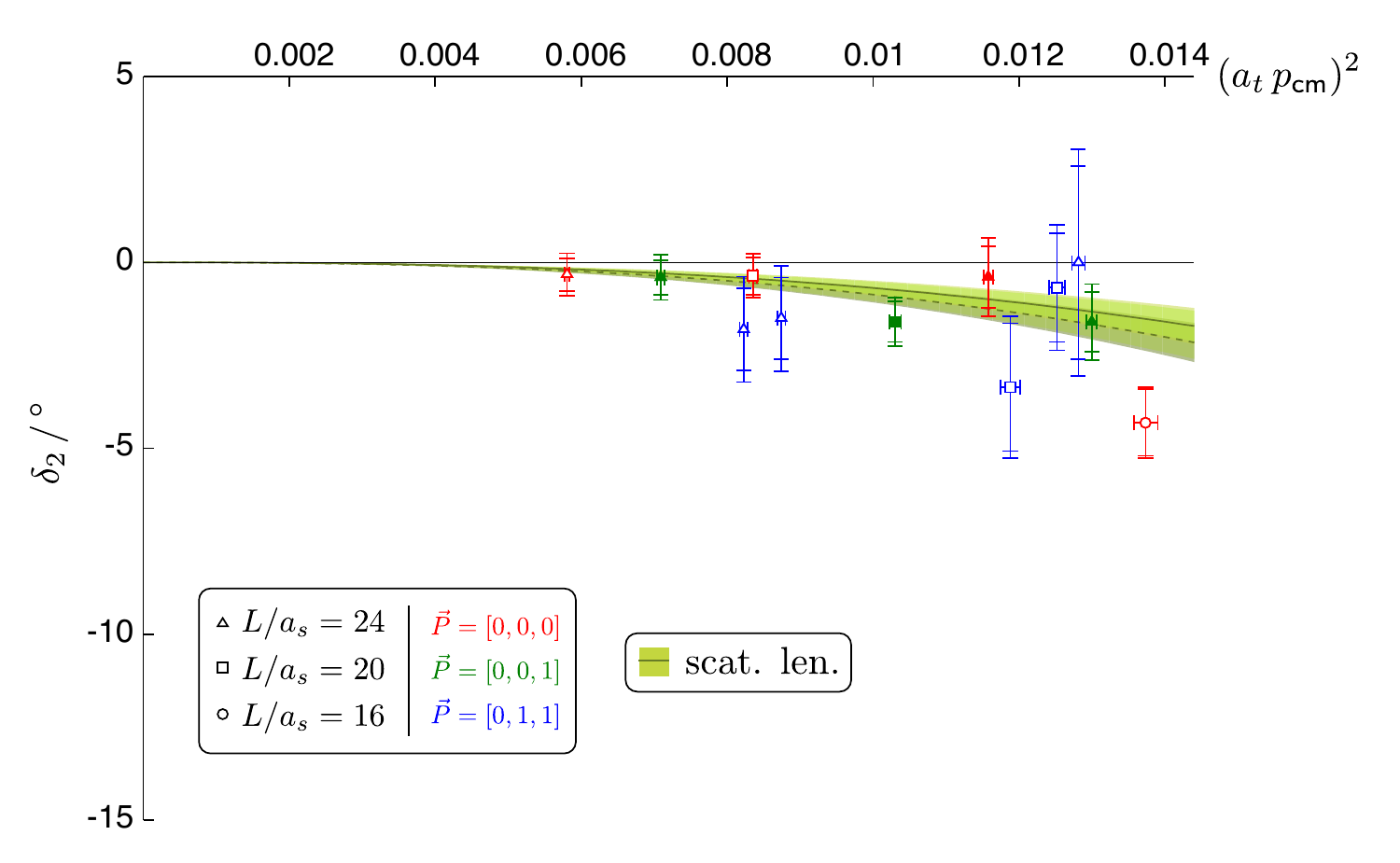}
\caption{Extracted $I=2$ $\pi\pi$ elastic scattering phase-shift in $D$-wave, $\delta_2(p_\mathsf{cm})$, as obtained from analysis of finite-volume spectra with $m_\pi = 396\,\mathrm{MeV}$. Center-of-momentum frame scattering momentum expressed in units of the temporal lattice spacing. Momentum region plotted is entirely elastic, with the $4\pi$ threshold opening at $(a_t p_\mathsf{cm})^2 = 0.014$. Colored points correspond to an analysis treating energy regions locally as described earlier in the manuscript. The inner errorbar is the statistical uncertainty, and the outer errorbar reflects the combined statistical uncertainty and uncertainty in $a_t m_\pi$, $\xi$ and the value of $\delta_4$ (errors added in quadrature). Curves indicate a global analysis of all energy levels describing the phase-shift by a scattering length.
\label{delta2summary}}
\end{figure*}

Figure \ref{delta2summary} shows the $\pi\pi$ $\ell=2$ elastic scattering phase shift which can be well described by a scattering length of $m_\pi^5 \cdot a_{\ell=2} = (-1.89 \pm 0.53)\times 10^{-6}$. Statistically significant signals for elastic scattering in the $\ell=4$ wave were not observed and we estimate that $ m_\pi^9 \cdot |a_{\ell=4}| \lesssim 1 \times 10^{-4}$.

\begin{figure*}
\includegraphics[width=\textwidth
]{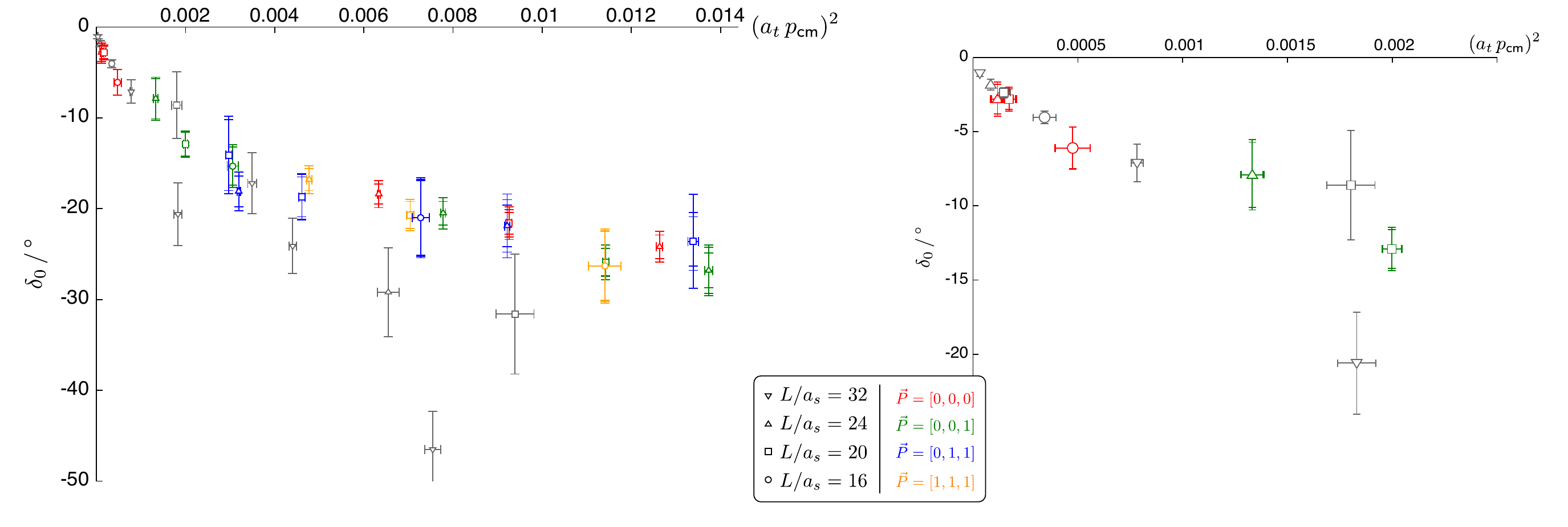} 
\caption{Our $\delta_0$ extraction (colored points) compared with those of NPLQCD (grey points) over the elastic region (left) and zoomed in to small scattering momentum (right).
\label{NPLQCD}}
\end{figure*}

We note here that the same $L/a_s = 16,20,24$ lattice ensembles (plus a larger $L/a_s = 32$ ensemble) were used by NPLQCD to extract $\delta_0(p_\mathsf{cm})$ in \cite{Beane:2011sc}. They considered many of the same frames, but limited themselves to the ``scalar" irreps ($A_1^+$ for $\vec{P}=[0,0,0]$ and $A_1$ for $\vec{P} \neq [0,0,0]$), and they did not use a variational basis of operators. A comparison of results is shown in Figure \ref{NPLQCD} where low-lying levels are observed to have energies (and hence phase-shifts) that agree well, but where discrepancies appear at higher energies. The most significantly discrepant points (at $(a_t p_\mathsf{cm})^2 \sim 0.0017$ and $\sim 0.008$) in the NPLQCD analysis correspond to levels which are either nearly degenerate with another level (the $\vec{P}=[0,1,1]$ ground state\footnote{see Figure \ref{P110}}) or are highly excited ($\vec{P}=[0,0,0]$ second excited level). Since in our analysis we see no such discrepancies it may be that the variational method more reliably determines energies in cases where orthogonality of states is important.

\begin{figure*}
\includegraphics[width=0.8\textwidth
]{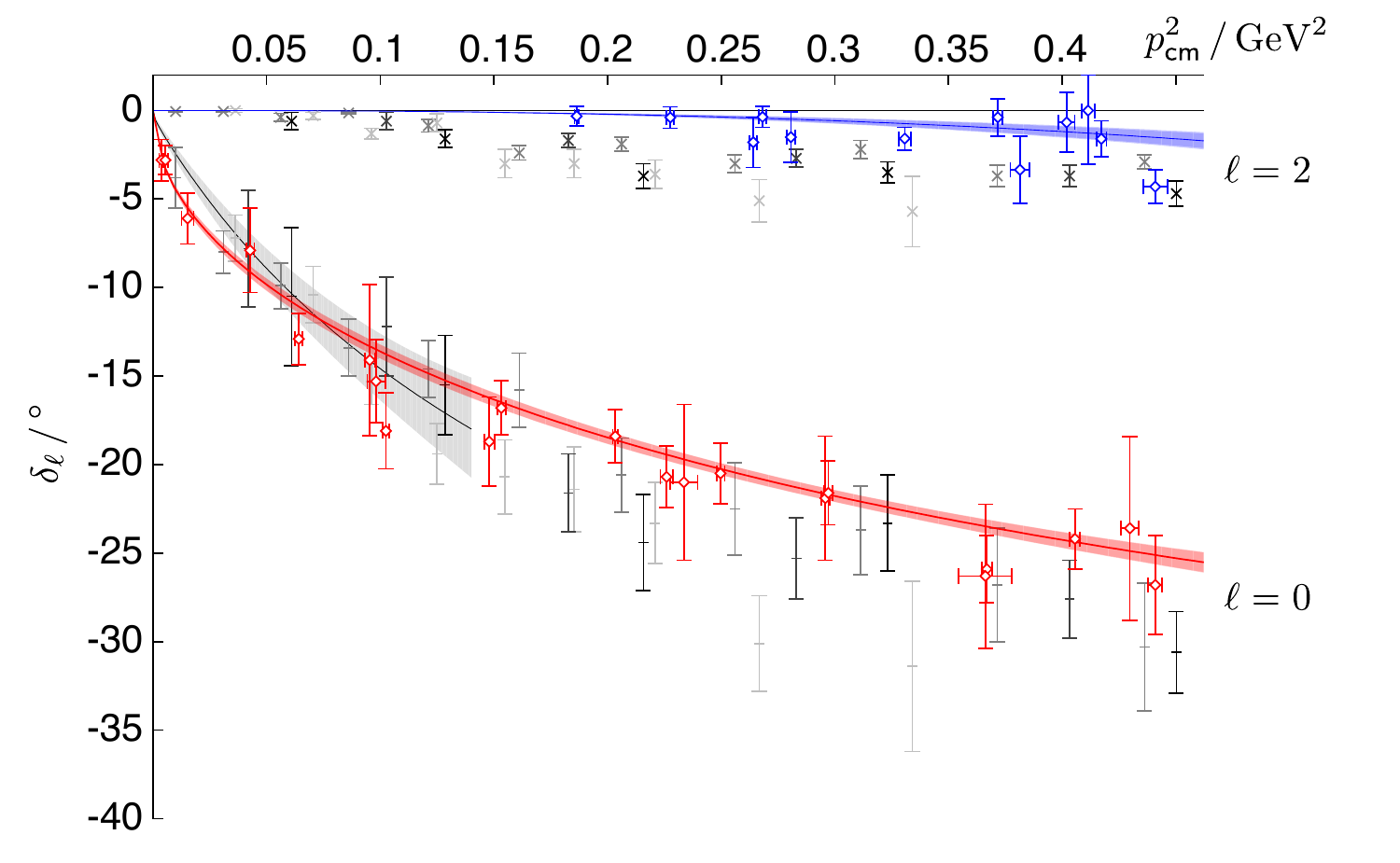} 
\caption{Extracted $I=2$ $\pi\pi$ elastic $S$-wave(red), $D$-wave(blue) scattering phase shift (for $m_\pi = 396\,\mathrm{MeV}$, all errors combined). Shown in grey the experimental data from \cite{Hoogland:1977kt,Cohen:1973yx,Zieminski:1974ex,Durusoy:1973aj} and the constrained analysis using Roy equations \cite{Roy:1971tc,Colangelo:2001df} (black line, grey band). For the heavy pion mass the entire region plotted is elastic while for the experimental data only $p_\mathsf{cm}^2 <  0.058\,\mathrm{GeV}$ is elastic. 
\label{expt}}
\end{figure*}

In our somewhat limited previous analysis of $\pi\pi$ $I=2$ scattering \cite{Dudek:2010ew}, we considered three pion masses and observed no significant dependence of the energy variation of $\delta_0$ on the pion mass, which appeared to agree rather well with the experimental data. In Figure \ref{expt} we show our $\delta_{0,2}(p_\mathsf{cm})$ obtained at $m_\pi = 396\,\mathrm{MeV}$ along with the experimental data taken from \cite{Hoogland:1977kt,Cohen:1973yx,Zieminski:1974ex,Durusoy:1973aj}. Our data points have the absolute energy scale of the scattering momentum set using the $\Omega$-baryon mass procedure suggested in \cite{Lin:2008pr}, $p_\mathsf{cm} = (a_t p_\mathsf{cm}) \frac{m_\Omega^\mathrm{phys}}{(a_t m_\Omega)}$ with $(a_t m_\Omega) = 0.2951(22)$ on these lattices \cite{Edwards:2011jj}. Also shown is the $\pi\pi$ $I=2$ $\ell=0$ phase-shift obtained using experimental information in multiple channels from a constrained analysis provided by the Roy equations, which implement manifestly crossing symmetry and the chiral behavior of the scattering amplitudes \cite{Roy:1971tc,Colangelo:2001df}. 

\section{Summary}\label{sec:summary}

A crucial step in the extraction of hadronic resonance properties is the determination of their resonant scattering behavior. Within a Euclidean quantum field theory, the relevant elastic scattering matrix elements can be inferred indirectly through a systematic study of the spectrum within a finite volume. 
 In this paper, we extend our previous study~\cite{Dudek:2010ew} determining the $\ell=0$ and $\ell=2$ wave phase shifts in the $\pi\pi$ $I=2$ systems, investigating more thoroughly the effects of operator basis, finite temporal extent, as well as the role of higher partial waves.

With access to only modest lattice volumes, in order to map out the energy dependence with a significant number of points, we determined the excited state spectrum in moving frames. This was achieved by constructing a basis of $\pi\pi$ operators transforming irreducibly under the reduced symmetry of a moving particle in a cubic box. Variational analysis of matrices of correlators built in this operator basis leads to extraction of excited state energies with high precision.

The construction of a basis of operators with suitable quantum numbers corresponding to the $\pi\pi$ system in-flight  
is a significant extension beyond the previous work, and has allowed for the determination of the phase shifts at many discrete energies below the $4\pi$ inelastic threshold. This increased operator basis, covering many irreducible representations, allows for more constraints on the contributions of higher partial waves.
However, the weakness of $\pi\pi$ scattering in the isospin-2 channel presents a particular challenge to extraction from finite-volume methods. The changes in energy with respect to non-interacting pions determine the phase-shift and since these are small it is important to take care over systematic effects that may be small in absolute terms but which could be large on the scale of the energy shifts.

We reduced the contribution of excited pion-like meson states to our $\pi\pi$ correlators by using optimised pion operators. These operators are constructed from a linear combination of composite QCD operators with pion quantum numbers and their important property is that they relax to the ground state faster than any single simple operator construction. The reduced contribution of $\pi\pi^\star$ states to our correlators allows analysis at earlier Euclidean times.

At larger Euclidean times, the effect of the finite temporal extent of the lattice can be observed, distorting the time-dependence from the desired sum of exponentials corresponding to discrete state energies. We have explicitly accounted for the largest unwanted finite-$T$ effects leaving sub-leading effects which are somewhat smaller than the statistical uncertainty. 

The reduced symmetry of a cubic box at rest is such that $\delta_0$ always appears with some sensitivity to $\delta_4$, but the very small value of $\delta_4$ throughout the elastic region is such that the rest-frame spectrum is mostly independent of $\delta_4$. On the other hand, the symmetry of a cubic box is further reduced when placed in-flight and $\delta_0$ extractions become sensitive to the value of $\delta_2$, which is not necessarily negligibly small. We investigated the effects that non-zero values of $\delta_{2,4}$ can have on the finite-volume spectrum using a toy model showing that some energy levels can show significant sensitivity. 

We attempted to account for the effects of higher partial waves on the extraction of $\delta_{0,2}$, finding that they are generally small (except in a limited number of sensitive cases identified in the toy model analysis). We associated a systematic error with our imperfect knowledge of them that was found to be always smaller than the statistical uncertainty. We found that the finite volume energies could be well described by a scattering length parameterization in both $\ell=0$ and $\ell=2$ over the elastic region. The fit could be moderately improved by adding an effective range in $\ell=0$, albeit with a significant correlation between the effective range and scattering length. The fits did not indicate the need for significant strength in the $\ell=4$ wave. 

The calculations reported in this paper were performed at only a single pion mass of 396 MeV. While they demonstrate that the procedure outlined can indeed determine scattering phase shifts with a high degree of confidence, the obtained results cannot be directly compared with experimental data. Future calculations using lighter pion masses will be required, as will eventual consideration of other systematic effects such as the lattice spacing dependence. The results presented in this paper supersede those presented in \cite{Dudek:2010ew} which considered only rest-frame correlators using un-optimised pion operators and where finite-$T$ effects were not fully accounted for.

The techniques developed in this calculation are a necessary ingredient to future investigations of resonances in hadron-hadron scattering that arise from the strong interactions. At unphysical pion masses, the phase space available for decays can be small as seen in studies of the $I=1$ $\pi\pi$ sector~\cite{Aoki:2007rd, Feng:2010es, Lang:2011mn, Aoki:2011yj} giving rise to a rapid variation of phase-shift with energy. Thus, the formalism and construction of operators in-flight developed in this work will be necessary to compute a sufficient number of energies within the resonance region to allow for a reliable determination of resonance parameters. To compute these energies, the operator basis used in the variational method will feature both single and multi-hadron constructions. Annihilation diagrams will arise, which as shown in the isoscalar meson sector~\cite{Dudek:2011tt}, can be efficiently constructed using the ``distillation'' method.

 
\begin{acknowledgments}

We thank our colleagues within the Hadron Spectrum Collaboration. {\tt Chroma}~\cite{Edwards:2004sx} and {\tt QUDA}~\cite{Clark:2009wm,Babich:2010mu} were used to perform this work on clusters at Jefferson Laboratory under the USQCD Initiative and the LQCD ARRA project. Gauge configurations were generated using resources awarded from the U.S. Department of Energy INCITE program at Oak Ridge National Lab, the NSF Teragrid at the Texas Advanced Computer Center and the Pittsburgh Supercomputer Center, as well as at Jefferson Lab. RGE and JJD acknowledge support from U.S. Department of Energy contract DE-AC05-06OR23177, under which Jefferson Science Associates, LLC, manages and operates Jefferson Laboratory. JJD also acknowledges the support of the Jeffress Memorial Fund and the U.S. Department of Energy Early Career award contract DE-SC0006765.  CET acknowledges support from a Marie Curie International Incoming Fellowship, PIIF-GA-2010-273320, within the 7th European Community Framework Programme.

\end{acknowledgments}

\bibliography{bibliography} 

\begin{thebibliography}{44}%
\makeatletter
\providecommand \@ifxundefined [1]{%
 \@ifx{#1\undefined}
}%
\providecommand \@ifnum [1]{%
 \ifnum #1\expandafter \@firstoftwo
 \else \expandafter \@secondoftwo
 \fi
}%
\providecommand \@ifx [1]{%
 \ifx #1\expandafter \@firstoftwo
 \else \expandafter \@secondoftwo
 \fi
}%
\providecommand \natexlab [1]{#1}%
\providecommand \enquote  [1]{``#1''}%
\providecommand \bibnamefont  [1]{#1}%
\providecommand \bibfnamefont [1]{#1}%
\providecommand \citenamefont [1]{#1}%
\providecommand \href@noop [0]{\@secondoftwo}%
\providecommand \href [0]{\begingroup \@sanitize@url \@href}%
\providecommand \@href[1]{\@@startlink{#1}\@@href}%
\providecommand \@@href[1]{\endgroup#1\@@endlink}%
\providecommand \@sanitize@url [0]{\catcode `\\12\catcode `\$12\catcode
  `\&12\catcode `\#12\catcode `\^12\catcode `\_12\catcode `\%12\relax}%
\providecommand \@@startlink[1]{}%
\providecommand \@@endlink[0]{}%
\providecommand \url  [0]{\begingroup\@sanitize@url \@url }%
\providecommand \@url [1]{\endgroup\@href {#1}{\urlprefix }}%
\providecommand \urlprefix  [0]{URL }%
\providecommand \Eprint [0]{\href }%
\providecommand \doibase [0]{http://dx.doi.org/}%
\providecommand \selectlanguage [0]{\@gobble}%
\providecommand \bibinfo  [0]{\@secondoftwo}%
\providecommand \bibfield  [0]{\@secondoftwo}%
\providecommand \translation [1]{[#1]}%
\providecommand \BibitemOpen [0]{}%
\providecommand \bibitemStop [0]{}%
\providecommand \bibitemNoStop [0]{.\EOS\space}%
\providecommand \EOS [0]{\spacefactor3000\relax}%
\providecommand \BibitemShut  [1]{\csname bibitem#1\endcsname}%
\let\auto@bib@innerbib\@empty
\bibitem [{\citenamefont {DeWitt}(1956)}]{DeWitt:1956be}%
  \BibitemOpen
  \bibfield  {author} {\bibinfo {author} {\bibfnamefont {B.~S.}\ \bibnamefont
  {DeWitt}},\ }\href {\doibase 10.1103/PhysRev.103.1565} {\bibfield  {journal}
  {\bibinfo  {journal} {Phys. Rev.}\ }\textbf {\bibinfo {volume} {103}},\
  \bibinfo {pages} {1565} (\bibinfo {year} {1956})}\BibitemShut {NoStop}%
\bibitem [{\citenamefont {Luscher}(1991)}]{Luscher:1991cf}%
  \BibitemOpen
  \bibfield  {author} {\bibinfo {author} {\bibfnamefont {M.}~\bibnamefont
  {Luscher}},\ }\href {\doibase 10.1016/0550-3213(91)90584-K} {\bibfield
  {journal} {\bibinfo  {journal} {Nucl. Phys.}\ }\textbf {\bibinfo {volume}
  {B364}},\ \bibinfo {pages} {237} (\bibinfo {year} {1991})}\BibitemShut
  {NoStop}%
\bibitem [{\citenamefont {Hoogland}\ \emph {et~al.}(1977)\citenamefont
  {Hoogland}, \citenamefont {Peters}, \citenamefont {Grayer}, \citenamefont
  {Hyams}, \citenamefont {Weilhammer} \emph {et~al.}}]{Hoogland:1977kt}%
  \BibitemOpen
  \bibfield  {author} {\bibinfo {author} {\bibfnamefont {W.}~\bibnamefont
  {Hoogland}}, \bibinfo {author} {\bibfnamefont {S.}~\bibnamefont {Peters}},
  \bibinfo {author} {\bibfnamefont {G.}~\bibnamefont {Grayer}}, \bibinfo
  {author} {\bibfnamefont {B.}~\bibnamefont {Hyams}}, \bibinfo {author}
  {\bibfnamefont {P.}~\bibnamefont {Weilhammer}},  \emph {et~al.},\ }\href
  {\doibase 10.1016/0550-3213(77)90154-7} {\bibfield  {journal} {\bibinfo
  {journal} {Nucl.Phys.}\ }\textbf {\bibinfo {volume} {B126}},\ \bibinfo
  {pages} {109} (\bibinfo {year} {1977})}\BibitemShut {NoStop}%
\bibitem [{\citenamefont {Cohen}\ \emph {et~al.}(1973)\citenamefont {Cohen},
  \citenamefont {Ferbel}, \citenamefont {Slattery},\ and\ \citenamefont
  {Werner}}]{Cohen:1973yx}%
  \BibitemOpen
  \bibfield  {author} {\bibinfo {author} {\bibfnamefont {D.~H.}\ \bibnamefont
  {Cohen}}, \bibinfo {author} {\bibfnamefont {T.}~\bibnamefont {Ferbel}},
  \bibinfo {author} {\bibfnamefont {P.}~\bibnamefont {Slattery}}, \ and\
  \bibinfo {author} {\bibfnamefont {B.}~\bibnamefont {Werner}},\ }\href
  {\doibase 10.1103/PhysRevD.7.661} {\bibfield  {journal} {\bibinfo  {journal}
  {Phys.Rev.}\ }\textbf {\bibinfo {volume} {D7}},\ \bibinfo {pages} {661}
  (\bibinfo {year} {1973})}\BibitemShut {NoStop}%
\bibitem [{\citenamefont {Zieminski}\ \emph {et~al.}(1974)\citenamefont
  {Zieminski}, \citenamefont {Chaloupka}, \citenamefont {Dobrzynski},
  \citenamefont {Ferrando}, \citenamefont {Losty} \emph
  {et~al.}}]{Zieminski:1974ex}%
  \BibitemOpen
  \bibfield  {author} {\bibinfo {author} {\bibfnamefont {A.}~\bibnamefont
  {Zieminski}}, \bibinfo {author} {\bibfnamefont {V.}~\bibnamefont
  {Chaloupka}}, \bibinfo {author} {\bibfnamefont {L.}~\bibnamefont
  {Dobrzynski}}, \bibinfo {author} {\bibfnamefont {A.}~\bibnamefont
  {Ferrando}}, \bibinfo {author} {\bibfnamefont {M.}~\bibnamefont {Losty}},
  \emph {et~al.},\ }\href {\doibase 10.1016/0550-3213(74)90451-9} {\bibfield
  {journal} {\bibinfo  {journal} {Nucl.Phys.}\ }\textbf {\bibinfo {volume}
  {B69}},\ \bibinfo {pages} {502} (\bibinfo {year} {1974})}\BibitemShut
  {NoStop}%
\bibitem [{\citenamefont {Durusoy}\ \emph {et~al.}(1973)\citenamefont
  {Durusoy}, \citenamefont {Baubillier}, \citenamefont {George}, \citenamefont
  {Goldberg}, \citenamefont {Touchard} \emph {et~al.}}]{Durusoy:1973aj}%
  \BibitemOpen
  \bibfield  {author} {\bibinfo {author} {\bibfnamefont {N.}~\bibnamefont
  {Durusoy}}, \bibinfo {author} {\bibfnamefont {M.}~\bibnamefont {Baubillier}},
  \bibinfo {author} {\bibfnamefont {R.}~\bibnamefont {George}}, \bibinfo
  {author} {\bibfnamefont {M.}~\bibnamefont {Goldberg}}, \bibinfo {author}
  {\bibfnamefont {A.}~\bibnamefont {Touchard}},  \emph {et~al.},\ }\href
  {\doibase 10.1016/0370-2693(73)90658-8} {\bibfield  {journal} {\bibinfo
  {journal} {Phys.Lett.}\ }\textbf {\bibinfo {volume} {B45}},\ \bibinfo {pages}
  {517} (\bibinfo {year} {1973})}\BibitemShut {NoStop}%
\bibitem [{\citenamefont {Dudek}\ \emph
  {et~al.}(2011{\natexlab{a}})\citenamefont {Dudek}, \citenamefont {Edwards},
  \citenamefont {Peardon}, \citenamefont {Richards},\ and\ \citenamefont
  {Thomas}}]{Dudek:2010ew}%
  \BibitemOpen
  \bibfield  {author} {\bibinfo {author} {\bibfnamefont {J.~J.}\ \bibnamefont
  {Dudek}}, \bibinfo {author} {\bibfnamefont {R.~G.}\ \bibnamefont {Edwards}},
  \bibinfo {author} {\bibfnamefont {M.~J.}\ \bibnamefont {Peardon}}, \bibinfo
  {author} {\bibfnamefont {D.~G.}\ \bibnamefont {Richards}}, \ and\ \bibinfo
  {author} {\bibfnamefont {C.~E.}\ \bibnamefont {Thomas}},\ }\href {\doibase
  10.1103/PhysRevD.83.071504} {\bibfield  {journal} {\bibinfo  {journal} {Phys.
  Rev.}\ }\textbf {\bibinfo {volume} {D83}},\ \bibinfo {pages} {071504}
  (\bibinfo {year} {2011}{\natexlab{a}})},\ \Eprint
  {http://arxiv.org/abs/1011.6352} {arXiv:1011.6352 [hep-ph]} \BibitemShut
  {NoStop}%
\bibitem [{\citenamefont {Michael}(1985)}]{Michael:1985ne}%
  \BibitemOpen
  \bibfield  {author} {\bibinfo {author} {\bibfnamefont {C.}~\bibnamefont
  {Michael}},\ }\href {\doibase 10.1016/0550-3213(85)90297-4} {\bibfield
  {journal} {\bibinfo  {journal} {Nucl. Phys.}\ }\textbf {\bibinfo {volume}
  {B259}},\ \bibinfo {pages} {58} (\bibinfo {year} {1985})}\BibitemShut
  {NoStop}%
\bibitem [{\citenamefont {Luscher}\ and\ \citenamefont
  {Wolff}(1990)}]{Luscher:1990ck}%
  \BibitemOpen
  \bibfield  {author} {\bibinfo {author} {\bibfnamefont {M.}~\bibnamefont
  {Luscher}}\ and\ \bibinfo {author} {\bibfnamefont {U.}~\bibnamefont
  {Wolff}},\ }\href {\doibase 10.1016/0550-3213(90)90540-T} {\bibfield
  {journal} {\bibinfo  {journal} {Nucl. Phys.}\ }\textbf {\bibinfo {volume}
  {B339}},\ \bibinfo {pages} {222} (\bibinfo {year} {1990})}\BibitemShut
  {NoStop}%
\bibitem [{\citenamefont {Blossier}\ \emph {et~al.}(2009)\citenamefont
  {Blossier}, \citenamefont {Della~Morte}, \citenamefont {von Hippel},
  \citenamefont {Mendes},\ and\ \citenamefont {Sommer}}]{Blossier:2009kd}%
  \BibitemOpen
  \bibfield  {author} {\bibinfo {author} {\bibfnamefont {B.}~\bibnamefont
  {Blossier}}, \bibinfo {author} {\bibfnamefont {M.}~\bibnamefont
  {Della~Morte}}, \bibinfo {author} {\bibfnamefont {G.}~\bibnamefont {von
  Hippel}}, \bibinfo {author} {\bibfnamefont {T.}~\bibnamefont {Mendes}}, \
  and\ \bibinfo {author} {\bibfnamefont {R.}~\bibnamefont {Sommer}},\ }\href
  {\doibase 10.1088/1126-6708/2009/04/094} {\bibfield  {journal} {\bibinfo
  {journal} {JHEP}\ }\textbf {\bibinfo {volume} {04}},\ \bibinfo {pages} {094}
  (\bibinfo {year} {2009})},\ \Eprint {http://arxiv.org/abs/0902.1265}
  {arXiv:0902.1265 [hep-lat]} \BibitemShut {NoStop}%
\bibitem [{\citenamefont {Dudek}\ \emph {et~al.}(2009)\citenamefont {Dudek},
  \citenamefont {Edwards}, \citenamefont {Peardon}, \citenamefont {Richards},\
  and\ \citenamefont {Thomas}}]{Dudek:2009qf}%
  \BibitemOpen
  \bibfield  {author} {\bibinfo {author} {\bibfnamefont {J.~J.}\ \bibnamefont
  {Dudek}}, \bibinfo {author} {\bibfnamefont {R.~G.}\ \bibnamefont {Edwards}},
  \bibinfo {author} {\bibfnamefont {M.~J.}\ \bibnamefont {Peardon}}, \bibinfo
  {author} {\bibfnamefont {D.~G.}\ \bibnamefont {Richards}}, \ and\ \bibinfo
  {author} {\bibfnamefont {C.~E.}\ \bibnamefont {Thomas}},\ }\href {\doibase
  10.1103/PhysRevLett.103.262001} {\bibfield  {journal} {\bibinfo  {journal}
  {Phys. Rev. Lett.}\ }\textbf {\bibinfo {volume} {103}},\ \bibinfo {pages}
  {262001} (\bibinfo {year} {2009})},\ \Eprint {http://arxiv.org/abs/0909.0200}
  {arXiv:0909.0200 [hep-ph]} \BibitemShut {NoStop}%
\bibitem [{\citenamefont {Dudek}\ \emph {et~al.}(2010)\citenamefont {Dudek}
  \emph {et~al.}}]{Dudek:2010wm}%
  \BibitemOpen
  \bibfield  {author} {\bibinfo {author} {\bibfnamefont {J.~J.}\ \bibnamefont
  {Dudek}} \emph {et~al.},\ }\href {\doibase 10.1103/PhysRevD.82.034508}
  {\bibfield  {journal} {\bibinfo  {journal} {Phys. Rev.}\ }\textbf {\bibinfo
  {volume} {D82}},\ \bibinfo {pages} {034508} (\bibinfo {year} {2010})},\
  \Eprint {http://arxiv.org/abs/1004.4930} {arXiv:1004.4930 [hep-ph]}
  \BibitemShut {NoStop}%
\bibitem [{\citenamefont {Dudek}\ \emph
  {et~al.}(2011{\natexlab{b}})\citenamefont {Dudek}, \citenamefont {Edwards},
  \citenamefont {Joo}, \citenamefont {Peardon}, \citenamefont {Richards} \emph
  {et~al.}}]{Dudek:2011tt}%
  \BibitemOpen
  \bibfield  {author} {\bibinfo {author} {\bibfnamefont {J.~J.}\ \bibnamefont
  {Dudek}}, \bibinfo {author} {\bibfnamefont {R.~G.}\ \bibnamefont {Edwards}},
  \bibinfo {author} {\bibfnamefont {B.}~\bibnamefont {Joo}}, \bibinfo {author}
  {\bibfnamefont {M.~J.}\ \bibnamefont {Peardon}}, \bibinfo {author}
  {\bibfnamefont {D.~G.}\ \bibnamefont {Richards}},  \emph {et~al.},\ }\href
  {\doibase 10.1103/PhysRevD.83.111502} {\bibfield  {journal} {\bibinfo
  {journal} {Phys.Rev.}\ }\textbf {\bibinfo {volume} {D83}},\ \bibinfo {pages}
  {111502} (\bibinfo {year} {2011}{\natexlab{b}})},\ \Eprint
  {http://arxiv.org/abs/1102.4299} {arXiv:1102.4299 [hep-lat]} \BibitemShut
  {NoStop}%
\bibitem [{\citenamefont {Edwards}\ \emph {et~al.}(2011)\citenamefont
  {Edwards}, \citenamefont {Dudek}, \citenamefont {Richards},\ and\
  \citenamefont {Wallace}}]{Edwards:2011jj}%
  \BibitemOpen
  \bibfield  {author} {\bibinfo {author} {\bibfnamefont {R.~G.}\ \bibnamefont
  {Edwards}}, \bibinfo {author} {\bibfnamefont {J.~J.}\ \bibnamefont {Dudek}},
  \bibinfo {author} {\bibfnamefont {D.~G.}\ \bibnamefont {Richards}}, \ and\
  \bibinfo {author} {\bibfnamefont {S.~J.}\ \bibnamefont {Wallace}},\ }\href
  {\doibase 10.1103/PhysRevD.84.074508} {\bibfield  {journal} {\bibinfo
  {journal} {Phys.Rev.}\ }\textbf {\bibinfo {volume} {D84}},\ \bibinfo {pages}
  {074508} (\bibinfo {year} {2011})},\ \Eprint {http://arxiv.org/abs/1104.5152}
  {arXiv:1104.5152 [hep-ph]} \BibitemShut {NoStop}%
\bibitem [{\citenamefont {Dudek}\ and\ \citenamefont
  {Edwards}(2012)}]{Dudek:2012ag}%
  \BibitemOpen
  \bibfield  {author} {\bibinfo {author} {\bibfnamefont {J.~J.}\ \bibnamefont
  {Dudek}}\ and\ \bibinfo {author} {\bibfnamefont {R.~G.}\ \bibnamefont
  {Edwards}},\ }\href@noop {} {\bibfield  {journal} {\bibinfo  {journal} {to be
  published in Phys.Rev.D.}\ } (\bibinfo {year} {2012})},\ \Eprint
  {http://arxiv.org/abs/1201.2349} {arXiv:1201.2349 [hep-ph]} \BibitemShut
  {NoStop}%
\bibitem [{\citenamefont {Rummukainen}\ and\ \citenamefont
  {Gottlieb}(1995)}]{Rummukainen:1995vs}%
  \BibitemOpen
  \bibfield  {author} {\bibinfo {author} {\bibfnamefont {K.}~\bibnamefont
  {Rummukainen}}\ and\ \bibinfo {author} {\bibfnamefont {S.~A.}\ \bibnamefont
  {Gottlieb}},\ }\href {\doibase 10.1016/0550-3213(95)00313-H} {\bibfield
  {journal} {\bibinfo  {journal} {Nucl.Phys.}\ }\textbf {\bibinfo {volume}
  {B450}},\ \bibinfo {pages} {397} (\bibinfo {year} {1995})},\ \Eprint
  {http://arxiv.org/abs/hep-lat/9503028} {arXiv:hep-lat/9503028 [hep-lat]}
  \BibitemShut {NoStop}%
\bibitem [{\citenamefont {Kim}\ \emph {et~al.}(2005)\citenamefont {Kim},
  \citenamefont {Sachrajda},\ and\ \citenamefont {Sharpe}}]{Kim:2005gf}%
  \BibitemOpen
  \bibfield  {author} {\bibinfo {author} {\bibfnamefont {C.}~\bibnamefont
  {Kim}}, \bibinfo {author} {\bibfnamefont {C.}~\bibnamefont {Sachrajda}}, \
  and\ \bibinfo {author} {\bibfnamefont {S.~R.}\ \bibnamefont {Sharpe}},\
  }\href {\doibase 10.1016/j.nuclphysb.2005.08.029} {\bibfield  {journal}
  {\bibinfo  {journal} {Nucl.Phys.}\ }\textbf {\bibinfo {volume} {B727}},\
  \bibinfo {pages} {218} (\bibinfo {year} {2005})},\ \Eprint
  {http://arxiv.org/abs/hep-lat/0507006} {arXiv:hep-lat/0507006 [hep-lat]}
  \BibitemShut {NoStop}%
\bibitem [{\citenamefont {Christ}\ \emph {et~al.}(2005)\citenamefont {Christ},
  \citenamefont {Kim},\ and\ \citenamefont {Yamazaki}}]{Christ:2005gi}%
  \BibitemOpen
  \bibfield  {author} {\bibinfo {author} {\bibfnamefont {N.~H.}\ \bibnamefont
  {Christ}}, \bibinfo {author} {\bibfnamefont {C.}~\bibnamefont {Kim}}, \ and\
  \bibinfo {author} {\bibfnamefont {T.}~\bibnamefont {Yamazaki}},\ }\href
  {\doibase 10.1103/PhysRevD.72.114506} {\bibfield  {journal} {\bibinfo
  {journal} {Phys.Rev.}\ }\textbf {\bibinfo {volume} {D72}},\ \bibinfo {pages}
  {114506} (\bibinfo {year} {2005})},\ \Eprint
  {http://arxiv.org/abs/hep-lat/0507009} {arXiv:hep-lat/0507009 [hep-lat]}
  \BibitemShut {NoStop}%
\bibitem [{\citenamefont {Leskovec}\ and\ \citenamefont
  {Prelovsek}(2012)}]{Leskovec:2012gb}%
  \BibitemOpen
  \bibfield  {author} {\bibinfo {author} {\bibfnamefont {L.}~\bibnamefont
  {Leskovec}}\ and\ \bibinfo {author} {\bibfnamefont {S.}~\bibnamefont
  {Prelovsek}},\ }\href@noop {} {\  (\bibinfo {year} {2012})},\ \Eprint
  {http://arxiv.org/abs/1202.2145} {arXiv:1202.2145 [hep-lat]} \BibitemShut
  {NoStop}%
\bibitem [{\citenamefont {Edwards}\ \emph {et~al.}(2008)\citenamefont
  {Edwards}, \citenamefont {Joo},\ and\ \citenamefont {Lin}}]{Edwards:2008ja}%
  \BibitemOpen
  \bibfield  {author} {\bibinfo {author} {\bibfnamefont {R.~G.}\ \bibnamefont
  {Edwards}}, \bibinfo {author} {\bibfnamefont {B.}~\bibnamefont {Joo}}, \ and\
  \bibinfo {author} {\bibfnamefont {H.-W.}\ \bibnamefont {Lin}},\ }\href
  {\doibase 10.1103/PhysRevD.78.054501} {\bibfield  {journal} {\bibinfo
  {journal} {Phys. Rev.}\ }\textbf {\bibinfo {volume} {D78}},\ \bibinfo {pages}
  {054501} (\bibinfo {year} {2008})},\ \Eprint
  {http://arxiv.org/abs/arXiv:0803.3960} {arXiv:arXiv:0803.3960 [hep-lat]}
  \BibitemShut {NoStop}%
\bibitem [{\citenamefont {Lin}\ \emph {et~al.}(2009)\citenamefont {Lin} \emph
  {et~al.}}]{Lin:2008pr}%
  \BibitemOpen
  \bibfield  {author} {\bibinfo {author} {\bibfnamefont {H.-W.}\ \bibnamefont
  {Lin}} \emph {et~al.} (\bibinfo {collaboration} {Hadron Spectrum}),\ }\href
  {\doibase 10.1103/PhysRevD.79.034502} {\bibfield  {journal} {\bibinfo
  {journal} {Phys. Rev.}\ }\textbf {\bibinfo {volume} {D79}},\ \bibinfo {pages}
  {034502} (\bibinfo {year} {2009})},\ \Eprint
  {http://arxiv.org/abs/arXiv:0810.3588} {arXiv:arXiv:0810.3588 [hep-lat]}
  \BibitemShut {NoStop}%
\bibitem [{\citenamefont {Beane}\ \emph {et~al.}(2012)\citenamefont {Beane}
  \emph {et~al.}}]{Beane:2011sc}%
  \BibitemOpen
  \bibfield  {author} {\bibinfo {author} {\bibfnamefont {S.}~\bibnamefont
  {Beane}} \emph {et~al.} (\bibinfo {collaboration} {NPLQCD Collaboration}),\
  }\href {\doibase 10.1103/PhysRevD.85.034505} {\bibfield  {journal} {\bibinfo
  {journal} {Phys.Rev.}\ }\textbf {\bibinfo {volume} {D85}},\ \bibinfo {pages}
  {034505} (\bibinfo {year} {2012})},\ \bibinfo {note} {22 pages, 16 figures},\
  \Eprint {http://arxiv.org/abs/1107.5023} {arXiv:1107.5023 [hep-lat]}
  \BibitemShut {NoStop}%
\bibitem [{\citenamefont {Beane}\ \emph {et~al.}(2008)\citenamefont {Beane}
  \emph {et~al.}}]{Beane:2007xs}%
  \BibitemOpen
  \bibfield  {author} {\bibinfo {author} {\bibfnamefont {S.~R.}\ \bibnamefont
  {Beane}} \emph {et~al.},\ }\href {\doibase 10.1103/PhysRevD.77.014505}
  {\bibfield  {journal} {\bibinfo  {journal} {Phys. Rev.}\ }\textbf {\bibinfo
  {volume} {D77}},\ \bibinfo {pages} {014505} (\bibinfo {year} {2008})},\
  \Eprint {http://arxiv.org/abs/0706.3026} {arXiv:0706.3026 [hep-lat]}
  \BibitemShut {NoStop}%
\bibitem [{\citenamefont {Sasaki}\ and\ \citenamefont
  {Ishizuka}(2008)}]{Sasaki:2008sv}%
  \BibitemOpen
  \bibfield  {author} {\bibinfo {author} {\bibfnamefont {K.}~\bibnamefont
  {Sasaki}}\ and\ \bibinfo {author} {\bibfnamefont {N.}~\bibnamefont
  {Ishizuka}},\ }\href {\doibase 10.1103/PhysRevD.78.014511} {\bibfield
  {journal} {\bibinfo  {journal} {Phys. Rev.}\ }\textbf {\bibinfo {volume}
  {D78}},\ \bibinfo {pages} {014511} (\bibinfo {year} {2008})},\ \Eprint
  {http://arxiv.org/abs/0804.2941} {arXiv:0804.2941 [hep-lat]} \BibitemShut
  {NoStop}%
\bibitem [{\citenamefont {Feng}\ \emph {et~al.}(2010)\citenamefont {Feng},
  \citenamefont {Jansen},\ and\ \citenamefont {Renner}}]{Feng:2009ij}%
  \BibitemOpen
  \bibfield  {author} {\bibinfo {author} {\bibfnamefont {X.}~\bibnamefont
  {Feng}}, \bibinfo {author} {\bibfnamefont {K.}~\bibnamefont {Jansen}}, \ and\
  \bibinfo {author} {\bibfnamefont {D.~B.}\ \bibnamefont {Renner}},\ }\href
  {\doibase 10.1016/j.physletb.2010.01.018} {\bibfield  {journal} {\bibinfo
  {journal} {Phys. Lett.}\ }\textbf {\bibinfo {volume} {B684}},\ \bibinfo
  {pages} {268} (\bibinfo {year} {2010})},\ \Eprint
  {http://arxiv.org/abs/0909.3255} {arXiv:0909.3255 [hep-lat]} \BibitemShut
  {NoStop}%
\bibitem [{\citenamefont {Moore}\ and\ \citenamefont
  {Fleming}(2006{\natexlab{a}})}]{Moore:2005dw}%
  \BibitemOpen
  \bibfield  {author} {\bibinfo {author} {\bibfnamefont {D.~C.}\ \bibnamefont
  {Moore}}\ and\ \bibinfo {author} {\bibfnamefont {G.~T.}\ \bibnamefont
  {Fleming}},\ }\href {\doibase 10.1103/PhysRevD.73.014504} {\bibfield
  {journal} {\bibinfo  {journal} {Phys. Rev.}\ }\textbf {\bibinfo {volume}
  {D73}},\ \bibinfo {pages} {014504} (\bibinfo {year} {2006}{\natexlab{a}})},\
  \bibinfo {note} {erratum-ibid. \textbf{D74}, 079905 (2006)},\ \Eprint
  {http://arxiv.org/abs/hep-lat/0507018} {arXiv:hep-lat/0507018} \BibitemShut
  {NoStop}%
\bibitem [{\citenamefont {Thomas}\ \emph {et~al.}(2012)\citenamefont {Thomas},
  \citenamefont {Edwards},\ and\ \citenamefont {Dudek}}]{Thomas:2011rh}%
  \BibitemOpen
  \bibfield  {author} {\bibinfo {author} {\bibfnamefont {C.~E.}\ \bibnamefont
  {Thomas}}, \bibinfo {author} {\bibfnamefont {R.~G.}\ \bibnamefont {Edwards}},
  \ and\ \bibinfo {author} {\bibfnamefont {J.~J.}\ \bibnamefont {Dudek}},\
  }\href {\doibase 10.1103/PhysRevD.85.014507} {\bibfield  {journal} {\bibinfo
  {journal} {Phys. Rev.}\ }\textbf {\bibinfo {volume} {D85}},\ \bibinfo {pages}
  {014507} (\bibinfo {year} {2012})},\ \Eprint {http://arxiv.org/abs/1107.1930}
  {arXiv:1107.1930 [hep-lat]} \BibitemShut {NoStop}%
\bibitem [{\citenamefont {Basak}\ \emph {et~al.}(2005)\citenamefont {Basak}
  \emph {et~al.}}]{Basak:2005aq}%
  \BibitemOpen
  \bibfield  {author} {\bibinfo {author} {\bibfnamefont {S.}~\bibnamefont
  {Basak}} \emph {et~al.},\ }\href {\doibase 10.1103/PhysRevD.72.094506}
  {\bibfield  {journal} {\bibinfo  {journal} {Phys. Rev.}\ }\textbf {\bibinfo
  {volume} {D72}},\ \bibinfo {pages} {094506} (\bibinfo {year} {2005})},\
  \Eprint {http://arxiv.org/abs/hep-lat/0506029} {arXiv:hep-lat/0506029}
  \BibitemShut {NoStop}%
\bibitem [{\citenamefont {Moore}\ and\ \citenamefont
  {Fleming}(2006{\natexlab{b}})}]{Moore:2006ng}%
  \BibitemOpen
  \bibfield  {author} {\bibinfo {author} {\bibfnamefont {D.~C.}\ \bibnamefont
  {Moore}}\ and\ \bibinfo {author} {\bibfnamefont {G.~T.}\ \bibnamefont
  {Fleming}},\ }\href {\doibase 10.1103/PhysRevD.74.054504} {\bibfield
  {journal} {\bibinfo  {journal} {Phys. Rev.}\ }\textbf {\bibinfo {volume}
  {D74}},\ \bibinfo {pages} {054504} (\bibinfo {year} {2006}{\natexlab{b}})},\
  \Eprint {http://arxiv.org/abs/hep-lat/0607004} {arXiv:hep-lat/0607004}
  \BibitemShut {NoStop}%
\bibitem [{\citenamefont {Peardon}\ \emph {et~al.}(2009)\citenamefont {Peardon}
  \emph {et~al.}}]{Peardon:2009gh}%
  \BibitemOpen
  \bibfield  {author} {\bibinfo {author} {\bibfnamefont {M.}~\bibnamefont
  {Peardon}} \emph {et~al.} (\bibinfo {collaboration} {Hadron Spectrum}),\
  }\href {\doibase 10.1103/PhysRevD.80.054506} {\bibfield  {journal} {\bibinfo
  {journal} {Phys. Rev.}\ }\textbf {\bibinfo {volume} {D80}},\ \bibinfo {pages}
  {054506} (\bibinfo {year} {2009})},\ \Eprint {http://arxiv.org/abs/0905.2160}
  {arXiv:0905.2160 [hep-lat]} \BibitemShut {NoStop}%
\bibitem [{\citenamefont {Beane}\ \emph {et~al.}(2009)\citenamefont {Beane}
  \emph {et~al.}}]{Beane:2009kya}%
  \BibitemOpen
  \bibfield  {author} {\bibinfo {author} {\bibfnamefont {S.~R.}\ \bibnamefont
  {Beane}} \emph {et~al.},\ }\href {\doibase 10.1103/PhysRevD.79.114502}
  {\bibfield  {journal} {\bibinfo  {journal} {Phys. Rev.}\ }\textbf {\bibinfo
  {volume} {D79}},\ \bibinfo {pages} {114502} (\bibinfo {year} {2009})},\
  \Eprint {http://arxiv.org/abs/0903.2990} {arXiv:0903.2990 [hep-lat]}
  \BibitemShut {NoStop}%
\bibitem [{\citenamefont {Detmold}\ and\ \citenamefont
  {Smigielski}(2011)}]{Detmold:2011kw}%
  \BibitemOpen
  \bibfield  {author} {\bibinfo {author} {\bibfnamefont {W.}~\bibnamefont
  {Detmold}}\ and\ \bibinfo {author} {\bibfnamefont {B.}~\bibnamefont
  {Smigielski}},\ }\href {\doibase 10.1103/PhysRevD.84.014508} {\bibfield
  {journal} {\bibinfo  {journal} {Phys.Rev.}\ }\textbf {\bibinfo {volume}
  {D84}},\ \bibinfo {pages} {014508} (\bibinfo {year} {2011})},\ \Eprint
  {http://arxiv.org/abs/1103.4362} {arXiv:1103.4362 [hep-lat]} \BibitemShut
  {NoStop}%
\bibitem [{\citenamefont {Dudek}\ \emph {et~al.}(2008)\citenamefont {Dudek},
  \citenamefont {Edwards}, \citenamefont {Mathur},\ and\ \citenamefont
  {Richards}}]{Dudek:2007wv}%
  \BibitemOpen
  \bibfield  {author} {\bibinfo {author} {\bibfnamefont {J.~J.}\ \bibnamefont
  {Dudek}}, \bibinfo {author} {\bibfnamefont {R.~G.}\ \bibnamefont {Edwards}},
  \bibinfo {author} {\bibfnamefont {N.}~\bibnamefont {Mathur}}, \ and\ \bibinfo
  {author} {\bibfnamefont {D.~G.}\ \bibnamefont {Richards}},\ }\href {\doibase
  10.1103/PhysRevD.77.034501} {\bibfield  {journal} {\bibinfo  {journal} {Phys.
  Rev.}\ }\textbf {\bibinfo {volume} {D77}},\ \bibinfo {pages} {034501}
  (\bibinfo {year} {2008})},\ \Eprint {http://arxiv.org/abs/arXiv:0707.4162}
  {arXiv:arXiv:0707.4162 [hep-lat]} \BibitemShut {NoStop}%
\bibitem [{\citenamefont {Roy}(1971)}]{Roy:1971tc}%
  \BibitemOpen
  \bibfield  {author} {\bibinfo {author} {\bibfnamefont {S.}~\bibnamefont
  {Roy}},\ }\href {\doibase 10.1016/0370-2693(71)90724-6} {\bibfield  {journal}
  {\bibinfo  {journal} {Phys.Lett.}\ }\textbf {\bibinfo {volume} {B36}},\
  \bibinfo {pages} {353} (\bibinfo {year} {1971})}\BibitemShut {NoStop}%
\bibitem [{\citenamefont {Colangelo}\ \emph {et~al.}(2001)\citenamefont
  {Colangelo}, \citenamefont {Gasser},\ and\ \citenamefont
  {Leutwyler}}]{Colangelo:2001df}%
  \BibitemOpen
  \bibfield  {author} {\bibinfo {author} {\bibfnamefont {G.}~\bibnamefont
  {Colangelo}}, \bibinfo {author} {\bibfnamefont {J.}~\bibnamefont {Gasser}}, \
  and\ \bibinfo {author} {\bibfnamefont {H.}~\bibnamefont {Leutwyler}},\ }\href
  {\doibase 10.1016/S0550-3213(01)00147-X} {\bibfield  {journal} {\bibinfo
  {journal} {Nucl.Phys.}\ }\textbf {\bibinfo {volume} {B603}},\ \bibinfo
  {pages} {125} (\bibinfo {year} {2001})},\ \Eprint
  {http://arxiv.org/abs/hep-ph/0103088} {arXiv:hep-ph/0103088 [hep-ph]}
  \BibitemShut {NoStop}%
\bibitem [{\citenamefont {Aoki}\ \emph {et~al.}(2007)\citenamefont {Aoki} \emph
  {et~al.}}]{Aoki:2007rd}%
  \BibitemOpen
  \bibfield  {author} {\bibinfo {author} {\bibfnamefont {S.}~\bibnamefont
  {Aoki}} \emph {et~al.} (\bibinfo {collaboration} {CP-PACS}),\ }\href
  {\doibase 10.1103/PhysRevD.76.094506} {\bibfield  {journal} {\bibinfo
  {journal} {Phys. Rev.}\ }\textbf {\bibinfo {volume} {D76}},\ \bibinfo {pages}
  {094506} (\bibinfo {year} {2007})},\ \Eprint {http://arxiv.org/abs/0708.3705}
  {arXiv:0708.3705 [hep-lat]} \BibitemShut {NoStop}%
\bibitem [{\citenamefont {Feng}\ \emph {et~al.}(2011)\citenamefont {Feng},
  \citenamefont {Jansen},\ and\ \citenamefont {Renner}}]{Feng:2010es}%
  \BibitemOpen
  \bibfield  {author} {\bibinfo {author} {\bibfnamefont {X.}~\bibnamefont
  {Feng}}, \bibinfo {author} {\bibfnamefont {K.}~\bibnamefont {Jansen}}, \ and\
  \bibinfo {author} {\bibfnamefont {D.~B.}\ \bibnamefont {Renner}},\ }\href
  {\doibase 10.1103/PhysRevD.83.094505} {\bibfield  {journal} {\bibinfo
  {journal} {Phys.Rev.}\ }\textbf {\bibinfo {volume} {D83}},\ \bibinfo {pages}
  {094505} (\bibinfo {year} {2011})},\ \Eprint {http://arxiv.org/abs/1011.5288}
  {arXiv:1011.5288 [hep-lat]} \BibitemShut {NoStop}%
\bibitem [{\citenamefont {Lang}\ \emph {et~al.}(2011)\citenamefont {Lang},
  \citenamefont {Mohler}, \citenamefont {Prelovsek},\ and\ \citenamefont
  {Vidmar}}]{Lang:2011mn}%
  \BibitemOpen
  \bibfield  {author} {\bibinfo {author} {\bibfnamefont {C.}~\bibnamefont
  {Lang}}, \bibinfo {author} {\bibfnamefont {D.}~\bibnamefont {Mohler}},
  \bibinfo {author} {\bibfnamefont {S.}~\bibnamefont {Prelovsek}}, \ and\
  \bibinfo {author} {\bibfnamefont {M.}~\bibnamefont {Vidmar}},\ }\href@noop {}
  {\bibfield  {journal} {\bibinfo  {journal} {Phys.Rev.}\ }\textbf {\bibinfo
  {volume} {D84}},\ \bibinfo {pages} {054503} (\bibinfo {year} {2011})},\
  \Eprint {http://arxiv.org/abs/1105.5636} {arXiv:1105.5636 [hep-lat]}
  \BibitemShut {NoStop}%
\bibitem [{\citenamefont {Aoki}\ \emph {et~al.}(2011)\citenamefont {Aoki} \emph
  {et~al.}}]{Aoki:2011yj}%
  \BibitemOpen
  \bibfield  {author} {\bibinfo {author} {\bibfnamefont {S.}~\bibnamefont
  {Aoki}} \emph {et~al.} (\bibinfo {collaboration} {CS Collaboration}),\
  }\href@noop {} {\bibfield  {journal} {\bibinfo  {journal} {Phys.Rev.}\
  }\textbf {\bibinfo {volume} {D84}},\ \bibinfo {pages} {094505} (\bibinfo
  {year} {2011})},\ \Eprint {http://arxiv.org/abs/1106.5365} {arXiv:1106.5365
  [hep-lat]} \BibitemShut {NoStop}%
\bibitem [{\citenamefont {Edwards}\ and\ \citenamefont
  {Joo}(2005)}]{Edwards:2004sx}%
  \BibitemOpen
  \bibfield  {author} {\bibinfo {author} {\bibfnamefont {R.~G.}\ \bibnamefont
  {Edwards}}\ and\ \bibinfo {author} {\bibfnamefont {B.}~\bibnamefont {Joo}},\
  }\href@noop {} {\bibfield  {journal} {\bibinfo  {journal} {Nucl. Phys. B.
  Proc. Suppl.}\ }\textbf {\bibinfo {volume} {140}},\ \bibinfo {pages} {832}
  (\bibinfo {year} {2005})}\BibitemShut {NoStop}%
\bibitem [{\citenamefont {Clark}\ \emph {et~al.}(2010)\citenamefont {Clark}
  \emph {et~al.}}]{Clark:2009wm}%
  \BibitemOpen
  \bibfield  {author} {\bibinfo {author} {\bibfnamefont {M.~A.}\ \bibnamefont
  {Clark}} \emph {et~al.},\ }\href {\doibase 10.1016/j.cpc.2010.05.002}
  {\bibfield  {journal} {\bibinfo  {journal} {Comput. Phys. Commun.}\ }\textbf
  {\bibinfo {volume} {181}},\ \bibinfo {pages} {1517} (\bibinfo {year}
  {2010})},\ \Eprint {http://arxiv.org/abs/0911.3191} {arXiv:0911.3191
  [hep-lat]} \BibitemShut {NoStop}%
\bibitem [{\citenamefont {Babich}\ \emph {et~al.}(2010)\citenamefont {Babich},
  \citenamefont {Clark},\ and\ \citenamefont {Joo}}]{Babich:2010mu}%
  \BibitemOpen
  \bibfield  {author} {\bibinfo {author} {\bibfnamefont {R.}~\bibnamefont
  {Babich}}, \bibinfo {author} {\bibfnamefont {M.~A.}\ \bibnamefont {Clark}}, \
  and\ \bibinfo {author} {\bibfnamefont {B.}~\bibnamefont {Joo}},\ }\href
  {\doibase 10.1109/SC.2010.40} {\bibfield  {journal} {\bibinfo  {journal}
  {{ACM/IEEE Int. Conf. High Performance Computing, Networking, Storage and
  Analysis, New Orleans}}\ } (\bibinfo {year} {2010}),\ 10.1109/SC.2010.40},\
  \Eprint {http://arxiv.org/abs/1011.0024} {arXiv:1011.0024 [hep-lat]}
  \BibitemShut {NoStop}%
\bibitem [{Sup()}]{SupplementaryMaterial}%
  \BibitemOpen
  \href@noop {} {}\bibinfo {note} {Supplementary material included as an
  ancillary file with the arXiv version of this article}\BibitemShut {NoStop}%
\bibitem [{\citenamefont {Cornwell}(1986)}]{CornwellGroupTheory}%
  \BibitemOpen
  \bibfield  {author} {\bibinfo {author} {\bibfnamefont {J.~F.}\ \bibnamefont
  {Cornwell}},\ }\href@noop {} {\emph {\bibinfo {title} {Group Theory in
  Physics: Volume I}}}\ (\bibinfo  {publisher} {Academic Press},\ \bibinfo
  {year} {1986})\BibitemShut {NoStop}%
\end{thebibliography}%

 \clearpage
\appendix

\section{Multi-particle operators}\label{app:operators}

In this appendix we give a generalisation of and further details for our multi-particle operator construction.  In Section \ref{app:construction} we describe our general multi-particle operator construction and in Section \ref{app:inducedrep} discuss the method used to calculate the Clebsch-Gordan coefficients.  In Section \ref{app:conventions} we state our conventions for the lattice irreps and choices of rotations before giving some example Clebsch-Gordan coefficients in Section \ref{app:CGs}.  Further Clebsch-Gordan coefficients are given in supplementary material~\cite{SupplementaryMaterial}.

\subsection{Operator construction}
\label{app:construction}

Here we describe our construction of multi-particle operators, generalising the discussion in Section \ref{sec:multi}.  We will consider two-particle operators but the procedure can be applied iteratively to construct multi-particle operators for a larger number of particles.

In Refs.~\cite{Dudek:2010wm,Dudek:2009qf} we discussed how operators with a definite continuum $J^P$ and spin-component $M$, $\mathcal{O}^{J^P,M}(\vec{p}=\vec{0})$, can be constructed out of gauge-covariant derivatives and Dirac gamma matrices.  The appropriate lattice operators were formed by \emph{subducing} these continuum operators into octahedral group irreps,
$$\left[\mathcal{O}^{[J^P]}_{\Lambda^P,\mu}(\vec{0})\right]^\dagger = \sum_{M} \mathcal{S}^{\Lambda,\mu}_{J,M} \left[\mathcal{O}^{J^P}(\vec{0})\right]^\dagger ~,$$
where $\Lambda^P$ is an irrep of $\text{O}^{\text{D}}_h$, $\mu$ is the irrep row and $\mathcal{S}^{\Lambda,\mu}_{J,M}$ are subduction coefficients discussed in Refs.~\cite{Dudek:2010wm,Dudek:2009qf,Edwards:2011jj}.

In Ref.~\cite{Thomas:2011rh} we discussed how to construct \emph{helicity operators}\footnote{here we give expressions for \emph{creation} operators},
\begin{equation}
\label{equ:helop}
\left[\mathbb{O}^{J^P,\lambda}(\vec{p})\right]^{\dagger} = \sum_{m} D^{(J)}_{m \lambda}(R) ~ \left[\mathcal{O}^{J^P,m}(\vec{p})\right]^{\dagger} ~,
\end{equation}
where $\lambda$ is the helicity, $J^P$ refers to the spin and parity of the operator with $\vec{p}=\vec{0}$ and $D$ is a Wigner-D matrix; a summary of our conventions is given in Appendix \ref{app:conventions} and we refer to Ref.~\cite{Thomas:2011rh} for more details, for example, on the choice of $R$.  From these we constructed \emph{subduced helicity operators}, 
$$\left[\mathbb{O}^{[J^P,|\lambda|]}_{\Lambda,\mu}(\vec{p})\right]^{\dagger} = \sum_{\hat{\lambda}=\pm|\lambda|} \mathcal{S}_{\Lambda,\mu}^{\tilde{\eta},\hat{\lambda}} \left[\mathbb{O}^{J^P,\hat{\lambda}}(\vec{p})\right]^{\dagger} ~,$$
where $\Lambda$ is an irrep of $\LG(\vec{p})$ (the little group for momentum $\vec{p}$), $\mu$ is the irrep row and $\tilde{\eta} \equiv P(-1)^J$.  The subduction coefficients, $\mathcal{S}_{\Lambda,\mu}^{\tilde{\eta},\hat{\lambda}}$, which were discussed in Ref.~\cite{Thomas:2011rh}\nopagebreak, are given in Table \ref{table:subductions}.  Note that here we give subductions coefficients for \emph{creation} helicity operators which transform like states, i.e. under a rotation by $\phi$ about the axis defined by $\vec{p}$ they transform as ${R}_{\phi} \left|\lambda \right> = e^{-i \phi \lambda} \left| \lambda \right>$.  In Ref.~\cite{Thomas:2011rh} we showed that these subduced helicity operators are useful for studying mesons with non-zero momentum on the lattice.

\begin{table}[b]
\begin{ruledtabular}
\begin{tabular}{c|c|c|c}
Group & $|\lambda|^{\tilde{\eta}}$ & $\Lambda(\mu)$ & $\mathcal{S}_{\Lambda,\mu}^{\tilde{\eta},\lambda}$ \\
\hline
\hline
\multirow{8}{*}{$\begin{matrix}\text{Dic}_{4} \\ [0,0,n]\end{matrix}$} 
 & $0^+$ & $A_1(1)$ & $1$ \\
 & $0^-$ & $A_2(1)$ & $1$ \\
 & $1$   & $E_2\left(\begin{smallmatrix}1 \\ 2\end{smallmatrix}\right)$ & $(\delta_{s,+} \pm \tilde{\eta} \delta_{s,-})/\sqrt{2}$ \\
 & $2$   & $B_1(1)$ & $(\delta_{s,+} + \tilde{\eta} \delta_{s,-})/\sqrt{2}$ \\
 & $2$   & $B_2(1)$ & $(\delta_{s,+} - \tilde{\eta} \delta_{s,-})/\sqrt{2}$ \\
 & $3$   & $E_2\left(\begin{smallmatrix}1 \\ 2\end{smallmatrix}\right)$ & $(\pm\delta_{s,+} + \tilde{\eta} \delta_{s,-})/\sqrt{2}$ \\
 & $4$   & $A_1(1)$ & $(\delta_{s,+} + \tilde{\eta} \delta_{s,-})/\sqrt{2}$ \\
 & $4$   & $A_2(1)$ & $(\delta_{s,+} - \tilde{\eta} \delta_{s,-})/\sqrt{2}$ \\
\hline
\multirow{10}{*}{$\begin{matrix}\text{Dic}_{2} \\ [0,n,n]\end{matrix}$}
 & $0^+$ & $A_1(1)$ & $1$ \\
 & $0^-$ & $A_2(1)$ & $1$ \\
 & $1$   & $B_1(1)$ & $(\delta_{s,+} + \tilde{\eta} \delta_{s,-})/\sqrt{2}$ \\
 & $1$   & $B_2(1)$ & $(\delta_{s,+} - \tilde{\eta} \delta_{s,-})/\sqrt{2}$ \\
 & $2$   & $A_1(1)$ & $(\delta_{s,+} + \tilde{\eta} \delta_{s,-})/\sqrt{2}$ \\
 & $2$   & $A_2(1)$ & $(\delta_{s,+} - \tilde{\eta} \delta_{s,-})/\sqrt{2}$ \\
 & $3$   & $B_1(1)$ & $(\delta_{s,+} + \tilde{\eta} \delta_{s,-})/\sqrt{2}$ \\
 & $3$   & $B_2(1)$ & $(\delta_{s,+} - \tilde{\eta} \delta_{s,-})/\sqrt{2}$ \\
 & $4$   & $A_1(1)$ & $(\delta_{s,+} + \tilde{\eta} \delta_{s,-})/\sqrt{2}$ \\
 & $4$   & $A_2(1)$ & $(\delta_{s,+} - \tilde{\eta} \delta_{s,-})/\sqrt{2}$ \\
\hline
\multirow{7}{*}{$\begin{matrix}\text{Dic}_{3} \\ [n,n,n]\end{matrix}$}
 & $0^+$ & $A_1(1)$ & $1$ \\
 & $0^-$ & $A_2(1)$ & $1$ \\
 & $1$   & $E_2\left(\begin{smallmatrix}1 \\ 2\end{smallmatrix}\right)$ & $(\delta_{s,+} \pm \tilde{\eta} \delta_{s,-})/\sqrt{2}$ \\
 & $2$   & $E_2\left(\begin{smallmatrix}1 \\ 2\end{smallmatrix}\right)$ & $(\pm\delta_{s,+} - \tilde{\eta} \delta_{s,-})/\sqrt{2}$ \\
 & $3$   & $A_1(1)$ & $(\delta_{s,+} - \tilde{\eta} \delta_{s,-})/\sqrt{2}$ \\
 & $3$   & $A_2(1)$ & $(\delta_{s,+} + \tilde{\eta} \delta_{s,-})/\sqrt{2}$ \\
 & $4$   & $E_2\left(\begin{smallmatrix}1 \\ 2\end{smallmatrix}\right)$ & $(\delta_{s,+} \mp \tilde{\eta} \delta_{s,-})/\sqrt{2}$ \\
\hline
\multirow{10}{*}{$\begin{matrix}\text{C}_{4} \\ [n,m,0] \\ [n,n,m]\end{matrix}$}
 & $0^+$ & $A_1(1)$ & $1$ \\
 & $0^-$ & $A_2(1)$ & $1$ \\
 & $1$   & $A_1(1)$ & $(\delta_{s,+} - \tilde{\eta} \delta_{s,-})/\sqrt{2}$ \\
 & $1$   & $A_2(1)$ & $(\delta_{s,+} + \tilde{\eta} \delta_{s,-})/\sqrt{2}$ \\
 & $2$   & $A_1(1)$ & $(\delta_{s,+} + \tilde{\eta} \delta_{s,-})/\sqrt{2}$ \\
 & $2$   & $A_2(1)$ & $(\delta_{s,+} - \tilde{\eta} \delta_{s,-})/\sqrt{2}$ \\
 & $3$   & $A_1(1)$ & $(\delta_{s,+} - \tilde{\eta} \delta_{s,-})/\sqrt{2}$ \\
 & $3$   & $A_2(1)$ & $(\delta_{s,+} + \tilde{\eta} \delta_{s,-})/\sqrt{2}$ \\
 & $4$   & $A_1(1)$ & $(\delta_{s,+} + \tilde{\eta} \delta_{s,-})/\sqrt{2}$ \\
 & $4$   & $A_2(1)$ & $(\delta_{s,+} - \tilde{\eta} \delta_{s,-})/\sqrt{2}$ \\
\end{tabular}
\end{ruledtabular}
\caption{Subduction coefficients, $\mathcal{S}_{\Lambda,\mu}^{\tilde{\eta},\lambda}$, for integer spin, $|\lambda| \leq 4$ and with $s \equiv \text{sign}(\lambda)$; other notation is defined in the text.}
\label{table:subductions}
\end{table}

In general, a two-particle creation operator with total momentum $\vec{P}$ can be constructed from the product of two single-particle operators,
\begin{widetext}
\begin{equation*}
\left[\mathbb{O}^{[\Lambda_1\{\vec{k}_1\!\}^{\!\star};\,\Lambda_2\{\vec{k}_2\!\}^{\!\star}]}_{\Lambda \mu}(\vec{P})\right]^{\dagger} = 
\sum_{\substack{\mu_1,\mu_2 \\ \vec{k}_1 \in \{\vec{k}_1\!\}^{\!\star} \\ \vec{k}_2 \in \{\vec{k}_2\!\}^{\!\star} \\ \vec{k}_1 + \vec{k}_2 = \vec{P}}}
~\mathcal{C}(\vec{P}\Lambda\mu; \vec{k}_1\Lambda_1\mu_1; \vec{k}_2\Lambda_2\mu_2 )~
\left[\mathbb{O}_{\Lambda_1\mu_1}(\vec{k}_1)\right]^{\dagger}~\left[\mathbb{O}_{\Lambda_2\mu_2}(\vec{k}_2)\right]^{\dagger} ,
\end{equation*}
where $\Lambda_{1,2}(\mu_{1,2})$ and $\Lambda(\mu)$ are respectively irreps(irrep rows) of $\LG(\vec{k}_{1,2})$ and $\LG(\vec{P})$.  $\mathcal{C}$ are the Clebsch-Gordan coefficients for $\Lambda_1 (\{\vec{k}_1\}^\star) \otimes \Lambda_2 (\{\vec{k}_2\}^\star) \rightarrow \Lambda (\vec{P})$ which we discuss in the following section.  The sum over $\vec{k}_{1,2}$ is a sum over all momenta in the \emph{stars} of $\vec{k}_{1,2}$, $\{\vec{k}_{1,2}\}^\star$, i.e. all momenta related to $\vec{k}_{1,2}$ by an allowed lattice rotation.  In other words, the sum is over $R\vec{k}_{1,2} ~ \forall ~ R \in \text{O}^{\text{D}}_h$; the restriction that $\vec{k}_1 + \vec{k}_2 = \vec{P}$ is equivalent to requiring $R \in \LG(\vec{P})$.

\end{widetext}
\subsection{Induced representation method for calculating Clebsch-Gordan coefficients}
\label{app:inducedrep}

We use the projection formula with the induced representation~\cite{CornwellGroupTheory} to construct the Clebsch-Gordan coefficients for $\Lambda_1 \otimes \Lambda_2 \rightarrow \Lambda$ where $\Lambda_1$, $\Lambda_2$ and $\Lambda$ are each irreps of, respectively, groups $\text{G}_1$, $\text{G}_2$ and $\text{G}$.  For our purposes, these groups will be the double cover of the octahedral group, $\text{O}^{\text{D}}_h$, or a little group $\subset \text{O}^{\text{D}}_h$.  

A group, G, can be partitioned into cosets by a subgroup $\text{H} \subset \text{G}$.  Two elements $x,y \in \text{G}$ are in the same left coset\footnote{note that if H is not a normal subgroup, the left and right cosets are different} if and only if $y^{-1} x \in \text{H}$.  One coset, containing the identity element, will be the subgroup H itself.  The number of cosets $n = |\text{G}|/|\text{H}|$.  A coset representative is one element from the coset; $\tilde{R}_1,\tilde{R}_2, \ldots, \tilde{R}_n$ are a set of $n$ coset representatives, one from each coset (a transversal).

If $\Lambda$ is a $|\Lambda|$-dimensional irrep of $H$, then $\Gamma = \Lambda(\text{H}) \uparrow \text{G}$ is a $(n|\Lambda|)$-dimensional unitary representation of G induced from irrep $\Lambda$ of H.  It is defined for $R \in \text{G}$ by
\begin{eqnarray}
\label{equ:inducedrep}
\Gamma(R)_{ir,js} = \left\{ \begin{array}{ll}
\Lambda(\tilde{R}_i^{-1} R \tilde{R}_j)_{rs} & \text{if $\tilde{R}_i^{-1} R \tilde{R}_j \in \text{H}$} \\
 0 & \text{otherwise}
\end{array}\right.
\\ \nonumber
\end{eqnarray}
Here $i,j$ label the coset and $r,s$ the rows and columns of the irrep $\Lambda$.

To make this more concrete, consider the double-cover octahedral group with parity, $\text{G}=\text{O}^{\text{D}}_h$ ($|\text{G}|=96$), and the little group for momentum $\pref = [0,0,1]$, $\text{H}=\text{Dic}_{4}$ ($|\text{H}|=16$).
There must be 6 left cosets with this little group and these correspond to the 6 momenta in $\{\pref\}^\star$.  If $R_1$ and $R_2$ are in the same coset, then $R_2 \pref = R_1R_1^{-1}R_2 \pref = R_1 R_H \pref$ where $R_H \in \text{H}$.  But from the definition of the little group, $R_H \pref = \pref$, and therefore $R_2 \pref = R_1 \pref$.  The converse can also be shown to be true.  Therefore, each left coset can be labelled by $\vec{p} \in \{\pref\}^\star$ with elements $R$ such that $R \pref = \vec{p}$. 

Therefore, in Eq.~\ref{equ:inducedrep}, the indices $i,j$ refer to a particular momentum direction, $\vec{p} \in \{\pref\}^\star$.  In effect the induced representation splits up $R$ into a piece $R_H$ in the little group [giving $\Lambda(R_H)$] and a piece which rotates the momentum direction from $\pref$ to $\vec{p}$.  The freedom to choose a particular coset representative for each coset is the same freedom as the choice of lattice rotation $R$ which rotates $\pref$ to $\vec{p}$ (see Ref.~\cite{Thomas:2011rh}).  It is not important which particular element is chosen as the representative but this choice should be made consistently; we discuss our conventions in Section \ref{app:conventions}.

Once the induced representations, $\Gamma_i = \Lambda_i(\text{G}_i) \uparrow \text{O}^{\text{D}}_h$ ($i=1,2$), have been constructed, the Clebsch-Gordan coefficients can be generated using the projection formula in the same way as Clebsch-Gordan coefficients for $\text{O}^{\text{D}}_h \otimes \text{O}^{\text{D}}_h \rightarrow \text{O}^{\text{D}}_h$.  The projection formula gives
\begin{widetext}
\begin{equation}
\left[\mathbb{O}_{\Lambda\mu}(\vec{P})\right]^{\dagger} = \frac{|\Lambda|}{|\text{G}|} \sum_{R \in \text{G}} \Lambda(R)^*_{\mu \mu'} 
\sum_{\substack{j_1, \mu_1'\\ j_2, \mu_2'}} 
\Gamma_1(R)_{j_1 \mu_1', i_1 \mu_1} \, \Gamma_2(R)_{j_2 \mu_2', i_2 \mu_2} 
\left[\mathbb{O}_{\Lambda_1\mu_1'}(\{\vec{p}_1\!\}^\star_{j_1})\right]^{\dagger} 
\left[\mathbb{O}_{\Lambda_2\mu_2'}(\{\vec{p}_2\!\}^\star_{j_2})\right]^{\dagger} ~,
\end{equation}
\end{widetext}
where $\vec{P} = \{\vec{p}_1\!\}^\star_{j_1} + \{\vec{p}_2\!\}^\star_{j_2}$ is fixed, $\mu,\mu_1^{(\prime)},\mu_2^{(\prime)}$ label irrep rows, and $i_1,j_1$ and $i_2,j_2$ label the different momenta in the sets $\{\vec{p}_1\!\}^\star$ and $\{\vec{p}_2\!\}^\star$ respectively.  Note that this is written for operators which transform as $\hat{R} \mathbb{O}^{\dagger}_{\Lambda\mu} = \sum_{\mu'} \Lambda(R)_{\mu' \mu} \mathbb{O}^{\dagger}_{\Lambda\mu'}$.  After forming the appropriate linearly independent combinations\footnote{linear combinations of $\mu',\mu_1,\mu_2$; it is sufficient to consider one particular $i_1$ and $i_2$}, the Clebsch-Gordan coefficients can be read off up to a phase choice and normalisation.  We choose the phase so that our resulting correlators are real.

When $\vec{P} = \{\vec{p}_1\!\}^\star_{j_1} + \{\vec{p}_2\!\}^\star_{j_2} = \vec{0}$, $\text{G} = \text{O}^{\text{D}}_h$ and the sum is over all $R \in \text{O}^{\text{D}}_h$.  In this case it is sufficient to consider $i_1=i_2=1$ to generate all the Clebsch-Gordan coefficients.

When $\vec{P} = \{\vec{p}_1\!\}^\star_{j_1} + \{\vec{p}_2\!\}^\star_{j_2} \neq \vec{0}$, G is a little group and the sum over $R$ is restricted to those elements in the little group.  A particular choice of $i_1$ and $i_2$ can be made such that $\vec{P} = \{\vec{p}_1\!\}^\star_{i_1} + \{\vec{p}_2\!\}^\star_{i_2}$; because $R$ is in the little group, the rotations of $\vec{p}_1$ and $\vec{p}_2$ will automatically ensure that $\vec{P}$ is fixed.


\subsection{Lattice and little group conventions and rotations}
\label{app:conventions}

As described in Ref.~\cite{Thomas:2011rh}, we break the (active) rotation $R$ appearing in Eq.~\ref{equ:helop} into two stages: $R = R_{\text{lat}} R_{\text{ref}}$.  The first rotation, $R_{\text{ref}}$, takes $(0,0,|\vec{p}|)$ into $\vec{p}_{\text{ref}}$, where $\vec{p}_{\text{ref}}$ is a reference direction for momenta of type $\vec{p}$ (i.e. for $\{\vec{p}\}^\star$).  In Table \ref{table:refrotations} we give the specific rotations that we use for $R_{\text{ref}}$.  We use the same convention as in Ref.~\cite{Thomas:2011rh}, namely that a rotation $R_{\phi,\theta,\psi} = e^{-i \phi \hat{J}_z} e^{-i \theta \hat{J}_y} e^{-i \psi \hat{J}_z}$ rotates around the $z$-axis by $\psi$, then around the $y$-axis by $\theta$ and finally around the $z$-axis by $\phi$ (with a fixed coordinate system).  In Table \ref{table:latrotations:00n} we give the rotations, $R_{\text{lat}}$, for each momentum of the form $[0,0,n]$; these correspond to the coset representatives discussed in Section \ref{app:inducedrep}.  Rotations, $R_{\text{lat}}$, for other types of momenta are given in supplementary material~\cite{SupplementaryMaterial}.


\begin{table}[b]
\begin{ruledtabular}
\begin{tabular}{cc|ccc}
Little Group & $\vec{p}_{\text{ref}}$ & $\phi$ & $\theta$ & $\psi$  \\
\hline
\hline
$\text{Dic}_4$ & $[0,0,n]$ & $0$ & $0$ & $0$ \\
$\text{Dic}_2$ & $[0,n,n]$ & $\pi/2$ & $\pi/4$ & $-\pi/2$ \\
$\text{Dic}_3$ & $[n,n,n]$ & $\pi/4$ & $\cos^{-1}(1/\sqrt{3})$ & $0$ \\
$\text{C}_4$   & $[0,n,2n]$ & $\pi/2$ & $\cos^{-1}(2/\sqrt{5})$ & $0$ \\
$\text{C}_4$   & $[n,n,2n]$ & $-3\pi/4$ & $-\cos^{-1}(\sqrt{2/3})$ & $0$ \\
\end{tabular}
\end{ruledtabular}
\caption{Rotations, $R_{\text{ref}}$, used, as described in the text; here $n$ is a non-zero integer.}
\label{table:refrotations}
\end{table}

\begin{table*}
\begin{ruledtabular}
\begin{tabular}{c c | c c c c c c c c}
Little Group & $\vec{p}_{\text{ref}}$ & $[n, 0, 0]$ & $[0, n, 0]$ & $[0, 0, n]$ & $[-n, 0, 0]$ & $[0, -n, 0]$ & $[0, 0, -n]$ \\
\hline
\hline
$\text{Dic}_4$ & $[0,0,n]$ & 
$\left(
\begin{smallmatrix}
 0 & 0 & 1 \\
 0 & 1 & 0 \\
 -1 & 0 & 0 \\
\end{smallmatrix}
\right)$ & 
 $\left(
\begin{smallmatrix}
 0 & -1 & 0 \\
 0 & 0 & 1 \\
 -1 & 0 & 0 \\
\end{smallmatrix}
\right)$ & 
 $\left(
\begin{smallmatrix}
 1 & 0 & 0 \\
 0 & 1 & 0 \\
 0 & 0 & 1 \\
\end{smallmatrix}
\right)$ & 
 $\left(
\begin{smallmatrix}
 0 & 0 & -1 \\
 0 & -1 & 0 \\
 -1 & 0 & 0 \\
\end{smallmatrix}
\right)$ & 
 $\left(
\begin{smallmatrix}
 0 & 1 & 0 \\
 0 & 0 & -1 \\
 -1 & 0 & 0 \\
\end{smallmatrix}
\right)$ & 
 $\left(
\begin{smallmatrix}
 -1 & 0 & 0 \\
 0 & 1 & 0 \\
 0 & 0 & -1 \\
\end{smallmatrix}
\right)$ \\
\end{tabular}
\end{ruledtabular}
\caption{Rotations, $R_{\text{lat}}$, used for momenta of type $[0,0,n]$, as described in the text; here $n$ is a non-zero integer.}
\label{table:latrotations:00n}
\end{table*}

In Tables \ref{table:rep_Dic4}, \ref{table:rep_Dic2}, \ref{table:rep_Dic3} and \ref{table:rep_C4} we give our choice of representation matrices for, respectively, the little groups $\text{Dic}_4$, $\text{Dic}_2$, $\text{Dic}_3$ and $\text{C}_4$.  For $\text{Dic}_2$, $\text{Dic}_3$ and $\text{C}_4$, the rotations and reflections refer to a coordinate system which has been transformed using $R_{\text{ref}}$, so that $\vec{p}$ defines the new $z$-axis.  Note that the convention is such that states in little group irrep $\Lambda$ (row $\mu$) transform as $\hat{R} \left| \Lambda \mu \right> = \sum_{\mu'} \Lambda_{\mu' \mu} \left| \Lambda \mu' \right>$.

\begin{table*}
\begin{ruledtabular}
\begin{tabular}{c|cccccccc}
Irrep & $I$ & $R(\pi)$ & $R(3\pi/2)$ & $R(\pi/2)$
& $\Pi$ & $R(\pi)\Pi$ & $R(\pi/2)\Pi$ & $R(3\pi/2)\Pi$ \\ 
\hline
\hline
$A_1$ & 1 & 1 & 1 & 1 & 1 & 1 & 1 & 1 \\
$A_2$ & 1 & 1 & 1 & 1 & -1 & -1 & -1 & -1 \\
$E_2$ &
$\left(
\begin{smallmatrix}
 1 & 0 \\
 0 & 1 \\
\end{smallmatrix}
\right)$ & 
$\left(
\begin{smallmatrix}
 -1 & 0 \\
 0 & -1 \\
\end{smallmatrix}
\right)$ & 
$\left(
\begin{smallmatrix}
 0 & i \\
 i & 0 \\
\end{smallmatrix}
\right)$ & 
$\left(
\begin{smallmatrix}
 0 & -i \\
 -i & 0 \\
\end{smallmatrix}
\right)$ & 
$\left(
\begin{smallmatrix}
 1 & 0 \\
 0 & -1 \\
\end{smallmatrix}
\right)$ & 
$\left(
\begin{smallmatrix}
 -1 & 0 \\
 0 & 1 \\
\end{smallmatrix}
\right)$ & 
$\left(
\begin{smallmatrix}
 0 & -i \\
 i & 0 \\
\end{smallmatrix}
\right)$ & 
$\left(
\begin{smallmatrix}
 0 & i \\
 -i & 0 \\
\end{smallmatrix}
\right)$ \\
$B_1$ & 1 & 1 & -1 & -1 & 1 & 1 & -1 & -1 \\
$B_2$ & 1 & 1 & -1 & -1 & -1 & -1 & 1 & 1 \\
\end{tabular}
\end{ruledtabular}
\caption{Choice of representation matrices for the $\text{Dic}_4$ little group. $I$ denotes the identify transformation, $R(\phi)$ denotes a rotation around the $z$-axis by $\phi$ and $\Pi$ denotes a reflection in the $yz$ plane ($x \rightarrow -x$).  Note that, because we are considering only irreps relevant for integer spin, the representation matrices for $R(\phi+2\pi)$ are the same as those for $R(\phi)$.}
\label{table:rep_Dic4}
\end{table*}

\begin{table}
\begin{ruledtabular}
\begin{tabular}{c|cccc}
Irrep & $I$ & $R(\pi)$ & $\Pi$ & $R(\pi)\Pi$ \\
\hline
\hline
$A_1$ & 1 & 1 & 1 & 1 \\
$A_2$ & 1 & 1 & -1 & -1 \\
$B_1$ & 1 & -1 & 1 & -1 \\
$B_2$ & 1 & -1 & -1 & 1 \\
\end{tabular}
\caption{As Table \ref{table:rep_Dic4} but for the $\text{Dic}_2$ little group.}
\label{table:rep_Dic2}
\end{ruledtabular}
\end{table}

\begin{table*}
\begin{ruledtabular}
\begin{tabular}{c|cccccc}
Irrep & $I$ & $R(2\pi/3)$ & $R(4\pi/3)$ & $R(\pi)\Pi$ & $R(\pi/3)\Pi$ & $R(5\pi/3)\Pi$ \\ 
\hline
\hline
$A_1$ & 1 & 1 & 1 & 1 & 1 & 1 \\
$A_2$ & 1 & 1 & 1 & -1 & -1 & -1 \\
$E_2$ &
$\left(
\begin{smallmatrix}
 1 & 0 \\
 0 & 1 \\
\end{smallmatrix}
\right)$ & 
$\left(
\begin{smallmatrix}
 -\frac{1}{2} & -\frac{i \sqrt{3}}{2} \\
 -\frac{i \sqrt{3}}{2} & -\frac{1}{2} \\
\end{smallmatrix}
\right)$ & 
$\left(
\begin{smallmatrix}
 -\frac{1}{2} & \frac{i \sqrt{3}}{2} \\
 \frac{i \sqrt{3}}{2} & -\frac{1}{2} \\
\end{smallmatrix}
\right)$ & 
$\left(
\begin{smallmatrix}
 -1 & 0 \\
 0 & 1 \\
\end{smallmatrix}
\right)$ & 
$\left(
\begin{smallmatrix}
 \frac{1}{2} & -\frac{i \sqrt{3}}{2} \\
 \frac{i \sqrt{3}}{2} & -\frac{1}{2} \\
\end{smallmatrix}
\right)$ & 
$\left(
\begin{smallmatrix}
 \frac{1}{2} & \frac{i \sqrt{3}}{2} \\
 -\frac{i \sqrt{3}}{2} & -\frac{1}{2} \\
\end{smallmatrix}
\right)$ \\
\end{tabular}
\end{ruledtabular}
\caption{As Table \ref{table:rep_Dic4} but for the $\text{Dic}_3$ little group.}
\label{table:rep_Dic3}
\end{table*}

\begin{table}
\begin{ruledtabular}
\begin{tabular}{c|cc}
Irrep & $I$ & $R(\pi)\Pi$ \\
\hline
\hline
$A_1$ & 1 & 1 \\
$A_2$ & 1 & -1 \\
\end{tabular}
\end{ruledtabular}
\caption{As Table \ref{table:rep_Dic4} but for the $\text{C}_4$ little group.}
\label{table:rep_C4}
\end{table}

\subsection{Clebsch-Gordan coefficients}
\label{app:CGs}

Clebsch-Gordan coefficients for zero total momentum ($\vec{P}=\vec{k}_1+\vec{k}_2=\vec{0}$) with $\vec{k}_1 = \vec{k}_2 = \vec{0}$ are given in Ref.~\cite{Basak:2005aq}.  Example Clebsch-Gordan coefficients for $\vec{P} = \vec{0}$ with $\vec{k}_1 = -\vec{k}_2 \neq \vec{0}$ are presented in Tables \ref{table:CGs:rest:n00}-\ref{table:CGs:rest:nnn}; others are given in supplementary material~\cite{SupplementaryMaterial}.  We show Clebsch-Gordan coefficients for $A_1 \otimes A_1 \rightarrow \Lambda$; those for $A_2 \otimes A_2$ are identical, and those for $A_1 \otimes A_2$ and $A_2 \otimes A_1$ follow by switching the target irrep's parity, $\Lambda^{\pm} \rightarrow \Lambda^{\mp}$.

The cases of non-zero total momentum where $\vec{k}_1 = \vec{0}$ and $\vec{P} = \vec{k}_2 \neq \vec{0}$ (or $\vec{k}_2 = \vec{0}$ and $\vec{P} = \vec{k}_1 \neq \vec{0}$) are trivial because there is only one momentum in the sum.  We have
$$\mathcal{C}\left( \Lambda(\mu)[\vec{P}] \otimes A_1^{+}[0,0,0] \rightarrow \Lambda(\mu)[\vec{P}] \right) = 1 ~,$$
where $\mu$ is the irrep row; those for other target irreps and rows are zero.  In addition, 
$$\mathcal{C}\left( \Lambda(\mu)[\vec{P}] \otimes A_1^{-}[0,0,0] \rightarrow \Lambda'(\mu')[\vec{P}] \right) = 1 ~,$$
where if $\Lambda = A_1, A_2, B_1, B_2$ then $\Lambda' = A_2, A_1, B_2, B_1$ respectively and if $\Lambda(\mu) = E_2(1), E_2(2)$ then $\Lambda'(\mu') = E_2(2), E_2(1)$ respectively (the rows are swapped around).  The coefficients for other target irreps and rows are zero.

For non-zero total momentum and $\vec{k}_1, \vec{k}_2 \neq \vec{0}$ we present some examples of Clebsch-Gordan coefficients in Table \ref{table:CGs:flight:001}; others are given in supplementary material~\cite{SupplementaryMaterial}.  We show Clebsch-Gordan coefficients for $A_1 \otimes A_1 \rightarrow \Lambda$; those for $A_2 \otimes A_2$ are identical, and those for $A_1 \otimes A_2$ and $A_2 \otimes A_1$ follow by replacing the target irrep $\Lambda(\mu) = A_1, A_2, B_1, B_2, E_2(1), E_2(2)$ by $\Lambda(\mu) = A_2, A_1, B_2, B_1, E_2(2), E_2(1)$ respectively.

\begin{table}
\begin{ruledtabular}
\begin{tabular}{cc|cccccc}
$\Lambda$ & $\mu$ & \begin{sideways}$[n, 0, 0][\text{-}n, 0, 0]$\end{sideways} & \begin{sideways}$[0, n, 0][0,\text{-}n, 0]$\end{sideways} & \begin{sideways}$[0, 0, n][0, 0,\text{-}n]$\end{sideways} & \begin{sideways}$[\text{-}n, 0, 0][n, 0, 0]$\end{sideways} & \begin{sideways}$[0,\text{-}n, 0][0, n, 0]$\end{sideways} & \begin{sideways}$[0, 0,\text{-}n][0, 0, n]$\end{sideways} \\
\hline
\hline
$A_1^+$ & 1 & $\frac{1}{\sqrt{6}}$ & $\frac{1}{\sqrt{6}}$ & $\frac{1}{\sqrt{6}}$ & $\frac{1}{\sqrt{6}}$ & $\frac{1}{\sqrt{6}}$ & $\frac{1}{\sqrt{6}}$ \\
\hline
\multirow{2}{*}{$E^+$}
 & 1 & $-\frac{1}{2 \sqrt{3}}$ & $-\frac{1}{2 \sqrt{3}}$ & $\frac{1}{\sqrt{3}}$ & $-\frac{1}{2 \sqrt{3}}$ & $-\frac{1}{2 \sqrt{3}}$ & $\frac{1}{\sqrt{3}}$ \\
 & 2 & $\frac{1}{2}$ & $-\frac{1}{2}$ & $0$ & $\frac{1}{2}$ & $-\frac{1}{2}$ & $0$ \\
\hline
\multirow{3}{*}{$T_1^-$}
 & 1 & $-\frac{1}{2}$ & $-\frac{i}{2}$ & $0$ & $\frac{1}{2}$ & $\frac{i}{2}$ & $0$ \\
 & 2 & $0$ & $0$ & $\frac{1}{\sqrt{2}}$ & $0$ & $0$ & $-\frac{1}{\sqrt{2}}$ \\
 & 3 & $\frac{1}{2}$ & $-\frac{i}{2}$ & $0$ & $-\frac{1}{2}$ & $\frac{i}{2}$ & $0$ \\
\end{tabular}
\end{ruledtabular}
\caption{Clebsch-Gordan coefficients for $A_1[0,0,n] \otimes A_1[0,0,n] \rightarrow \Lambda[0,0,0]$ where $\mu$ is the row of $\Lambda$ and $n$ is a non-zero integer.}
\label{table:CGs:rest:n00}
\end{table}

\begin{table*}
\begin{ruledtabular}
\begin{tabular}{cc|cccccccccccc}
  $\Lambda$ & $\mu$ & \begin{sideways}$[n, n, 0][\text{-}n,\text{-}n, 0]$\end{sideways} & \begin{sideways}$[0, n, n][0,\text{-}n,\text{-}n]$\end{sideways} & \begin{sideways}$[n, 0, n][\text{-}n, 0,\text{-}n]$\end{sideways} & \begin{sideways}$[n,\text{-}n, 0][\text{-}n, n, 0]$\end{sideways} & \begin{sideways}$[0, n,\text{-}n][0,\text{-}n, n]$\end{sideways} & \begin{sideways}$[\text{-}n, 0, n][n, 0,\text{-}n]$\end{sideways} & \begin{sideways}$[\text{-}n, n, 0][n,\text{-}n, 0]$\end{sideways} & \begin{sideways}$[0,\text{-}n, n][0, n,\text{-}n]$\end{sideways} & \begin{sideways}$[n, 0,\text{-}n][\text{-}n, 0, n]$\end{sideways} & \begin{sideways}$[\text{-}n,\text{-}n, 0][n, n, 0]$\end{sideways} & \begin{sideways}$[0,\text{-}n,\text{-}n][0, n, n]$\end{sideways} & \begin{sideways}$[\text{-}n, 0,\text{-}n][n, 0, n]$\end{sideways} \\
\hline
\hline
$A_1^+$ & 1 & $\frac{1}{\sqrt{12}}$ & $\frac{1}{\sqrt{12}}$ & $\frac{1}{\sqrt{12}}$ & $\frac{1}{\sqrt{12}}$ & $\frac{1}{\sqrt{12}}$ & $\frac{1}{\sqrt{12}}$ & $\frac{1}{\sqrt{12}}$ & $\frac{1}{\sqrt{12}}$ & $\frac{1}{\sqrt{12}}$ & $\frac{1}{\sqrt{12}}$ & $\frac{1}{\sqrt{12}}$ & $\frac{1}{\sqrt{12}}$ \\
\hline
\multirow{2}{*}{$E^+$}
 & 1 & $-\frac{1}{\sqrt{6}}$ & $\frac{1}{2 \sqrt{6}}$ & $\frac{1}{2 \sqrt{6}}$ & $-\frac{1}{\sqrt{6}}$ & $\frac{1}{2 \sqrt{6}}$ & $\frac{1}{2 \sqrt{6}}$ & $-\frac{1}{\sqrt{6}}$ & $\frac{1}{2 \sqrt{6}}$ & $\frac{1}{2 \sqrt{6}}$ & $-\frac{1}{\sqrt{6}}$ & $\frac{1}{2 \sqrt{6}}$ & $\frac{1}{2 \sqrt{6}}$ \\
 & 2 & $0$ & $-\frac{1}{2 \sqrt{2}}$ & $\frac{1}{2 \sqrt{2}}$ & $0$ & $-\frac{1}{2 \sqrt{2}}$ & $\frac{1}{2 \sqrt{2}}$ & $0$ & $-\frac{1}{2 \sqrt{2}}$ & $\frac{1}{2 \sqrt{2}}$ & $0$ & $-\frac{1}{2 \sqrt{2}}$ & $\frac{1}{2 \sqrt{2}}$ \\
\hline
\multirow{3}{*}{$T_2^+$}
 & 1 & $0$ & $-\frac{i}{2 \sqrt{2}}$ & $-\frac{1}{2 \sqrt{2}}$ & $0$ & $\frac{i}{2 \sqrt{2}}$ & $\frac{1}{2 \sqrt{2}}$ & $0$ & $\frac{i}{2 \sqrt{2}}$ & $\frac{1}{2 \sqrt{2}}$ & $0$ & $-\frac{i}{2 \sqrt{2}}$ & $-\frac{1}{2 \sqrt{2}}$ \\
 & 2 & $\frac{i}{2}$ & $0$ & $0$ & $-\frac{i}{2}$ & $0$ & $0$ & $-\frac{i}{2}$ & $0$ & $0$ & $\frac{i}{2}$ & $0$ & $0$ \\
 & 3 & $0$ & $-\frac{i}{2 \sqrt{2}}$ & $\frac{1}{2 \sqrt{2}}$ & $0$ & $\frac{i}{2 \sqrt{2}}$ & $-\frac{1}{2 \sqrt{2}}$ & $0$ & $\frac{i}{2 \sqrt{2}}$ & $-\frac{1}{2 \sqrt{2}}$ & $0$ & $-\frac{i}{2 \sqrt{2}}$ & $\frac{1}{2 \sqrt{2}}$ \\
\hline
\multirow{3}{*}{$T_1^-$}
 & 1 & $-\frac{1}{4}-\frac{i}{4}$ & $-\frac{i}{4}$ & $-\frac{1}{4}$ & $-\frac{1}{4}+\frac{i}{4}$ & $-\frac{i}{4}$ & $\frac{1}{4}$ & $\frac{1}{4}-\frac{i}{4}$ & $\frac{i}{4}$ & $-\frac{1}{4}$ & $\frac{1}{4}+\frac{i}{4}$ & $\frac{i}{4}$ & $\frac{1}{4}$ \\
 & 2 & $0$ & $\frac{1}{2 \sqrt{2}}$ & $\frac{1}{2 \sqrt{2}}$ & $0$ & $-\frac{1}{2 \sqrt{2}}$ & $\frac{1}{2 \sqrt{2}}$ & $0$ & $\frac{1}{2 \sqrt{2}}$ & $-\frac{1}{2 \sqrt{2}}$ & $0$ & $-\frac{1}{2 \sqrt{2}}$ & $-\frac{1}{2 \sqrt{2}}$ \\
 & 3 & $\frac{1}{4}-\frac{i}{4}$ & $-\frac{i}{4}$ & $\frac{1}{4}$ & $\frac{1}{4}+\frac{i}{4}$ & $-\frac{i}{4}$ & $-\frac{1}{4}$ & $-\frac{1}{4}-\frac{i}{4}$ & $\frac{i}{4}$ & $\frac{1}{4}$ & $-\frac{1}{4}+\frac{i}{4}$ & $\frac{i}{4}$ & $-\frac{1}{4}$ \\
\hline
\multirow{3}{*}{$T_2^-$}
 & 1 & $\frac{1}{4}+\frac{i}{4}$ & $-\frac{i}{4}$ & $-\frac{1}{4}$ & $\frac{1}{4}-\frac{i}{4}$ & $-\frac{i}{4}$ & $\frac{1}{4}$ & $-\frac{1}{4}+\frac{i}{4}$ & $\frac{i}{4}$ & $-\frac{1}{4}$ & $-\frac{1}{4}-\frac{i}{4}$ & $\frac{i}{4}$ & $\frac{1}{4}$ \\
 & 2 & $0$ & $\frac{1}{2 \sqrt{2}}$ & $-\frac{1}{2 \sqrt{2}}$ & $0$ & $-\frac{1}{2 \sqrt{2}}$ & $-\frac{1}{2 \sqrt{2}}$ & $0$ & $\frac{1}{2 \sqrt{2}}$ & $\frac{1}{2 \sqrt{2}}$ & $0$ & $-\frac{1}{2 \sqrt{2}}$ & $\frac{1}{2 \sqrt{2}}$ \\
 & 3 & $\frac{1}{4}-\frac{i}{4}$ & $\frac{i}{4}$ & $-\frac{1}{4}$ & $\frac{1}{4}+\frac{i}{4}$ & $\frac{i}{4}$ & $\frac{1}{4}$ & $-\frac{1}{4}-\frac{i}{4}$ & $-\frac{i}{4}$ & $-\frac{1}{4}$ & $-\frac{1}{4}+\frac{i}{4}$ & $-\frac{i}{4}$ & $\frac{1}{4}$ \\
\end{tabular}
\end{ruledtabular}
\caption{Clebsch-Gordan coefficients for $A_1[0,n,n] \otimes A_1[0,n,n] \rightarrow \Lambda[0,0,0]$ where $\mu$ is the row of $\Lambda$ and $n$ is a non-zero integer.}
\label{table:CGs:rest:nn0}
\end{table*}

\begin{table*}
\begin{ruledtabular}
\begin{tabular}{cc|cccccccc}
$\Lambda$ & $\mu$ & \begin{sideways}$[n, n, n][\text{-}n,\text{-}n,\text{-}n]$\end{sideways} & \begin{sideways}$[\text{-}n, n, n][n,\text{-}n,\text{-}n]$\end{sideways} & \begin{sideways}$[n,\text{-}n, n][\text{-}n, n,\text{-}n]$\end{sideways} & \begin{sideways}$[n, n,\text{-}n][\text{-}n,\text{-}n, n]$\end{sideways} & \begin{sideways}$[\text{-}n,\text{-}n, n][n, n,\text{-}n]$\end{sideways} & \begin{sideways}$[n,\text{-}n,\text{-}n][\text{-}n, n, n]$\end{sideways} & \begin{sideways}$[\text{-}n, n,\text{-}n][n,\text{-}n, n]$\end{sideways} & \begin{sideways}$[\text{-}n,\text{-}n,\text{-}n][n, n, n]$\end{sideways} \\
\hline
\hline
$A_1^+$ & 1 & $\frac{1}{\sqrt{8}}$ & $\frac{1}{\sqrt{8}}$ & $\frac{1}{\sqrt{8}}$ & $\frac{1}{\sqrt{8}}$ & $\frac{1}{\sqrt{8}}$ & $\frac{1}{\sqrt{8}}$ & $\frac{1}{\sqrt{8}}$ & $\frac{1}{\sqrt{8}}$ \\
\hline
\multirow{3}{*}{$T_2^+$}
 & 1 & $-\frac{1}{4}-\frac{i}{4}$ & $\frac{1}{4}-\frac{i}{4}$ & $-\frac{1}{4}+\frac{i}{4}$ & $\frac{1}{4}+\frac{i}{4}$ & $\frac{1}{4}+\frac{i}{4}$ & $\frac{1}{4}-\frac{i}{4}$ & $-\frac{1}{4}+\frac{i}{4}$ & $-\frac{1}{4}-\frac{i}{4}$ \\
 & 2 & $\frac{i}{2 \sqrt{2}}$ & $-\frac{i}{2 \sqrt{2}}$ & $-\frac{i}{2 \sqrt{2}}$ & $\frac{i}{2 \sqrt{2}}$ & $\frac{i}{2 \sqrt{2}}$ & $-\frac{i}{2 \sqrt{2}}$ & $-\frac{i}{2 \sqrt{2}}$ & $\frac{i}{2 \sqrt{2}}$ \\
 & 3 & $\frac{1}{4}-\frac{i}{4}$ & $-\frac{1}{4}-\frac{i}{4}$ & $\frac{1}{4}+\frac{i}{4}$ & $-\frac{1}{4}+\frac{i}{4}$ & $-\frac{1}{4}+\frac{i}{4}$ & $-\frac{1}{4}-\frac{i}{4}$ & $\frac{1}{4}+\frac{i}{4}$ & $\frac{1}{4}-\frac{i}{4}$ \\
\hline
$A_2^-$ & 1 & $\frac{i}{2 \sqrt{2}}$ & $-\frac{i}{2 \sqrt{2}}$ & $-\frac{i}{2 \sqrt{2}}$ & $-\frac{i}{2 \sqrt{2}}$ & $\frac{i}{2 \sqrt{2}}$ & $\frac{i}{2 \sqrt{2}}$ & $\frac{i}{2 \sqrt{2}}$ & $-\frac{i}{2 \sqrt{2}}$ \\
\hline
\multirow{3}{*}{$T_1^-$}
 & 1 & $-\frac{1}{4}-\frac{i}{4}$ & $\frac{1}{4}-\frac{i}{4}$ & $-\frac{1}{4}+\frac{i}{4}$ & $-\frac{1}{4}-\frac{i}{4}$ & $\frac{1}{4}+\frac{i}{4}$ & $-\frac{1}{4}+\frac{i}{4}$ & $\frac{1}{4}-\frac{i}{4}$ & $\frac{1}{4}+\frac{i}{4}$ \\
 & 2 & $\frac{1}{2 \sqrt{2}}$ & $\frac{1}{2 \sqrt{2}}$ & $\frac{1}{2 \sqrt{2}}$ & $-\frac{1}{2 \sqrt{2}}$ & $\frac{1}{2 \sqrt{2}}$ & $-\frac{1}{2 \sqrt{2}}$ & $-\frac{1}{2 \sqrt{2}}$ & $-\frac{1}{2 \sqrt{2}}$ \\
 & 3 & $\frac{1}{4}-\frac{i}{4}$ & $-\frac{1}{4}-\frac{i}{4}$ & $\frac{1}{4}+\frac{i}{4}$ & $\frac{1}{4}-\frac{i}{4}$ & $-\frac{1}{4}+\frac{i}{4}$ & $\frac{1}{4}+\frac{i}{4}$ & $-\frac{1}{4}-\frac{i}{4}$ & $-\frac{1}{4}+\frac{i}{4}$ \\
\end{tabular}
\caption{Clebsch-Gordan coefficients for $A_1[n,n,n] \otimes A_1[n,n,n] \rightarrow \Lambda[0,0,0]$ where $\mu$ is the row of $\Lambda$ and $n$ is a non-zero integer.}
\label{table:CGs:rest:nnn}
\end{ruledtabular}
\end{table*}

\begin{table*}
\parbox{.45\linewidth}{
\begin{ruledtabular}
\begin{tabular}{cc|cccc}
$\Lambda$ & $\mu$ & \begin{sideways}$[n, 0, 0][\text{-}n, 0,n]$\end{sideways} & \begin{sideways}$[0,n, 0][0,\text{-}n,n]$\end{sideways} & \begin{sideways}$[\text{-}n, 0, 0][n, 0,n]$\end{sideways} & \begin{sideways}$[0,\text{-}n, 0][0,n,n]$\end{sideways} \\
\hline
\hline
$A_1$ & 1 & $\frac{1}{2}$ & $\frac{1}{2}$ & $\frac{1}{2}$ & $\frac{1}{2}$ \\
\hline
\multirow{2}{*}{$E_2$}
 & 1 & $0$ & $-\frac{i}{\sqrt{2}}$ & $0$ & $\frac{i}{\sqrt{2}}$ \\
 & 2 & $-\frac{1}{\sqrt{2}}$ & $0$ & $\frac{1}{\sqrt{2}}$ & $0$ \\
\hline
$B_1$ & 1 & $-\frac{1}{2}$ & $\frac{1}{2}$ & $-\frac{1}{2}$ & $\frac{1}{2}$ \\
\end{tabular}
\end{ruledtabular}
}
\hfill
\parbox{.45\linewidth}{
\begin{ruledtabular}
\begin{tabular}{cc|cccc}
$\Lambda$ & $\mu$ & \begin{sideways}$[n, n, 0][\text{-}n,\text{-}n, n]$\end{sideways} & \begin{sideways}$[n,\text{-}n, 0][\text{-}n, n, n]$\end{sideways} & \begin{sideways}$[\text{-}n, n, 0][n,\text{-}n, n]$\end{sideways} & \begin{sideways}$[\text{-}n,\text{-}n, 0][n, n, n]$\end{sideways} \\
\hline
\hline
$A_1$ & 1 & $\frac{1}{2}$ & $\frac{1}{2}$ & $\frac{1}{2}$ & $\frac{1}{2}$ \\
\hline
\multirow{2}{*}{$E_2$}
 & 1 & $-\frac{i}{2}$ & $\frac{i}{2}$ & $-\frac{i}{2}$ & $\frac{i}{2}$ \\
 & 2 & $-\frac{1}{2}$ & $-\frac{1}{2}$ & $\frac{1}{2}$ & $\frac{1}{2}$ \\
\hline
$B_2$ & 1 & $\frac{i}{2}$ & $-\frac{i}{2}$ & $-\frac{i}{2}$ & $\frac{i}{2}$ \\
\end{tabular}
\end{ruledtabular}
}
\caption{Clebsch-Gordan coefficients for $A_1[0,-n,0] \otimes A_1[0,n,n] \rightarrow \Lambda[0,0,n]$ (left) and $A_1[-n,-n,0] \otimes A_1[n,n,n] \rightarrow \Lambda[0,0,n]$ (right) where $\mu$ is the row of $\Lambda$ and $n$ is a non-zero integer.}
\label{table:CGs:flight:001}
\end{table*}


\end{document}